\documentclass[acmsmall]{acmart}
\AtBeginDocument{%
  }

\usepackage{mathtools}   
\usepackage{bm}          
\DeclarePairedDelimiter{\norm}{\lVert}{\rVert} 

\usepackage{subfig}
\usepackage{multirow}
\begin{document}

\title{Enhanced Quality Aware-Scalable Underwater Image Compression}

\author{Linwei Zhu}
\affiliation{%
  \institution{Shenzhen Institutes of Advanced Technology, Chinese Academy of Sciences}
  \city{Shenzhen}
  \country{China}}
\email{lw.zhu@siat.ac.cn}

\author{Junhao Zhu}
\affiliation{
 \institution{Southern University of Science and Technology}
  \city{Shenzhen}
  \country{China}}
\email{jh.zhu1@siat.ac.cn}

\author{Xu Zhang}
\affiliation{
\institution{School of Computer Science, Wuhan University}
  \city{Wuhan}
  \country{China}
}
\email{zhangx0802@whu.edu.cn}

\author{Huan Zhang}
\affiliation{
\institution{School of Information Engineering, Guangdong University of Technology}
 \city{Guangzhou}
 \country{China}}
 \email{huanzhang2021@gdut.edu.cn}

\author{Ye Li}
\affiliation{
\institution{Shenzhen Institutes of Advanced Technology, Chinese Academy of Sciences}
  \city{Shenzhen}
  \country{China}}
  \email{ye.li@siat.ac.cn}

\author{Runmin Cong}
\affiliation{
\institution{School of Control Science and Engineering, Shandong University}
  \city{Jinan}
  \country{China}}
\email{rmcong@sdu.edu.cn}

\author{Sam Kwong}
\affiliation{
\institution{School of Data Sciences, Lingnan University}
  \city{Hongkong}
  \country{China}}
\email{samkwong@ln.edu.hk}

\renewcommand{\shortauthors}{L. Zhu et al.}

\begin{abstract}
Underwater imaging plays a pivotal role in marine exploration and ecological monitoring. However, it faces significant challenges of limited transmission bandwidth and severe distortion in the aquatic environment. In this work, to achieve the target of both underwater image compression and enhancement simultaneously, an enhanced quality-aware scalable underwater image compression framework is presented, which comprises a Base Layer (BL) and an Enhancement Layer (EL). In the BL, the underwater image is represented by controllable number of non-zero sparse coefficients for coding bits saving. Furthermore, the underwater image enhancement dictionary is derived with shared sparse coefficients to make reconstruction close to the enhanced version. In the EL, a dual-branch filter comprising rough filtering and detail refinement branches is designed to produce a pseudo-enhanced version for residual redundancy removal and to improve the quality of final reconstruction. Extensive experimental results demonstrate that the proposed scheme outperforms the state-of-the-art works under five large-scale underwater image datasets in terms of Underwater Image Quality Measure (UIQM).
\end{abstract}






\maketitle

\section{Introduction}
Underwater imaging \cite{10.1145/3656473} is pivotal for marine exploration and ecological monitoring. However, it faces significant challenges due to the aquatic environment's inherent limitations. On the one hand, unlike terrestrial settings where radio waves propagate freely, seawater severely attenuates electromagnetic signals, making them impractical beyond short distances. Consequently, acoustic communication becomes the primary method. Yet, sound propagation suffers from inherent bandwidth constraints due to absorption, scattering, and multi-path effects. This drastically restricts achievable data for underwater systems. On the other hand, optical imaging is severely hampered by light attenuation, which follows wavelength-dependent exponential decay. Longer wavelengths, particularly red light, are absorbed most rapidly, diminishing within meters. This leads to pronounced blue-green color shifts and a significant loss of contrast and detail in captured images. Furthermore, scattering by suspended particles causes blurring and haze. Together, these effects produce severe image distortion and degradation, complicating tasks like object recognition, color fidelity for biological studies, and precise measurement crucial for many underwater applications, compounding the challenges beyond bandwidth limitations. Therefore, underwater images require efficient compression and enhancement.

To effectively compress images/videos with smaller data size and better reconstruction quality, various codecs are developed in the last decades. The widely used image compression schemes are \textbf{Joint Photographic Experts Group} (\textbf{JPEG}) \cite{JPEG} and JPEG2000 \cite{JPEG2000}. The former is the established lossy image compression standard using \textbf{Discrete Cosine Transform} (\textbf{DCT}), while the latter utilizes wavelet transforms for better artifact handling. For the video compression, \textbf{High Efficiency Video Coding} (\textbf{HEVC}) \cite{HEVC} significantly improves compression efficiency over \textbf{Advanced Video Coding} (\textbf{AVC}) \cite{AVC} (typically 50\% bitrate reduction) using larger coding tree units and advanced prediction methods, while \textbf{Versatile Video Coding} (\textbf{VVC}) \cite{VVC, VVC2} further enhances efficiency (another 50\% bitrate reduction over HEVC \cite{HEVC}) through flexible partitioning and new coding tools for diverse contents and resolutions. With advances in deep learning, end-to-end image/video compression \cite{10.1145/3678472, 10949702,10891533,10247017,10.1145/3708347} has emerged, jointly optimizing the entire pipeline from pixel domain to bitstream, and reconstruction within unified neural architectures. Although these conventional and learning-based methods exhibit significant performance gains for natural images/videos, they are ill-suited for underwater imaging due to domain-specific characteristics (e.g., light attenuation, color distortion). Consequently, dedicated underwater compression research has expanded. Liu et al. \cite{10113175} proposed an autoencoder-based scheme with multistage training to enhance decoder robustness against channel degradation. In \cite{10445326}, it leveraged underwater T-map quantization and mixture entropy coding, aligning latent features with a comprehensive underwater dictionary. An extreme compression framework \cite{10058091} incorporated underwater priors to support dual human/machine vision scalability, prioritizing structural edges and high-level features critical for machine analysis. Li et al. \cite{9934922} developed an underwater physical prior-guided structure to adaptively adjust Gaussian feature distributions. Fang et al. \cite{9984191} utilized contrastive learning to enhance degraded features and compress machine-friendly features at low bitrates. However, compression-induced distortion in underwater images is not fully considered, which may be a challenge for its further enhancement due to the fact that additional artifacts are introduced.

\begin{figure}
    \centering
    \subfloat[]{
        \includegraphics[width=0.6\textwidth]{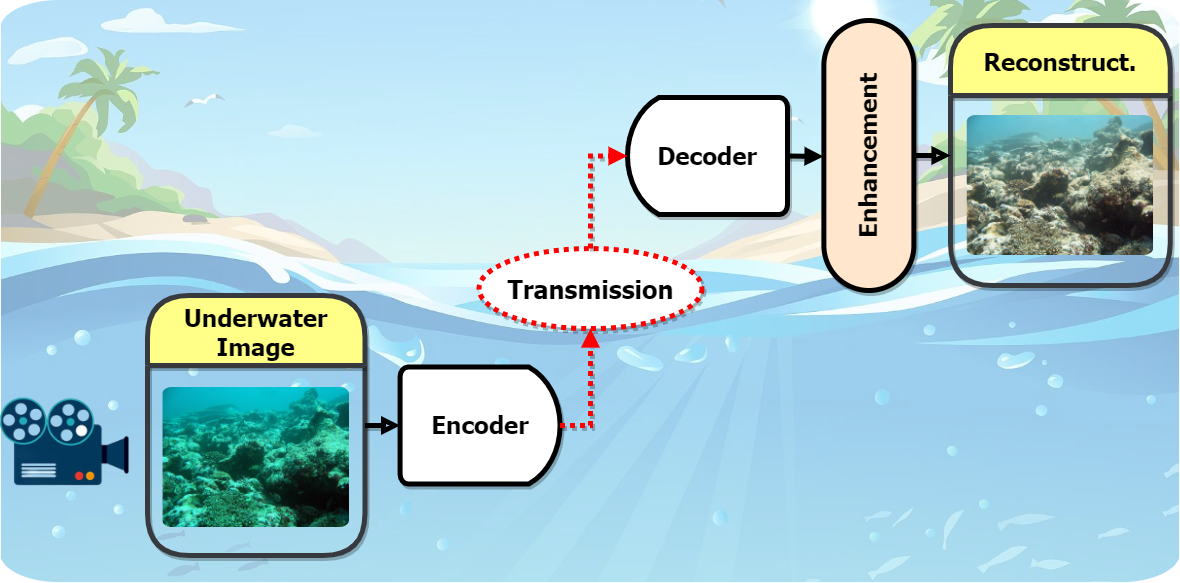}} \\
    \vspace{-0.3cm} 
    \subfloat[]{
        \includegraphics[width=0.6\textwidth]{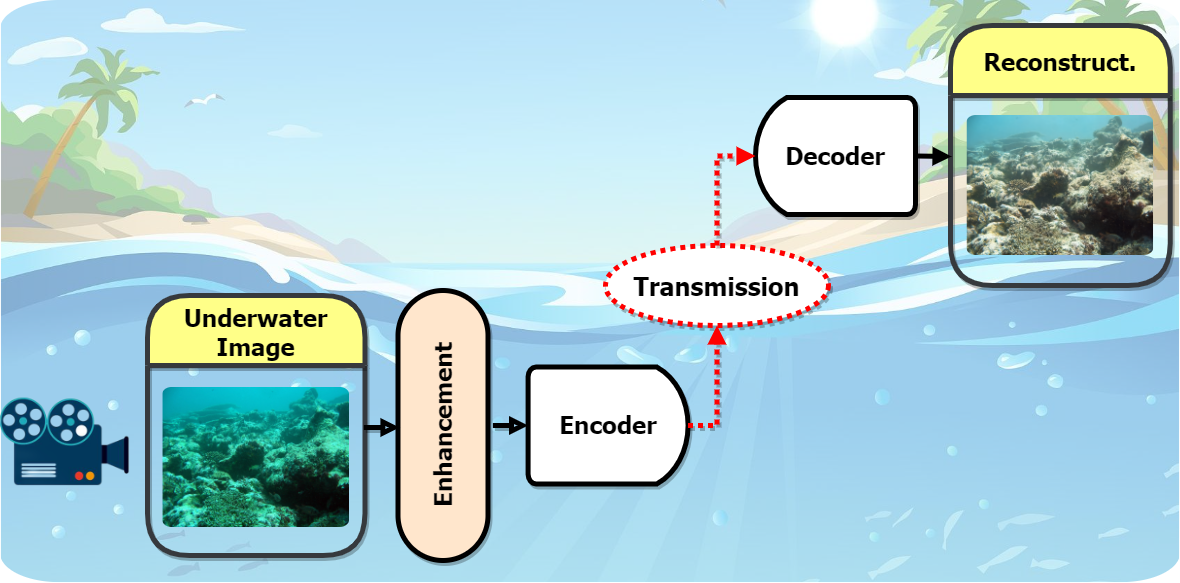}} \\
    \vspace{-0.3cm} 
    \subfloat[]{
        \includegraphics[width=0.6\textwidth]{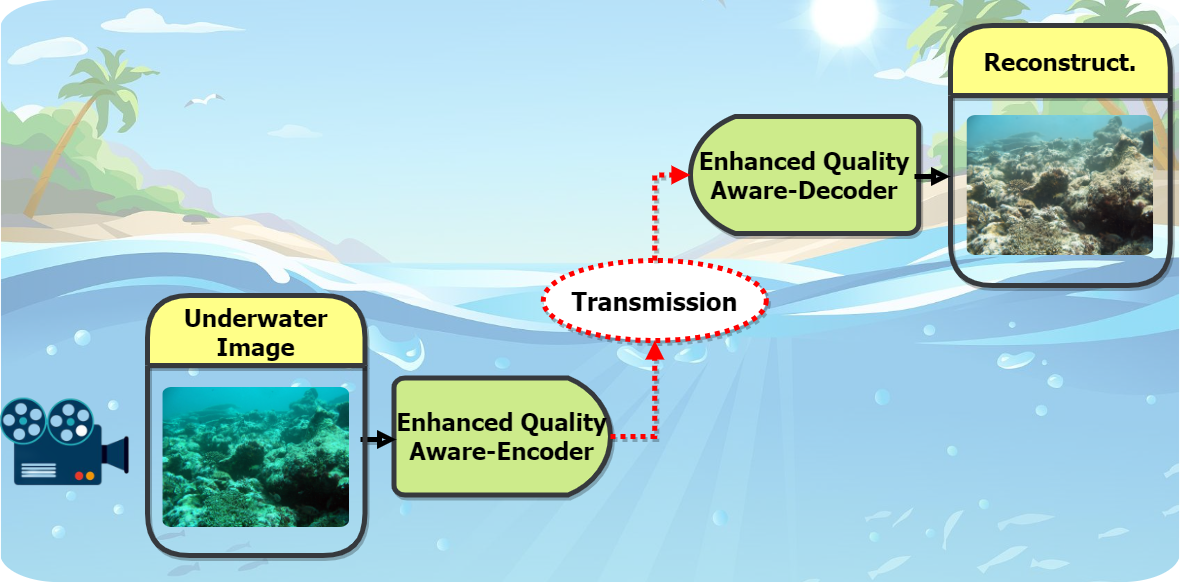}}
    \caption{Three frameworks of compression and enhancement for underwater image. (a) Compression first and enhancement second. (b) Enhancement first and compression second. (c) Proposed framework.  }
    \label{fig:framework_comparison}
\end{figure}

\begin{figure}[!t]
\centering
\includegraphics[width=0.4\textwidth]{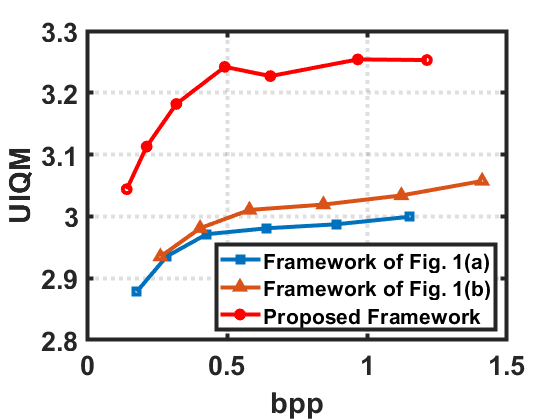}
\caption{Coding performance comparison of three frameworks in Fig. \ref{fig:framework_comparison}. The performance is evaluated by Bits Per Pixel (bpp) and Underwater Image Quality Measure (UIQM) \cite{UIQM}. In this experiment, the dataset UW2023 \cite{10945897} is used, which includes 1000 underwater images in the real-world. MLIC++ \cite{MLIC} and HCLR \cite{HCLR} are employed in Figs. \ref{fig:framework_comparison}(a) and (b) for underwater image compression and enhancement. The proposed framework aims to jointly compress and enhance the underwater image. }
\label{fig:performance_framework}
\end{figure}

For the underwater image enhancement \cite{10.1145/3489520, 10.1145/3688802, 10.1145/3511021}, it has been extensively studied in recent years. Li et al. \cite{7574330} proposed a systematic method combining dehazing and contrast enhancement algorithms. To address color deviation and low visibility caused by wavelength-dependent light absorption and scattering, an efficient and robust enhancement framework was introduced in \cite{9788535}. The first comprehensive perceptual analysis of underwater enhancement using large-scale real-world images was conducted by \cite{UIEB}, which established the \textbf{Underwater Image Enhancement Benchmark} (\textbf{UIEB}). In \cite{10196309}, a weighted wavelet visual perception fusion strategy was developed to integrate global and local contrast enhanced images by fusing multi-scale high- and low-frequency components. Further innovations include a few shot learning method operating in multi-color spaces \cite{10064015}, a two-phase domain adaptation network minimizing inter- and intra-domain gaps for task-oriented enhancement \cite{10048777}, and a real-time denoising diffusion probabilistic model outperforming generative adversarial network based approaches \cite{10250851}. Yu et al. \cite{10347022} designed an unsupervised framework disentangling latent representations based on machine vision utility. Additionally, \cite{10816000} proposed a diffusion color guided framework to correct color deviations in diffusion-based restoration, while \cite{10812849} leveraged global feature priors within a diffusion model architecture. However, more high frequency information will be produced after underwater image enhancement. Consequently, additional bits will be spent if the compression is followed.

To solve these two mentioned problems simultaneously, we present an underwater image compression method that jointly optimizes compression efficiency and enhanced image quality. The contributions of this work are summarized as follows.
\begin{enumerate}
  \item We propose an enhanced quality aware-scalable underwater image compression framework that integrates compression and enhancement into a unified process, consisting of a \textbf{Base Layer} (\textbf{BL}) and an \textbf{Enhancement Layer} (\textbf{EL}). To the best of our knowledge, this is the first work that enables joint optimization of underwater image compression and enhancement.
  \item In the BL, we introduce a controllable sparse representation mechanism, which encodes the underwater image using a limited number of non-zero sparse coefficients to reduce bit consumption. Moreover, an enhancement dictionary derived from shared sparse coefficients is constructed to make the reconstructed image visually closer to the enhanced version.
  \item In the EL, a dual-branch filtering structure is designed, comprising a coarse filtering branch and a detail refinement branch. This module generates a pseudo-enhanced version of underwater image to eliminate residual redundancy and further improve the final reconstruction quality.
\end{enumerate}

\begin{figure*}
    \centering
    \subfloat[Underwater image]{
        \includegraphics[height=0.16\textwidth]{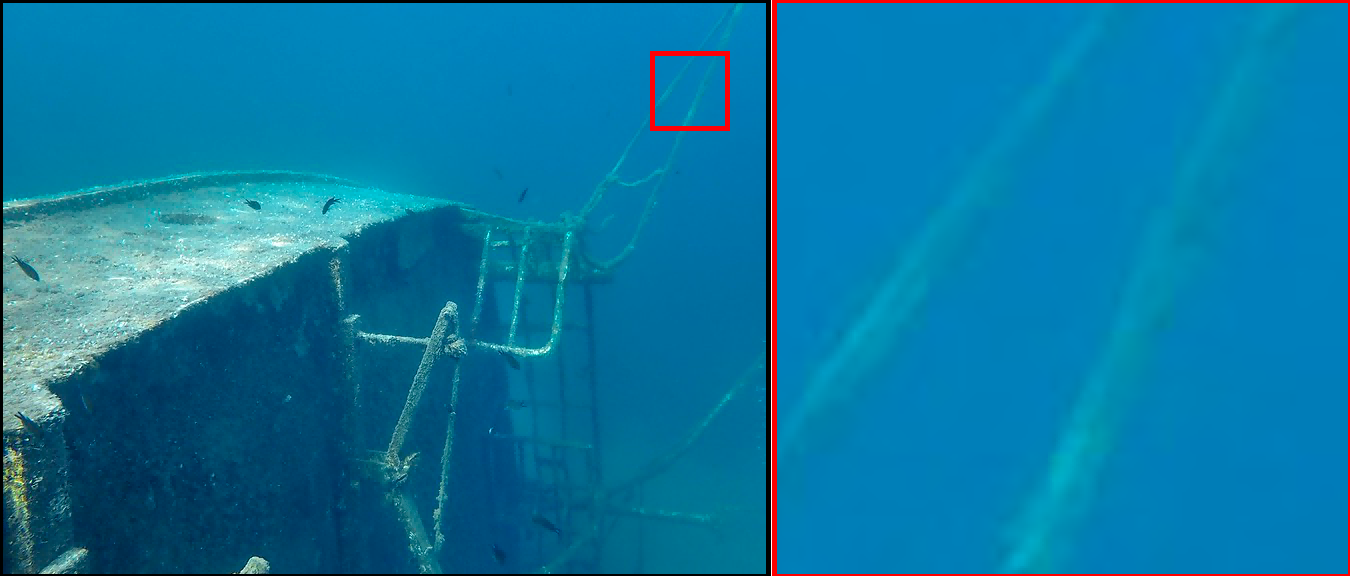}}
    \subfloat[Enhanced version]{
        \includegraphics[height=0.16\textwidth]{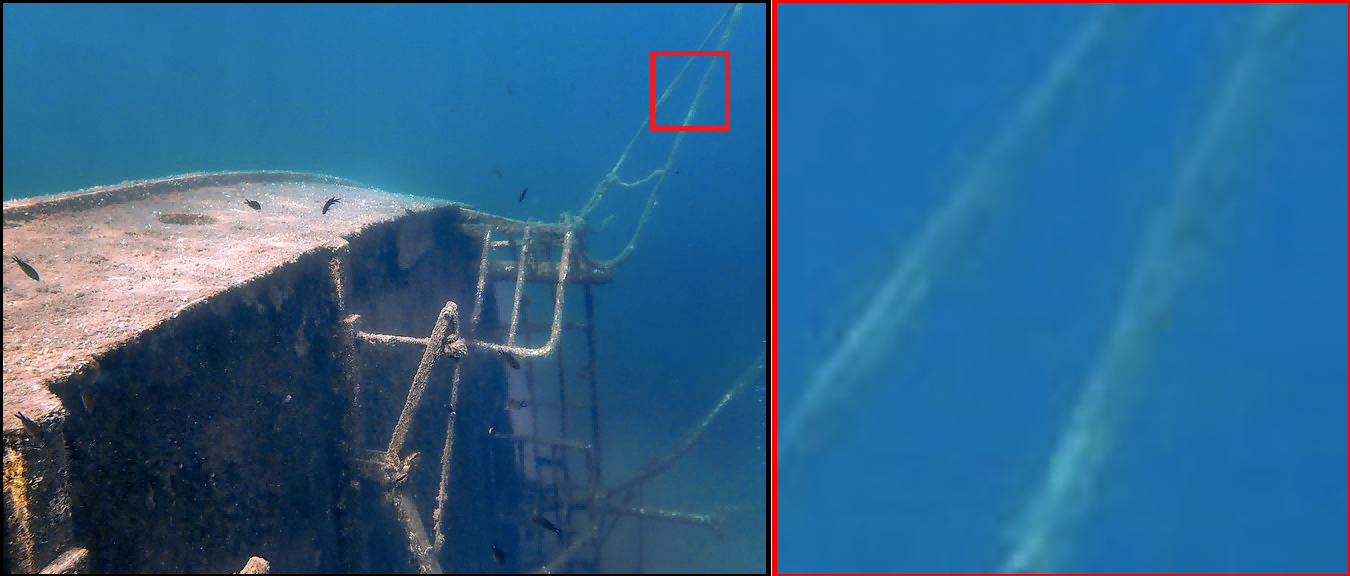}} \\
    \vspace{-0.3cm} 
    \subfloat[{L1: 0.04/1.629}]{
        \includegraphics[height=0.15\textwidth]{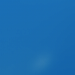}}
    \subfloat[{L2: 0.08/1.703}]{
        \includegraphics[height=0.15\textwidth]{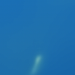}}
    \subfloat[{L3: 0.13/1.786}]{
        \includegraphics[height=0.15\textwidth]{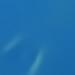}}
    \subfloat[{L4: 0.22/1.865}]{
        \includegraphics[height=0.15\textwidth]{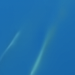}}
    \subfloat[{L5: 0.32/1.943}]{
        \includegraphics[height=0.15\textwidth]{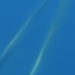}}
    \subfloat[{L6: 0.43/1.984}]{
        \includegraphics[height=0.15\textwidth]{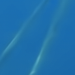}} \\
    \vspace{-0.3cm} 
    \subfloat[{L1: 0.06/1.722}]{
        \includegraphics[height=0.15\textwidth]{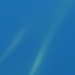}}
    \subfloat[{L2: 0.11/1.783}]{
        \includegraphics[height=0.15\textwidth]{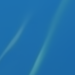}}
    \subfloat[{L3: 0.18/1.838}]{
        \includegraphics[height=0.15\textwidth]{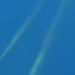}}
    \subfloat[{L4: 0.27/1.893}]{
        \includegraphics[height=0.15\textwidth]{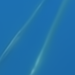}}
    \subfloat[{L5: 0.39/1.942}]{
        \includegraphics[height=0.15\textwidth]{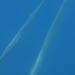}}
    \subfloat[{L6: 0.52/1.995}]{
        \includegraphics[height=0.15\textwidth]{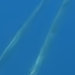}}
    \caption{Underwater image and enhanced versions under different cases. (a) Original underwater image. (b) Enhanced version of original underwater image without compression. (c)-(h) Reconstructed versions under framework of Fig. 1(a) with compression levels from Level 1 to Level 6.  (i)-(n) Reconstructed versions under framework of Fig. 1(b) with compression levels from Level 1 to Level 6. The values indicate bpp and UIQM, respectively.}
    \label{fig:framework_mot}
\end{figure*}

\section{Motivation and Problem Formulation}
There are two straightforward frameworks to achieve the target of both enhancement and compression simultaneously for underwater images. One is compression first and enhancement second, while the other is enhancement first and compression second, which are vividly illustrated in Figs. \ref{fig:framework_comparison}(a) and (b). For the first framework, as shown in Fig. \ref{fig:framework_comparison}(a), the underwater images are compressed during the stage of acquisition, and transmitted to the end user, where the module of enhancement is performed before watch and analysis. However, the compression distortion may make the enhancement failure, especially for the case of low bitrate. These compression distortions that have different levels are ignored by most of the enhancement algorithms. For the second framework, as shown in Fig. \ref{fig:framework_comparison}(b), the underwater images are directly enhanced after acquisition, then compressed and transmitted to the end user. However, in the enhanced underwater image, more detailed information is recovered, which will waste additional bits for compression. 

To make it more clear, extensive experiments are conducted. The state-of-the-art end to end image compression method MLIC++ \cite{MLIC} and underwater image enhancement method HCLR \cite{HCLR} are employed in these two frameworks. The performance is evaluated by \textbf{Bits Per Pixel} (\textbf{bpp}) and \textbf{Underwater Image Quality Measure} (\textbf{UIQM}) \cite{UIQM} for underwater image. In this experiment, two underwater image datasets are used, including UW2023 \cite{10945897} and EUVP-Dark \cite{EUVP}. The UW2023 dataset contains 1000 real-world underwater images without corresponding enhanced ground truth. The dataset of EUVP-Dark with 5500 underwater images has the enhanced versions of ground truth. The performance comparison of these frameworks is presented in Fig. \ref{fig:framework_mot} and Table \ref{table1}. Fig. \ref{fig:framework_mot} illustrates the reconstructed underwater images of the first and second frameworks under different compression levels. Moveover, the values of bpp and UIQM are provided as well. From Figs. \ref{fig:framework_mot}(c)-(h), it can be found that the compression distortion plays a negative role for the underwater image enhancement, especially at the low bitrate. While the second framework achieves better visual results at the same compression level, but it spends more bits, which can also be observed from Table \ref{table1}, 0.06$\sim$0.26 bits are additionally required. The reason is that there are more high frequency information in these underwater images after enhancement.

\begin{table}[t]\caption{Coding bits comparison of two frameworks under different compression levels in terms of bpp.} \label{table1}
\footnotesize
\begin{center}
    \begin{tabular}{|c|c|c|c|c|}
    \cline {1-5} 
    \multirow{2}{*}{Level}&\multicolumn{2}{c|}{UW2023}&\multicolumn{2}{c|}{EUVP-Dark}\\
    \cline {2-5}
    &{Fig.1(a)}&{Fig.1(b)}&{Fig.1(a)}&{Fig.1(b)}\\
    \hline
    {1}      &0.17&0.26&0.19&0.27\\
    {2}      &0.28&0.40&0.30&0.41\\
    {3}      &0.42&0.58&0.45&0.58\\
    {4}      &0.64&0.84&0.67&0.84\\
    {5}      &0.88&1.12&0.90&1.09\\
    {6}      &1.15&1.41&1.19&1.40\\
    \cline{1-5}
    {\textbf{AVG}}&\textbf{0.59}&\textbf{0.77}&\textbf{0.62}&\textbf{0.76}\\
    \hline
   \end{tabular}
\end{center}
\end{table}

Therefore, to address the limitations of the above two frameworks, the third one is proposed in this work, as shown in Fig. \ref{fig:framework_comparison}(c), which aims to encode and decode underwater image with enhanced quality-aware codec for performance improvement. Suppose the underwater image is $\textbf{I}_i$, $i \in [1, \mathcal{K}]$, this task can be formulated as follows,
\begin{align}
\{\mathbb{C}^*,\, \mathbb{D}^* \} \leftarrow \arg \min_{\{\mathbb{C},\, \mathbb{D}\}} 
\Biggl\{ 
\sum_{i=1}^\mathcal{K} \norm*{\mathbb{E}(\mathbf{I}_i) - \mathbf{O}_i}^2 
+ \lambda \sum_{i=1}^\mathcal{K} \mathcal{R}_i 
\Biggr\},
\end{align}
where $\mathbb{C}^*$() and $\mathbb{D}^*$() are the optimal enhanced quality-aware underwater image encoder and decoder, $\mathbb{E}$() is the underwater image enhancement operator, $\textbf{O}_i$ is the reconstructed underwater image under this proposed framework, $\mathcal{R}_i$ is the value of bpp after compression, $\lambda$ is a weight for the balance between bit and distortion, and $\mathcal{K}$ is the number of training underwater images. The stream $\textbf{S}_i$ and the value of bpp $\mathcal{R}_i$ can be achieved after compression, i.e., $\{\textbf{S}_i, \mathcal{R}_i\} \leftarrow \mathbb{C}(\textbf{I}_i, Q)$, $Q$ is the quantization parameter. The reconstructed underwater image $\textbf{O}_i$ can be achieved after decoding, i.e., $\textbf{O}_i \leftarrow \mathbb{D}(\textbf{S}_i)$.

\begin{figure*}[!t]
\centering
\includegraphics[width=0.98\textwidth]{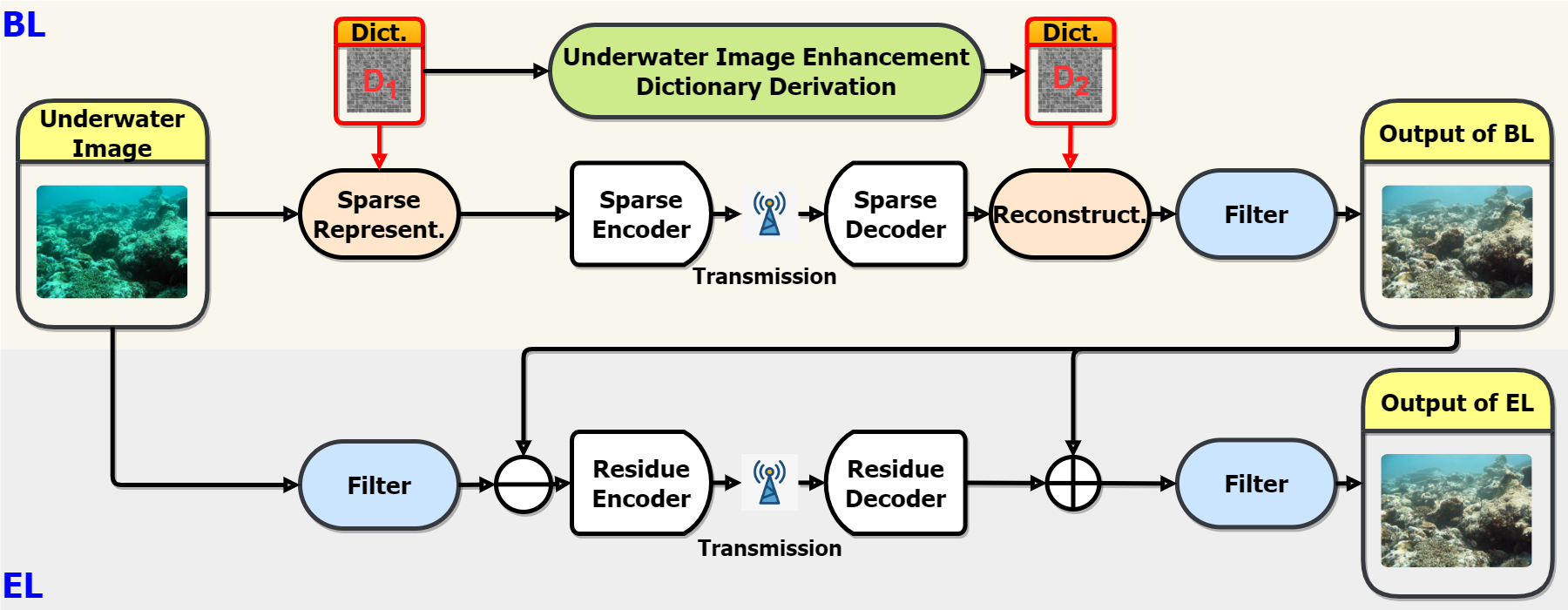}
\caption{Framework of the proposed enhanced quality aware-scalable underwater image compression method. ``EL" indicates enhancement layer and ``BL" indicates base layer. }
\label{fig:framework}
\end{figure*}

\section{Proposed Enhanced Quality Aware-Underwater Image Compression Method}

As shown in Fig. \ref{fig:framework}, it is the framework of the proposed enhanced quality aware-underwater image compression method. This is a scalable architecture to adapt the unstable transmission in the underwater environment, which includes BL and EL. In the BL, the input underwater image is firstly represented by sparse coefficients, where the number of non-zero sparse coefficients can be changed for the consideration of redundancy removal. These sparse coefficients are encoded by sparse codec and transmitted to the decoder side. At the decoder side, the underwater image is reconstructed with the decoded sparse coefficients and an enhanced dictionary. Moreover, a dual-branch filter is employed to improve the performance of reconstructed underwater image. In the EL, the filtering is firstly performed to generate a pseudo-enhanced version. Then, the residue between the output of BL and pseudo-enhanced version is encoded by residue codec and transmitted to the decoder side. The output of EL can be achieved by filtering the sum of decoded residue and output of BL. The modules in this architecture will be explained in detail as follows.

\begin{figure}[!t]
\centering
\includegraphics[width=0.7\textwidth]{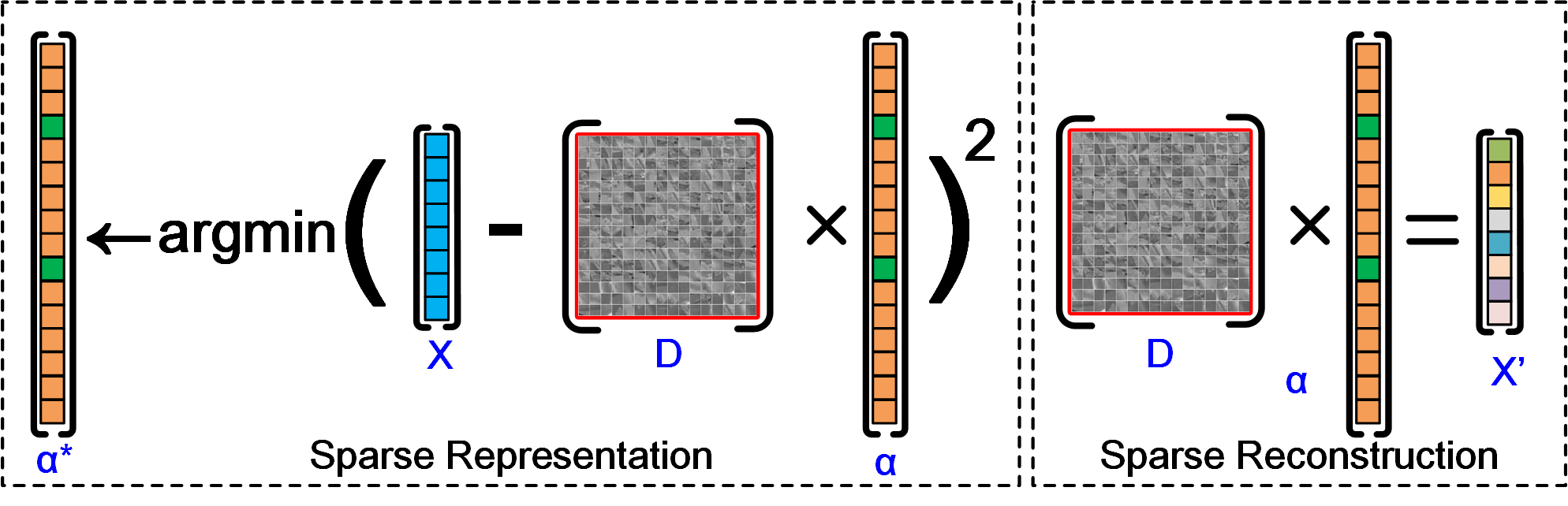}
\caption{Sparse representation and reconstruction. }
\label{fig:sparse_re}
\end{figure}

\begin{figure*}
    \centering
    \subfloat[Red Channel]{
        \includegraphics[width=0.32\textwidth]{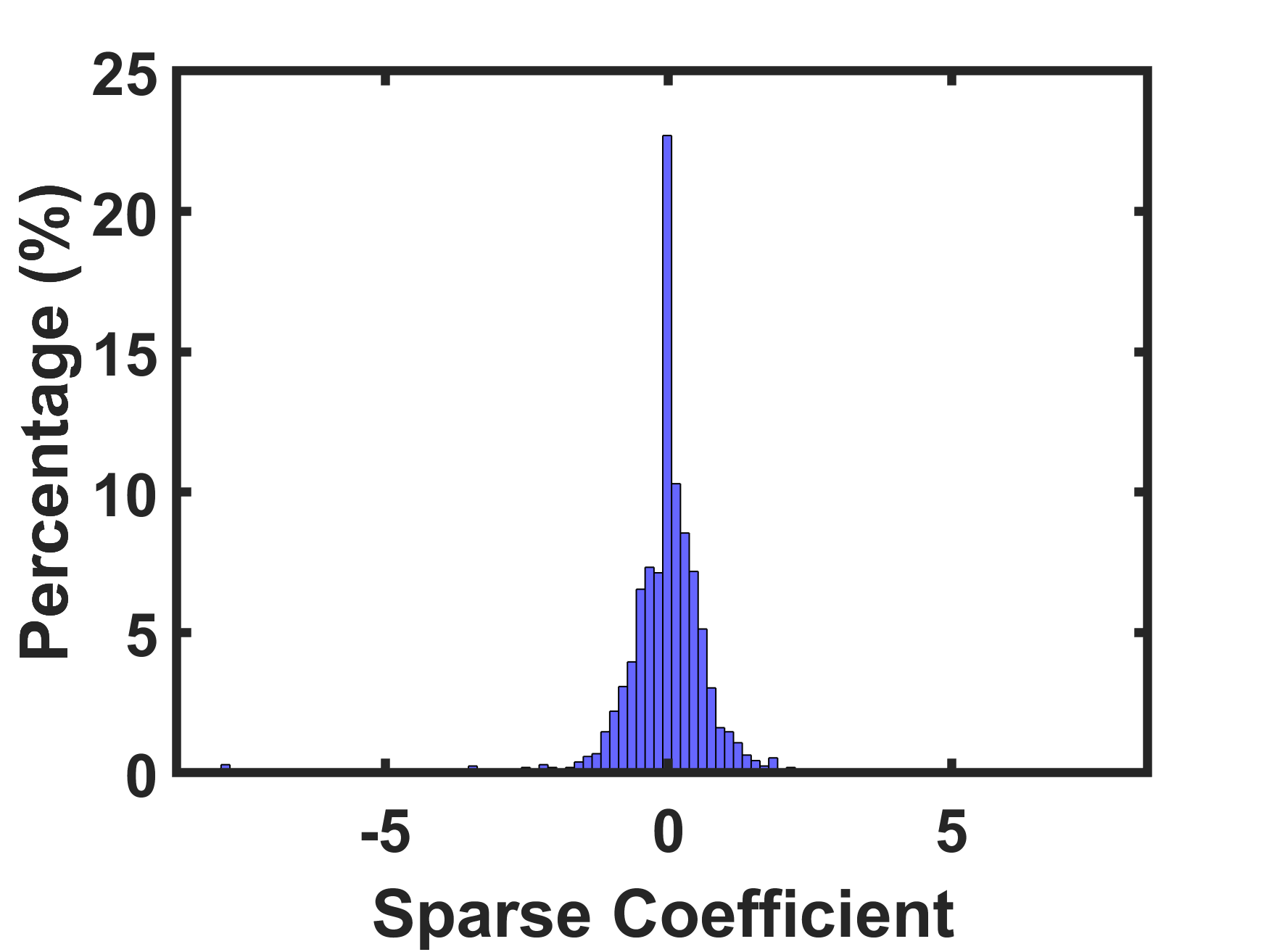}}
    \subfloat[Green Channel]{
        \includegraphics[width=0.32\textwidth]{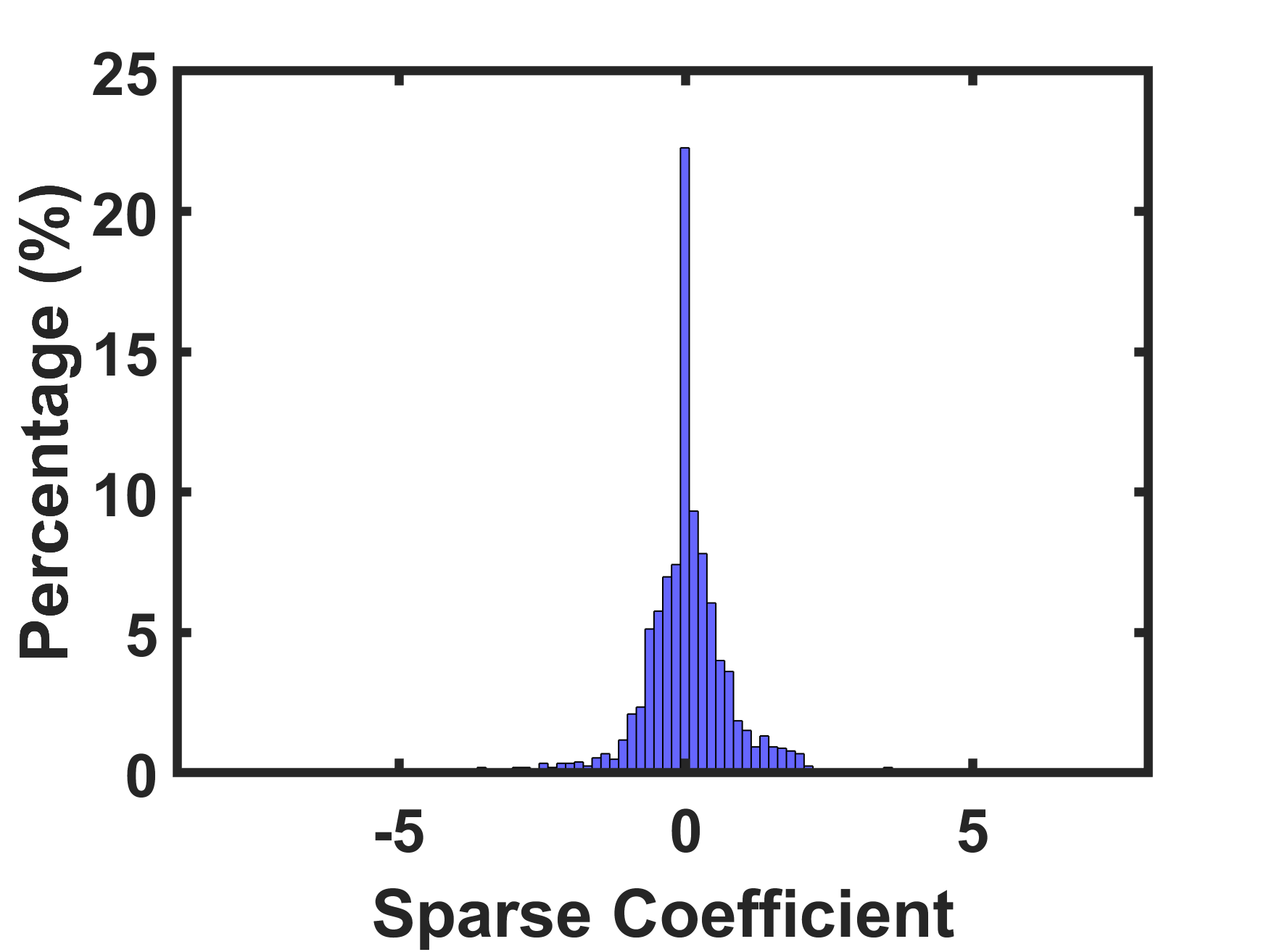}}
    \subfloat[Blue Channel]{
        \includegraphics[width=0.32\textwidth]{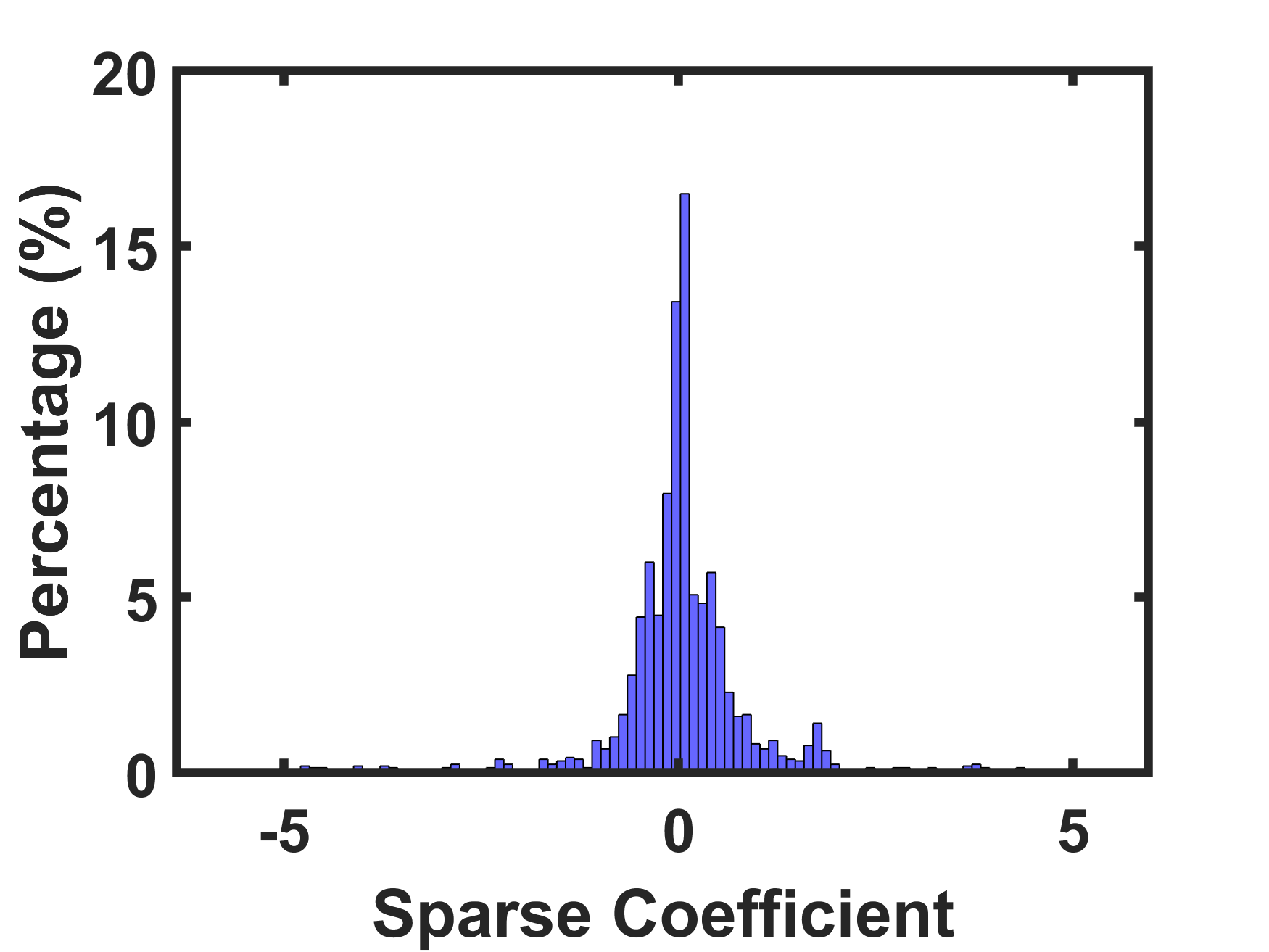}}
    \caption{Example of sparse coefficients distribution. The number of non-zero elements is set as 256 in the Red, Green, and Blue channels.}
    \label{fig:distribution}
\end{figure*}

\subsection{Sparse Representation and Reconstruction}
Firstly, the dictionary $\textbf{D}_1$ in Fig. \ref{fig:framework} is achieved under the traditional image dataset UCID \cite{UCID} following the equation,
\begin{equation}
\{\textbf{D}_1^*,  \alpha_i^* \}  \leftarrow \arg \min_{\{\textbf{D}_1,  \alpha_i \}}\Biggl\{ {\sum_{i=1}^\mathcal{K}||\textbf{X}_i - \textbf{D}_1\alpha_i||^2 + \eta \sum_{i=1}^\mathcal{K}||\alpha_i||_1}\Biggr\},
\end{equation}
where $\mathcal{K}$ is the number of training samples, $\textbf{X}_i$ is the $i^{th}$ training sample cropped from the traditional images, $\alpha_i$ is the sparse coefficient of the $i^{th}$ training sample, and $\eta$ is used to control the number of non-zero coefficients in $\alpha_i$. In this dictionary $\textbf{D}_1$, the size and number of dictionary base are set as $16\times16$ and 256, respectively. 

At the encoder side, the input underwater image is split into $16\times16$ blocks and sparse represented with this dictionary $\textbf{D}_1$, as shown in Fig. \ref{fig:sparse_re}, where the number of non-zero coefficients is pre-defined. One example of sparse coefficients distribution is illustrated in Fig. \ref{fig:distribution}. It can be observed that it follows the Gaussian distribution, which is appropriate for compression. Then, 256 coefficients in a $16\times16$ block, including the non-zero coefficients, are reshaped into $16\times16$ from one dimension to two dimensions. Suppose the size of input underwater image is $256\times256$ and the number of non-zero coefficients is 32 for a $16\times16$ block, thus most of the pixels will become zero, $1.0 - (32\times16\times16)/(256\times256)=87.5\%$, which can save a lot of coding bits.

At the decoder side, the decoded coefficients are employed to reconstruct the underwater image with the enhanced dictionary $\textbf{D}_2$, as shown in Fig. \ref{fig:sparse_re}. How to update the enhanced dictionary will be discussed in the following subsection.

\begin{figure}
    \centering
    \subfloat[$\textbf{D}_1$ Red Channel]{
        \includegraphics[width=0.23\textwidth]{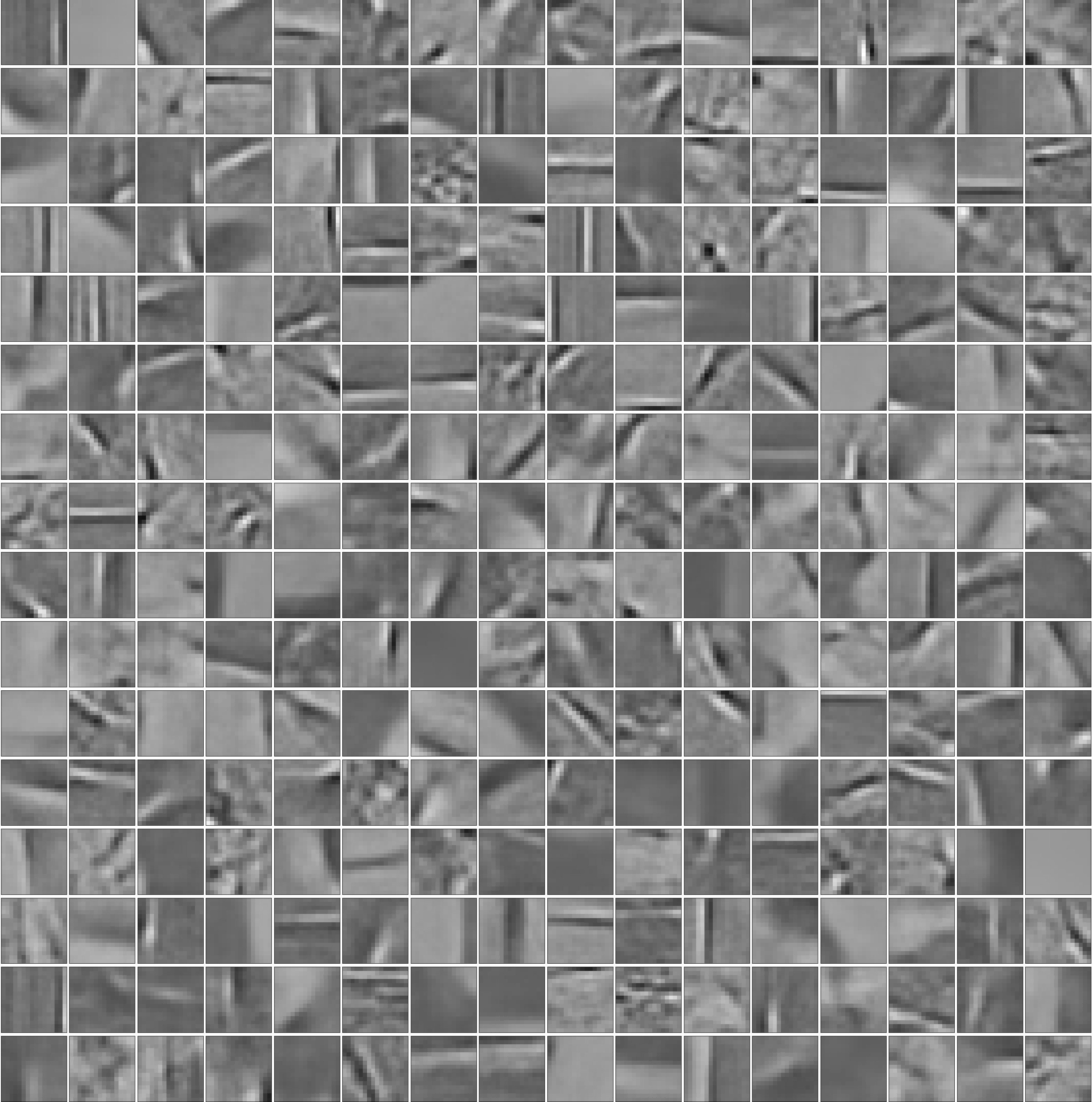}}
    \subfloat[$\textbf{D}_1$ Green Channel]{
        \includegraphics[width=0.23\textwidth]{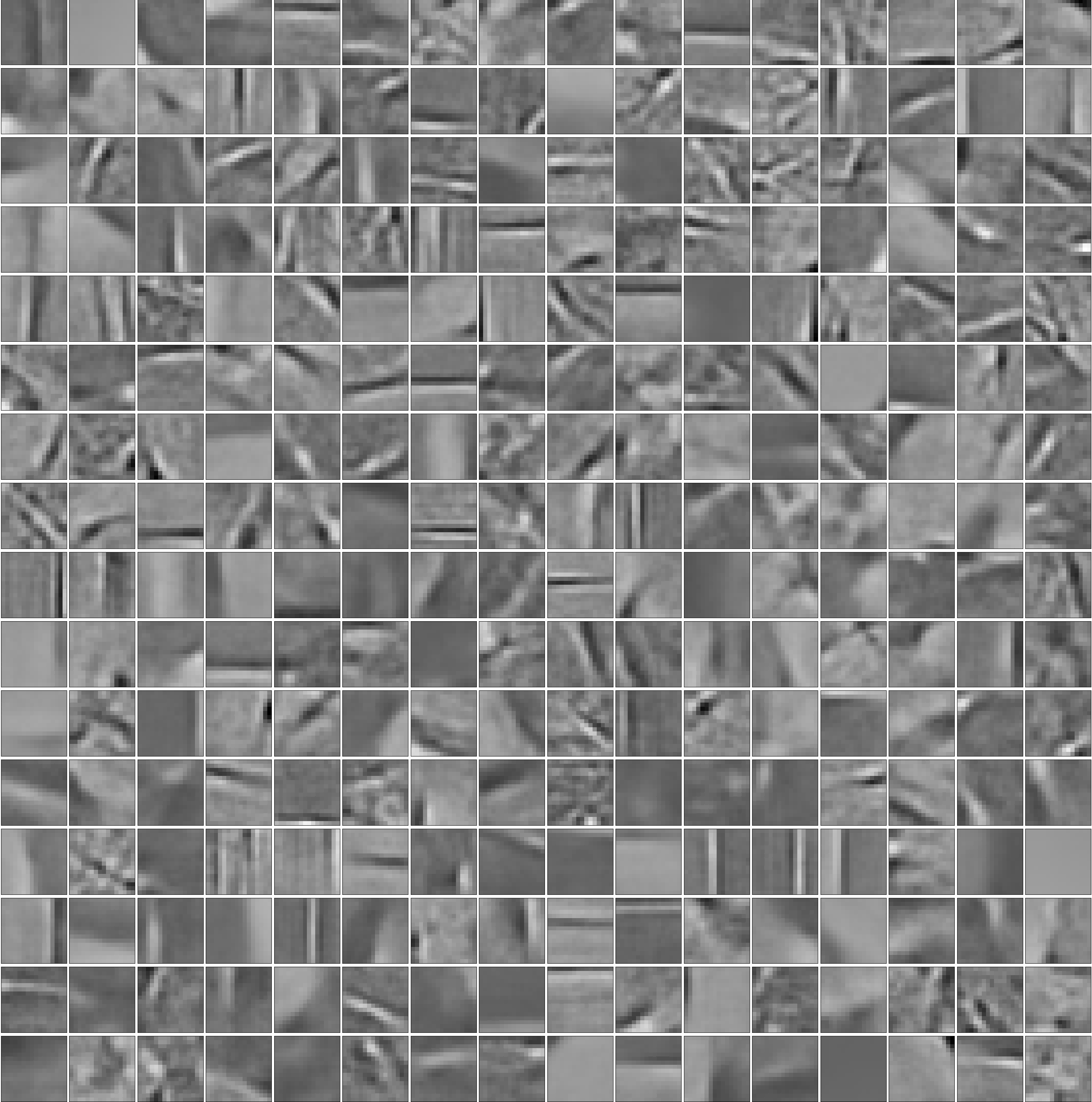}}
    \subfloat[$\textbf{D}_1$ Blue Channel]{
        \includegraphics[width=0.23\textwidth]{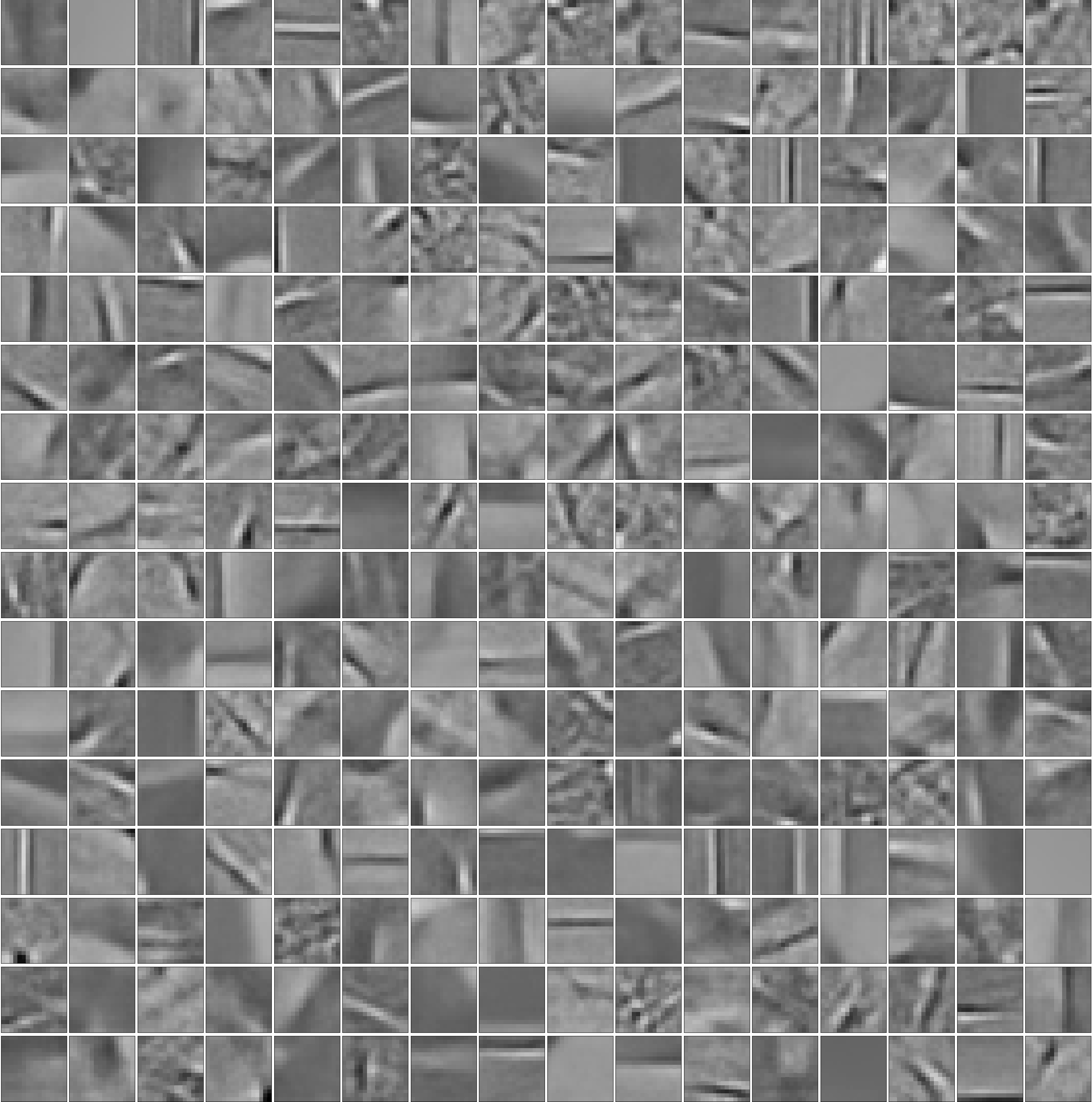}} \\
    \vspace{-0.3cm} 
    \subfloat[$\textbf{D}_2$ Red Channel]{
        \includegraphics[width=0.23\textwidth]{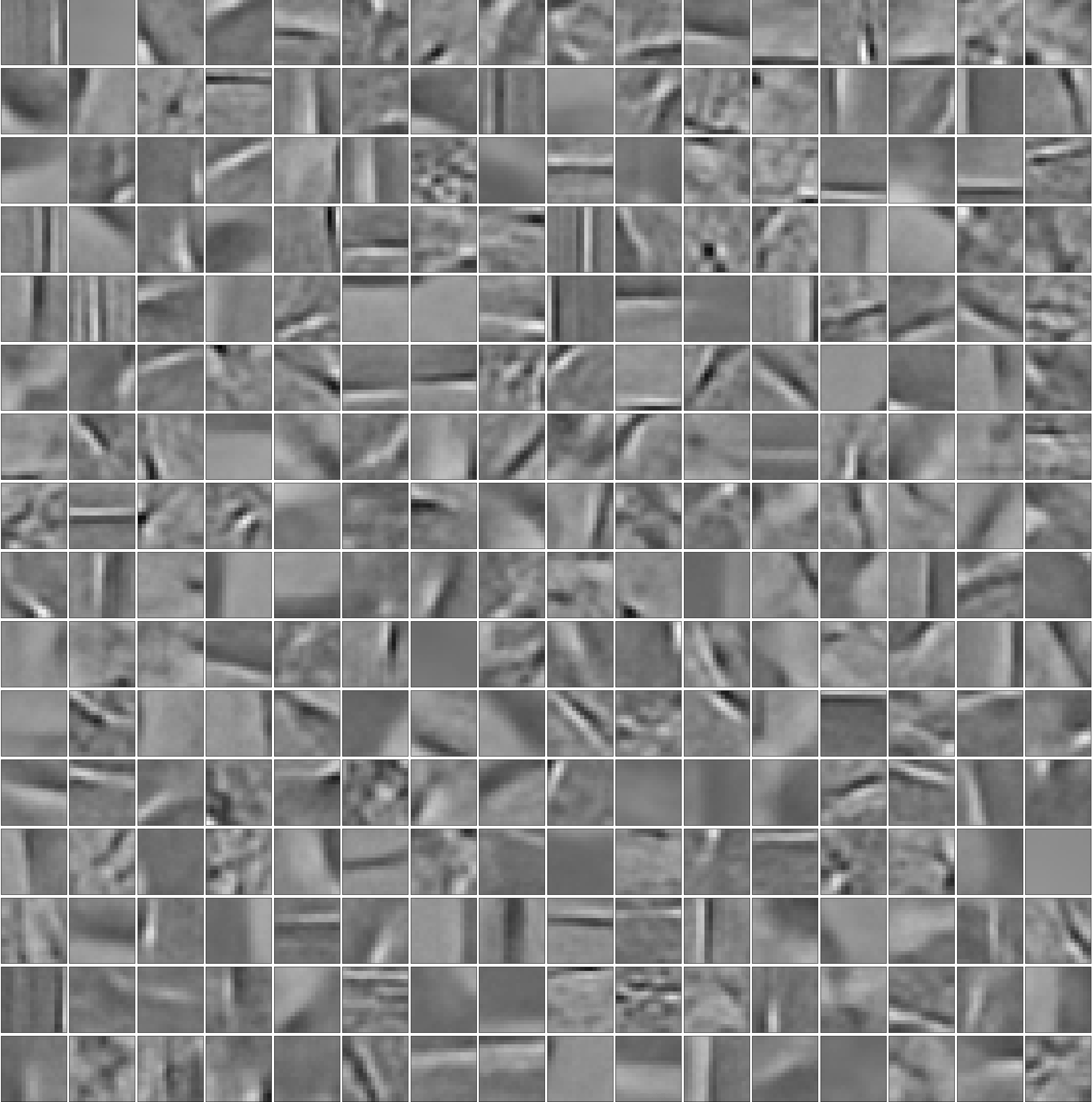}}
    \subfloat[$\textbf{D}_2$ Green Channel]{
        \includegraphics[width=0.23\textwidth]{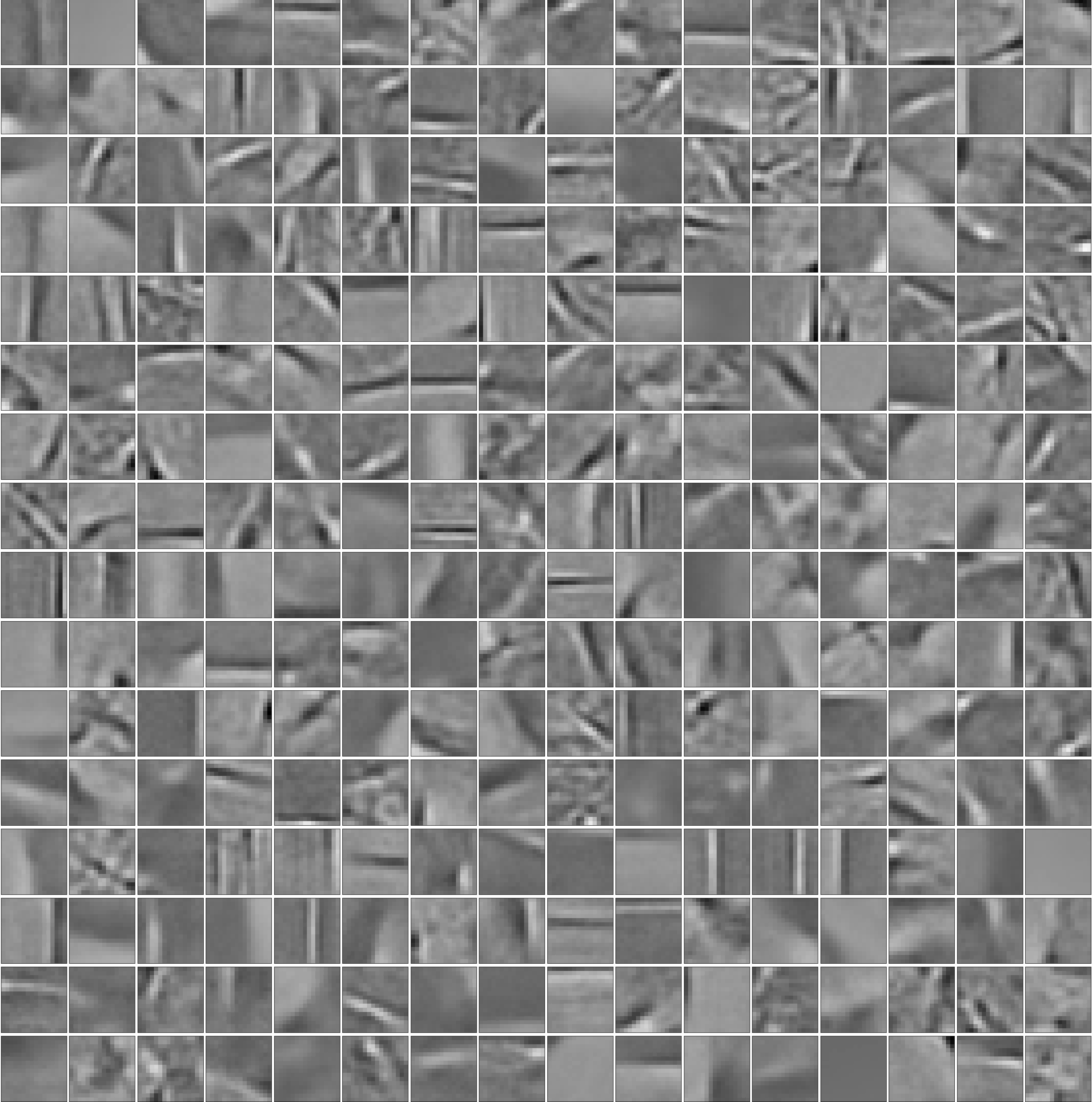}}
    \subfloat[$\textbf{D}_2$ Blue Channel]{
        \includegraphics[width=0.23\textwidth]{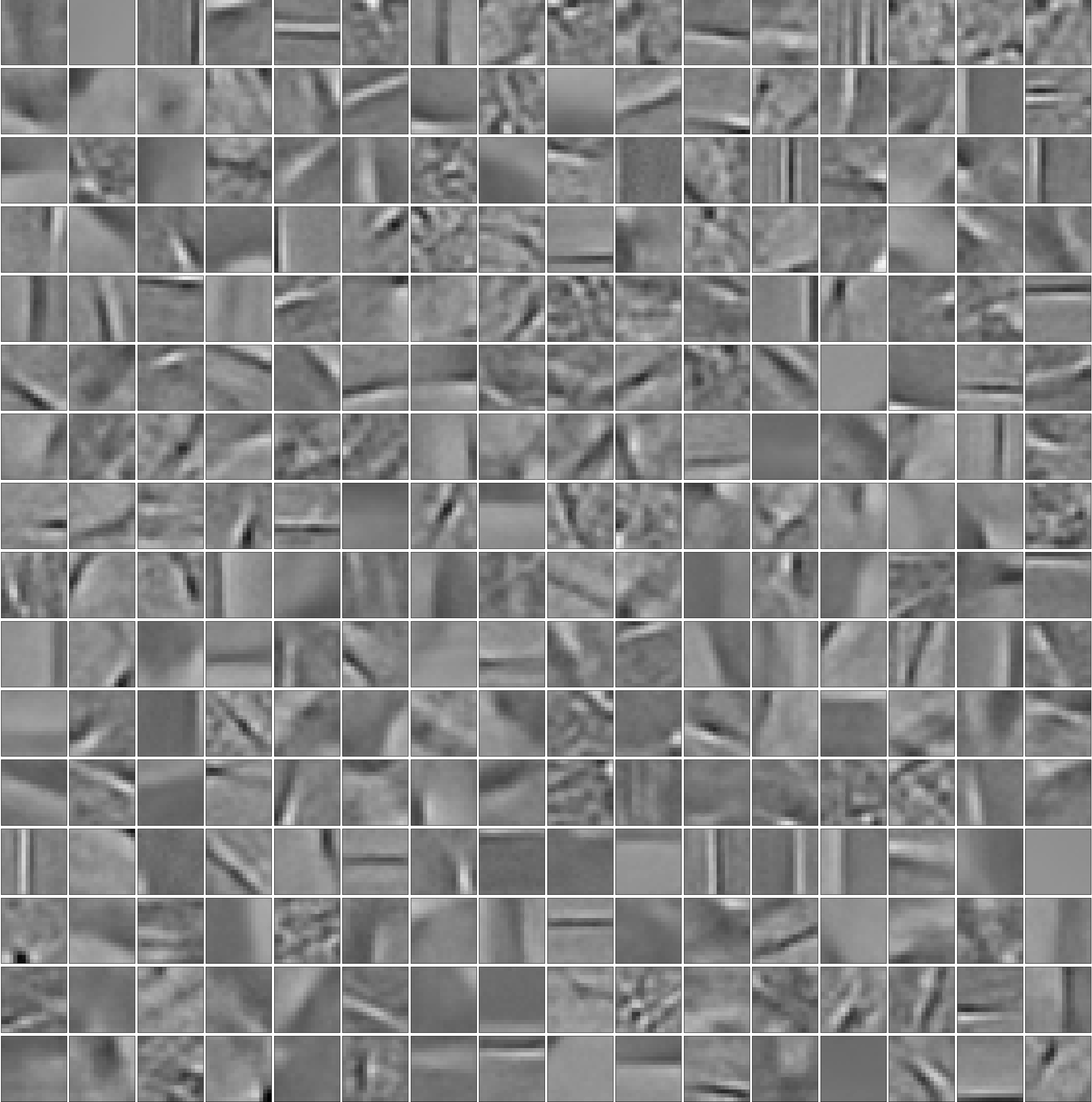}} \\
    \vspace{-0.3cm} 
    \subfloat[$\textbf{D}_1$-$\textbf{D}_2$ Red Channel]{
        \includegraphics[width=0.23\textwidth]{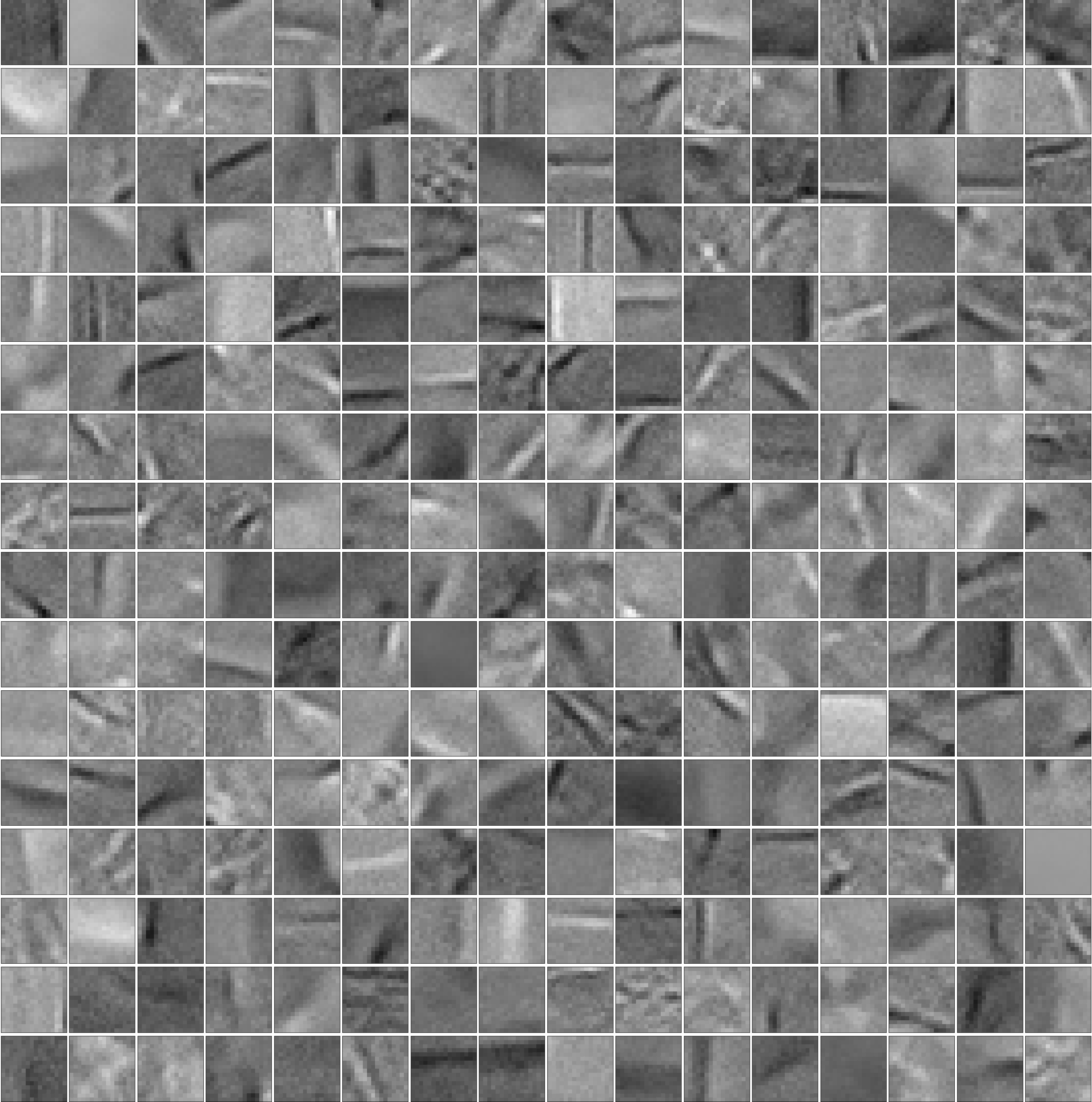}}
    \subfloat[$\textbf{D}_1$-$\textbf{D}_2$ Green Channel]{
        \includegraphics[width=0.23\textwidth]{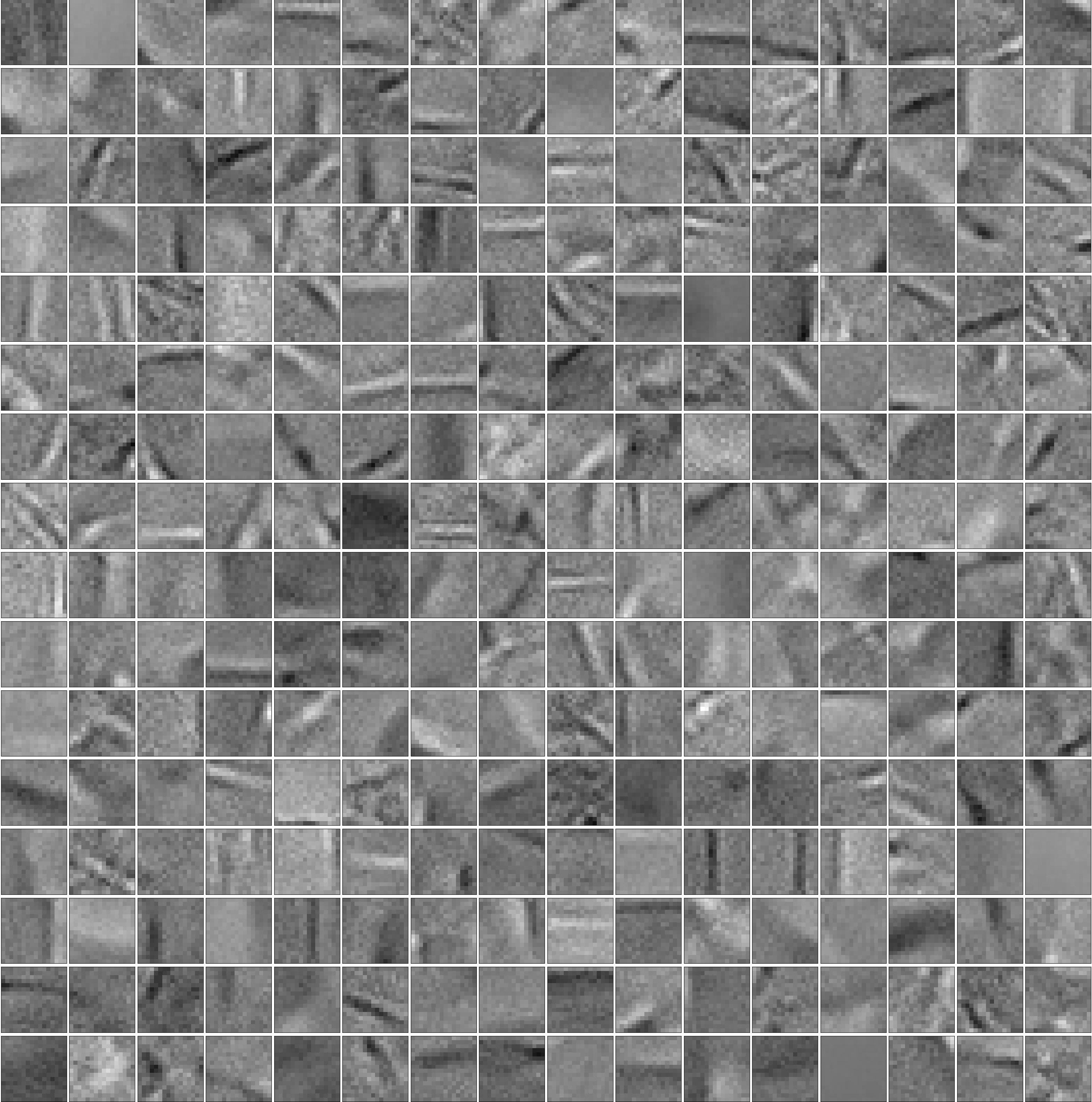}}
    \subfloat[$\textbf{D}_1$-$\textbf{D}_2$ Blue Channel]{
        \includegraphics[width=0.23\textwidth]{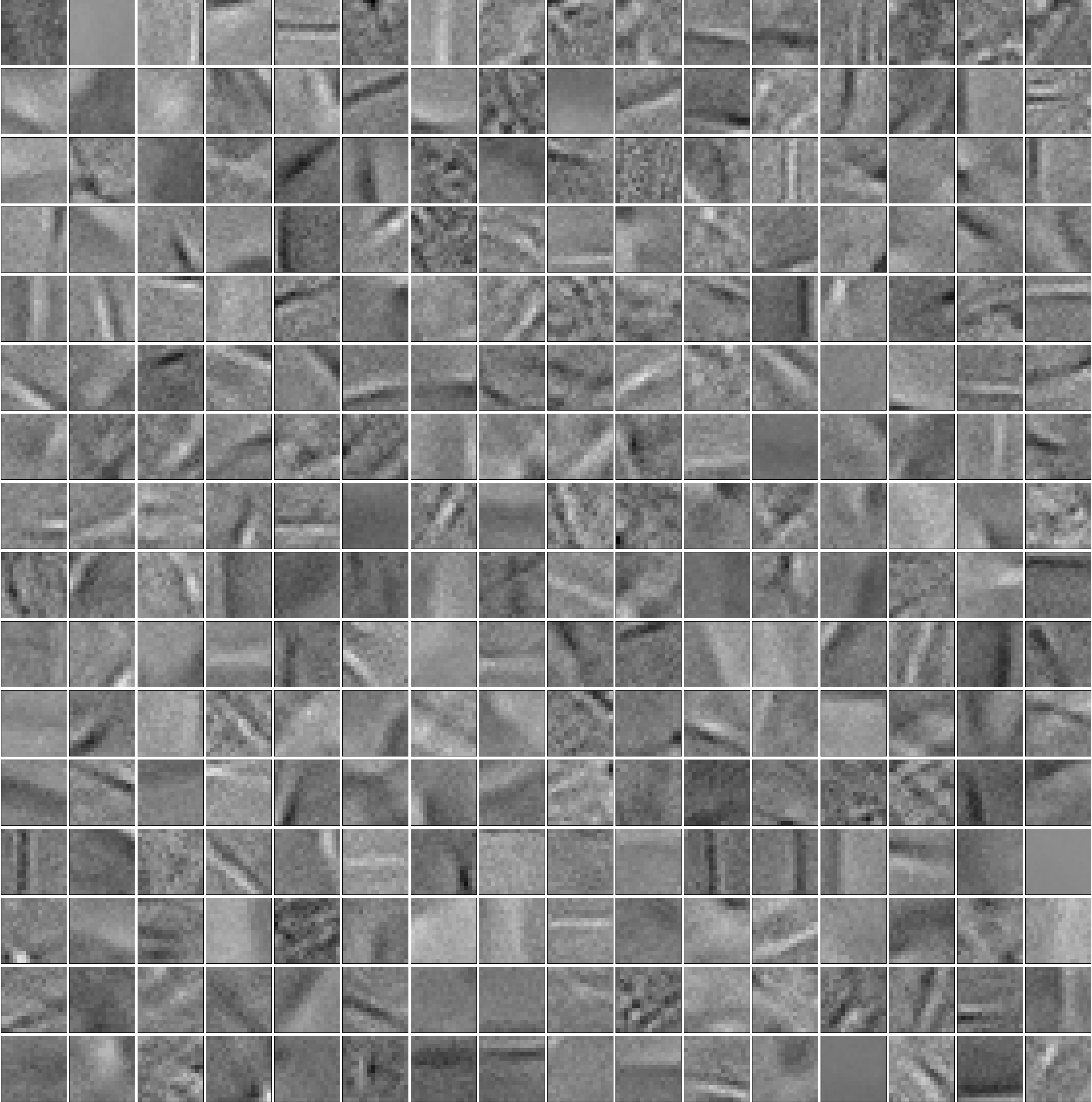}}
        
    \caption{Visualization of dictionaries $\textbf{D}_1$ and $\textbf{D}_2$.}
    \label{fig:visual_dict}
\end{figure}

\subsection{Underwater Image Enhancement Dictionary Derivation}
Here, the key idea of underwater image enhancement dictionary derivation aims to update the dictionary for achieving the output of BL close to the enhanced version, where the sparse coefficients are shared at the encoder and decoder sides. Another underwater image dataset EUVP-ImageNet \cite{EUVP}, which has the enhanced version of ground truth, is adopted for the training of the enhanced dictionary $\textbf{D}_2$. Suppose $\textbf{I}_i$ is the $i^{th}$ underwater image in this dataset, its associated sparse coefficients $\alpha_i^*$ can be calculated by
\begin{equation}
{\alpha}_i^* \leftarrow \arg\min_{{\alpha}_i} \Bigl\{ 
    \norm{\mathbf{I}_i - \mathbf{D}_1 {\alpha}_i}^2 
    + \eta \norm{{\alpha}_i}_1 
\Bigr\},
\end{equation}
where $\eta$ is employed to control the number of non-zero coefficients.
With these optimal sparse coefficients $\alpha_i^*$, the dictionary $\textbf{D}_2$ can be derived as follows,
\begin{equation}
\textbf{D}_2^*  \leftarrow \arg \min_{\mathbf{D}_2} \sum_{i=1}^{\mathcal{K}} \left\| \mathbb{E}(\mathbf{I}_i) - \mathbf{D}_2 {\alpha}_i^* \right\|^2,
\end{equation}
where $\mathbb{E}$() is the underwater image enhancement operator, $\mathcal{K}$ is the number of underwater images in the dataset EUVP-ImageNet.

Fig. \ref{fig:visual_dict} illustrates the visualization of these two dictionaries $\textbf{D}_1$ and $\textbf{D}_2$. Since it is difficult to detect the differences between $\textbf{D}_1$ and $\textbf{D}_2$ by human eyes, the difference maps of them are presented in the third row. Although the textures in the dictionary bases of $\textbf{D}_1$ and $\textbf{D}_2$ are similar, the values of intensity are different, which can be observed in the difference maps. The contribution of enhanced dictionary $\textbf{D}_2$ will be discussed in the experiment.

\begin{figure*}[!t]
\centering
\includegraphics[width=0.85\textwidth]{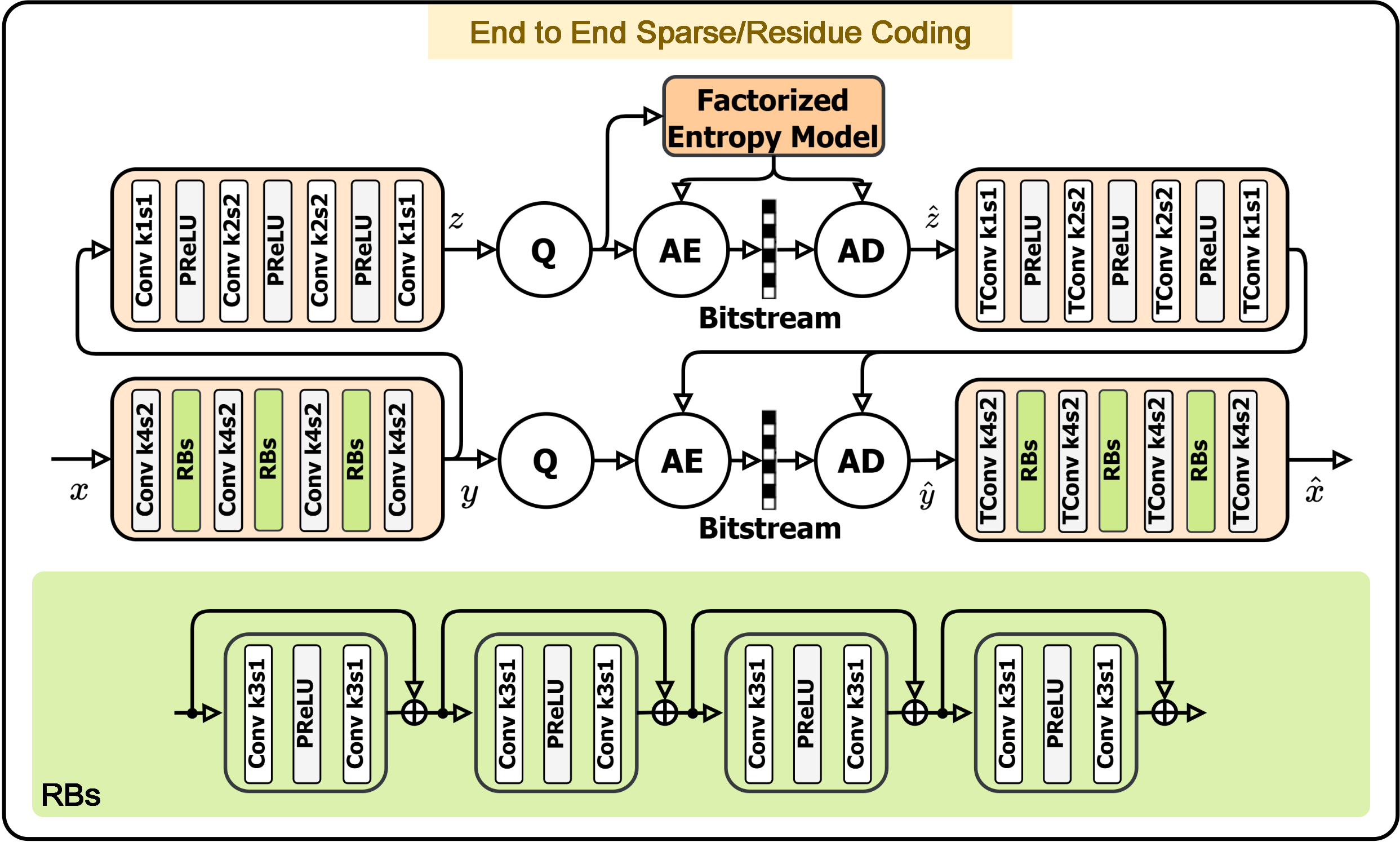}
\caption{End to end sparse/residue codec. ``Q" indicates quantization. ``AE" and ``AD" indicate arithmetic encoding and arithmetic decoding. ``Conv" indicates Convolution. ``TConv" indicates Transposed Convolution. ``k1s1" indicates that the kernel size is $1\times1$ and the stride size is 1. In this architecture, the activation function PReLU is used instead of ReLU.}
\label{fig:codec}
\end{figure*}

\subsection{Sparse Coefficients and Residues Compression}
Due to that the non-zero sparse coefficients are in the type of float and there are many negative values in the residues, traditional codec is inappropriate for compression, the end to end codec is adopted in this work. As shown in Fig. \ref{fig:codec}, it is the architecture of end to end codec for both sparse coefficients and residues compression. It should be noted that sparse coefficients codec and residues codec share the architecture, but they are equipped with different weights. 

This architecture is similar to the existing end to end image compression codec, which has the modules of Encoder, Decoder, Hyper Encoder, and Hyper Decoder. The Encoder aims to transfer the input signal $x \in \mathbb{R}^{W\times H\times C}$ to the latent space with small size, i.e., $y \in \mathbb{R}^{W/16\times H/16\times C}$. While the Decoder reconstructs the signal $\hat{y}$ to the original size, i.e., $\hat{x} \in \mathbb{R}^{W\times H\times C}$. It can be regarded as downsampling and upsampling operators. In the modules of Encoder and Decoder, three Residual Blocks (RBs) are equipped to exploit compact features in the latent space, which are placed between two convolutional layers. In the RBs, four residual networks are connected in series. The modules of Hyper Encoder and Hyper Decoder are used to provide the parameters $(\mu, \sigma^2)$ of Gaussian Model for high efficiency entropy coding/decoding in the AE and AD, i.e., the estimated distribution $\hat{P}_y(\hat{y}|\hat{z})$ follows $N(\mu, \sigma^2)$. Here, the bitrate of compact signal $y$ can be calculated as follows,
\begin{equation}
\mathcal{R}(y) = \mathrm{E}[-\log_{2}\hat{P}_y(\hat{y}|\hat{z})].
\end{equation}
The bitrate of signal $z$ can be calculated as follows,
\begin{equation}
\mathcal{R}(z) = \mathrm{E}[-\log_{2}\hat{P}_z(\hat{z})],
\end{equation}
where $\hat{P}_z(\hat{z})$ is estimated by the factorized entropy model.
As as a result, the total estimated bitrate $\mathcal{R}$ is the sum of $\mathcal{R}(y)$ and $\mathcal{R}(z)$, i.e.,
\begin{equation}
\mathcal{R} = \mathcal{R}(y) + \mathcal{R}(z) = \mathrm{E}[-\log_{2}\hat{P}_y(\hat{y}|\hat{z}) - \log_{2}\hat{P}_z(\hat{z})].
\end{equation}

\begin{figure*}[!t]
\centering
\includegraphics[width=0.85\textwidth]{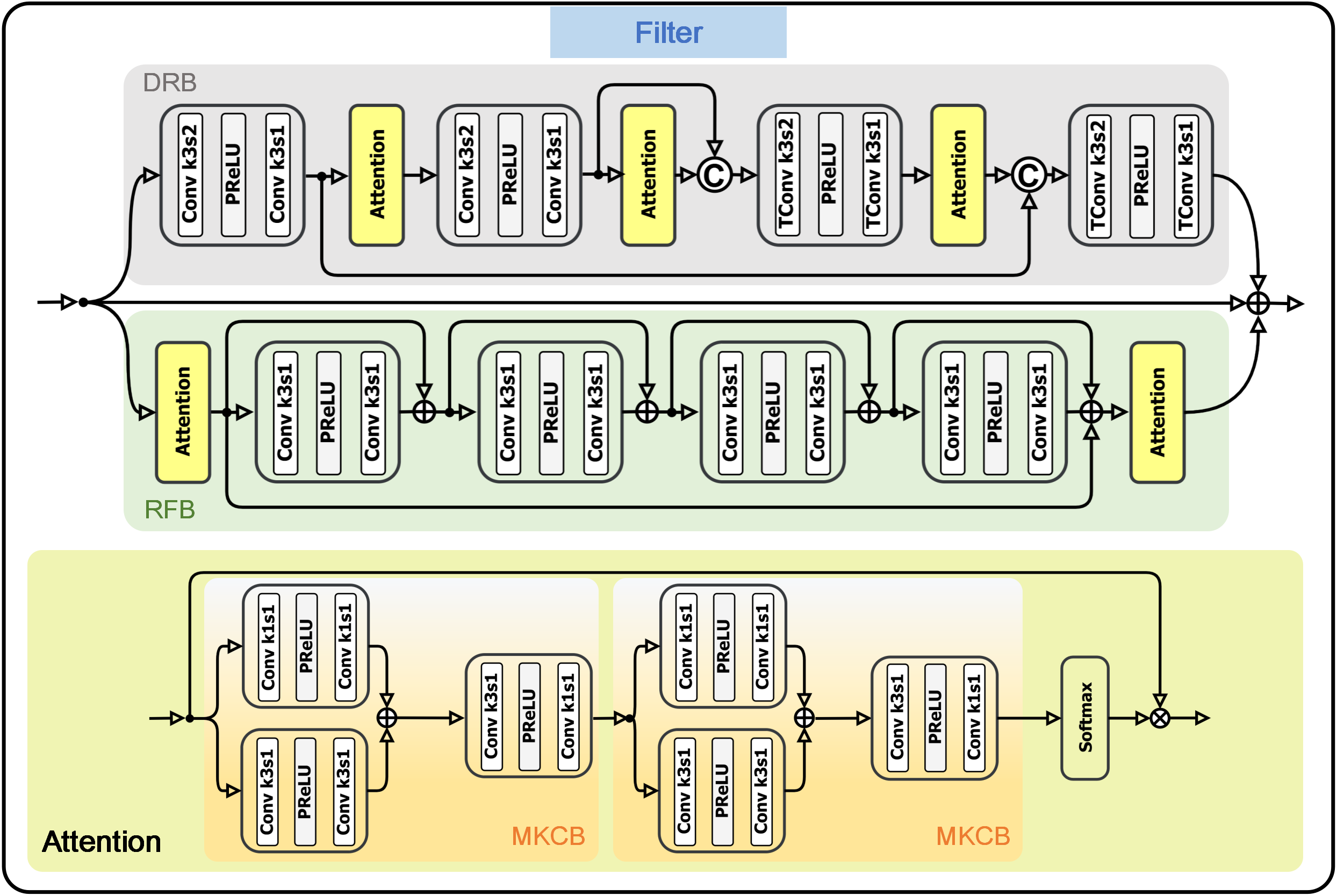}
\caption{Dual-branch filter. ``DFB" indicates the detail refinement branch, and ``RFB" indicates the rough filtering branch. ``MKCB" indicates multi-kernel convolutional block. ``Conv" indicates Convolution. ``TConv" indicates Transposed Convolution. ``k3s2" indicates that the kernel size is $3\times3$ and the stride size is 2. ``C" indicates concatenation. }
\label{fig:filter}
\end{figure*}

\subsection{Dual-branch Filter}
As shown in Fig. \ref{fig:filter}, it is the proposed filter in the framework of enhanced quality aware-underwater image compression, in which there are two branches, i.e., one is \textbf{Rough Filtering Branch} (\textbf{RFB}), and the other is \textbf{Detail Refinement Branch} (\textbf{DRB}). After the input signal is processed by these two branches individually, the outputs of RFB and DRB are added to produce the final output of the filter. 

\textbf{DRB}. The module of DRB can be regarded as multi-scale details extraction, in which there are two convolutional blocks (including Convk3s2, PReLU, and Convk3s1) and two transposed convolutional blocks (including TConvk3s2, PReLU, and TConvk3s1). The former one extracts the latent features at the low resolution and generates the output with half of spatial height and width when compared to those of its input, while the latter one extracts the latent features at the high resolution and produces the output with two times of spatial height and width when compared to those of its input. In addition, three attention modules are equipped between these convolutional and transposed convolutional blocks. Moreover, skip connection is added to transfer the output of first convolutional block to the input of the second transposed convolutional block, and transfer the output of the second convolutional block to the input of the first transposed convolutional block.

\textbf{RFB}. The module of RFB can be regarded as an improved residual block, where four convolutional blocks (including Convk3s1, PReLU, and Convk3s1) are connected in series. In specific, two attention modules are placed at the head and the tail of RFB respectively. Additionally, global residual learning is performed in the RFB. 

\textbf{Attention}. The attention module aims to exploit the important and non-important features with different weights in the feature space, where two \textbf{Multi-Kernel Convolutional Blocks} (\textbf{MKCBs}) are connected in series and followed by the activation function of Softmax. The final output is produced by the multiplication of the weight map and the initial input signal with skip connection.
In each MKCB, the input signal is processed by convolutional operators with different kernels, i.e., $3\times3$ and $1\times1$. Then, the individual outputs are added and processed by multi-kernel convolutions in series (including Convk3s1, PReLU, and Convk1s1).

\subsection{Loss Function and Neural Network Training}
The neural networks shown in Fig. \ref{fig:framework} are trained in an end to end manner, not a separate manner. During training, the enhanced version of ground truth is required, thus the dataset UIEB \cite{UIEB} that has 890 underwater images with resolution from $299\times168$ to $2180\times1447$ and the dataset UVEB \cite{UVEB} that has 1308 underwater videos with resolution from $960\times528$ to $3840\times2160$ are adopted. It should be mentioned that only the first frames of the underwater videos in UVEB are used. These underwater images are further downsampled to generate more training samples in the different scales. Therefore, there are 7152 images in total for training. The loss function can be written as follows,
\begin{equation}
\mathcal{L} = \mathcal{R}_{\mathrm{total}} + \lambda \sum_{i=1}^{4} \mu_i \mathcal{D}_i
\end{equation}
where $\mathcal{R}_{\mathrm{total}}$ indicates the total coding bits of spare coefficients and residues compression, i.e., $\mathcal{R}_{\mathrm{total}} = \mathcal{R}_{s}(y) + \mathcal{R}_{s}(z) + \mathcal{R}_{r}(y) + \mathcal{R}_{r}(z)$. $\lambda$ is the weight used to balance coding bits and distortion, while $\sum_{i=1}^4 \mu_i = 1, \mu_1 = 0.4, \mu_2 = 0.4, \mu_3 = 0.1, \mu_4 = 0.1$ are employed for the importance assignment of distortions. $\mathcal{D}_1, \mathcal{D}_2, \mathcal{D}_3$ represent the distortions of output of EL with respect to the ground truth, i.e., $\mathcal{D}_1 = \norm{\mathbf{G} - \mathbf{O}_{\mathrm{EL}}}^2$, $\mathcal{D}_2 = 1 - \mathbb{SSIM}(\textbf{G}, \textbf{O}_{\mathrm{EL}})$, $\mathcal{D}_3 = ||\mathbb{B}(\textbf{G}) - \mathbb{B}(\textbf{O}_{\mathrm{EL}})||^2$, $\textbf{G}$ indicates the ground truth, $\textbf{O}_{\mathrm{EL}}$ is the output of EL, $\mathbb{SSIM}()$ indicates the \textbf{Structural SIMilarity} (\textbf{SSIM}) \cite{SSIM} value calculation, $\mathbb{B}()$ indicates the boundary extraction. $\mathcal{D}_4$ represents the distortion of output of BL with respect to the ground truth, i.e., $\mathcal{D}_4 = ||\textbf{G} - \textbf{O}_{\mathrm{BL}}||^2$, $\textbf{O}_{\mathrm{BL}}$ is the output of BL.

In the  training stage, these training images are randomly cropped into $192\times192$ blocks for each epoch. The patch size, batch size, and initial learning rate are set as $192\times192$, 16, and $1\times10^{-4}$ respectively. The values of $\lambda$ in loss function are set as \{4, 8, 16, 32, 64, 128, 256\}. The number of non-zero sparse coefficients is randomly changed from 96 to 160 to adapt different cases.
The models are trained on the platform of NVIDIA GeForce RTX 3090 GPU and Intel i9-10900K CPU.

\begin{figure*}
    \centering
    \small
    \setlength{\tabcolsep}{1pt} 
    \begin{tabular}{c c c c c c}
      
      \multicolumn{1}{c}{\includegraphics[height=0.235\textwidth, width=0.165\textwidth]{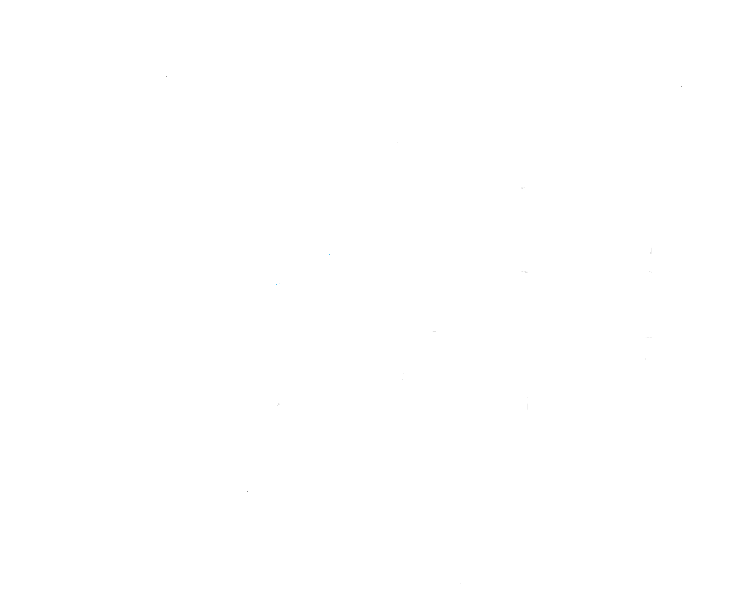}} &
      \multicolumn{2}{c}{\includegraphics[height=0.235\textwidth, width=0.33\textwidth]{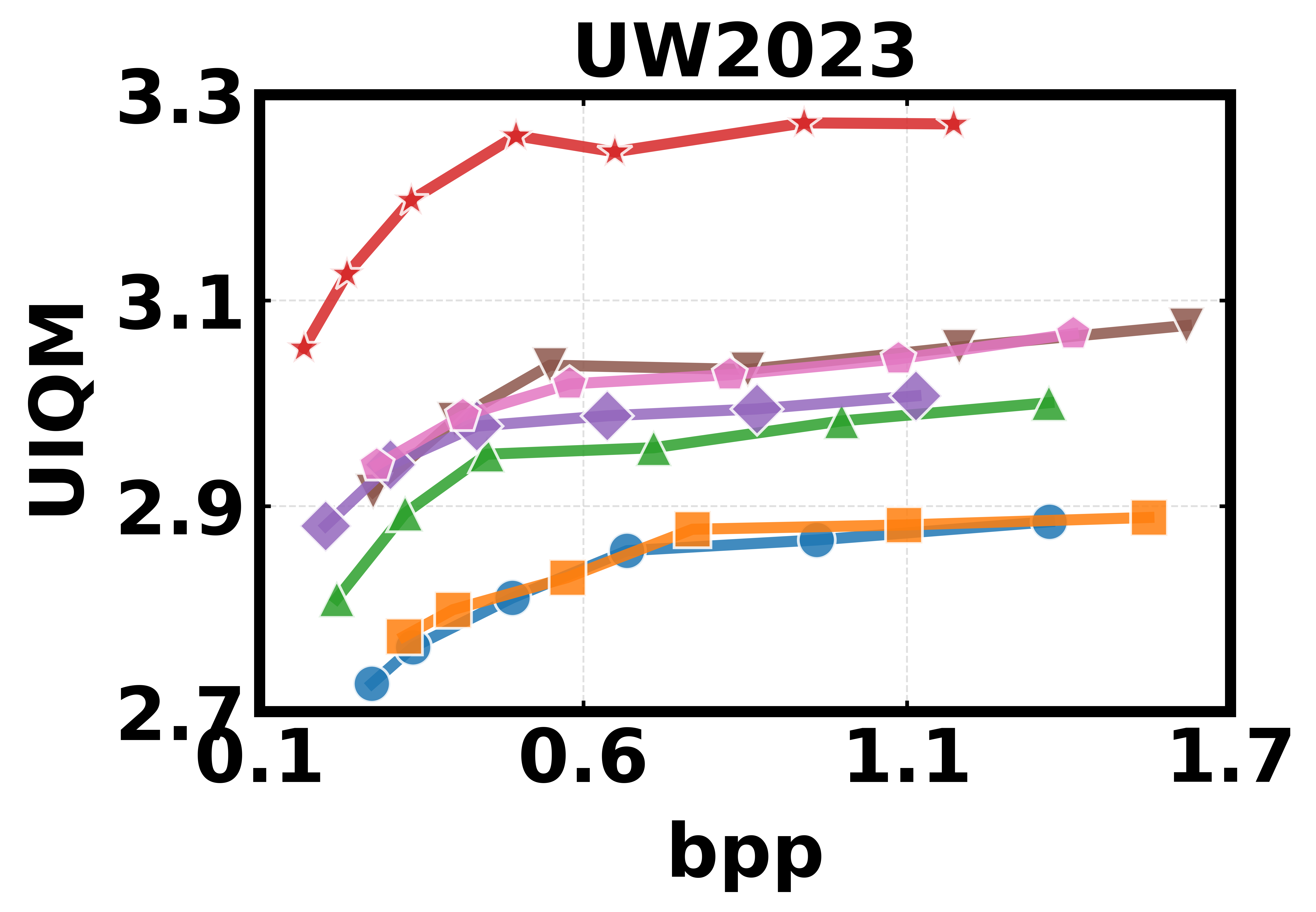}} &
      \multicolumn{2}{c}{\includegraphics[height=0.235\textwidth, width=0.33\textwidth]{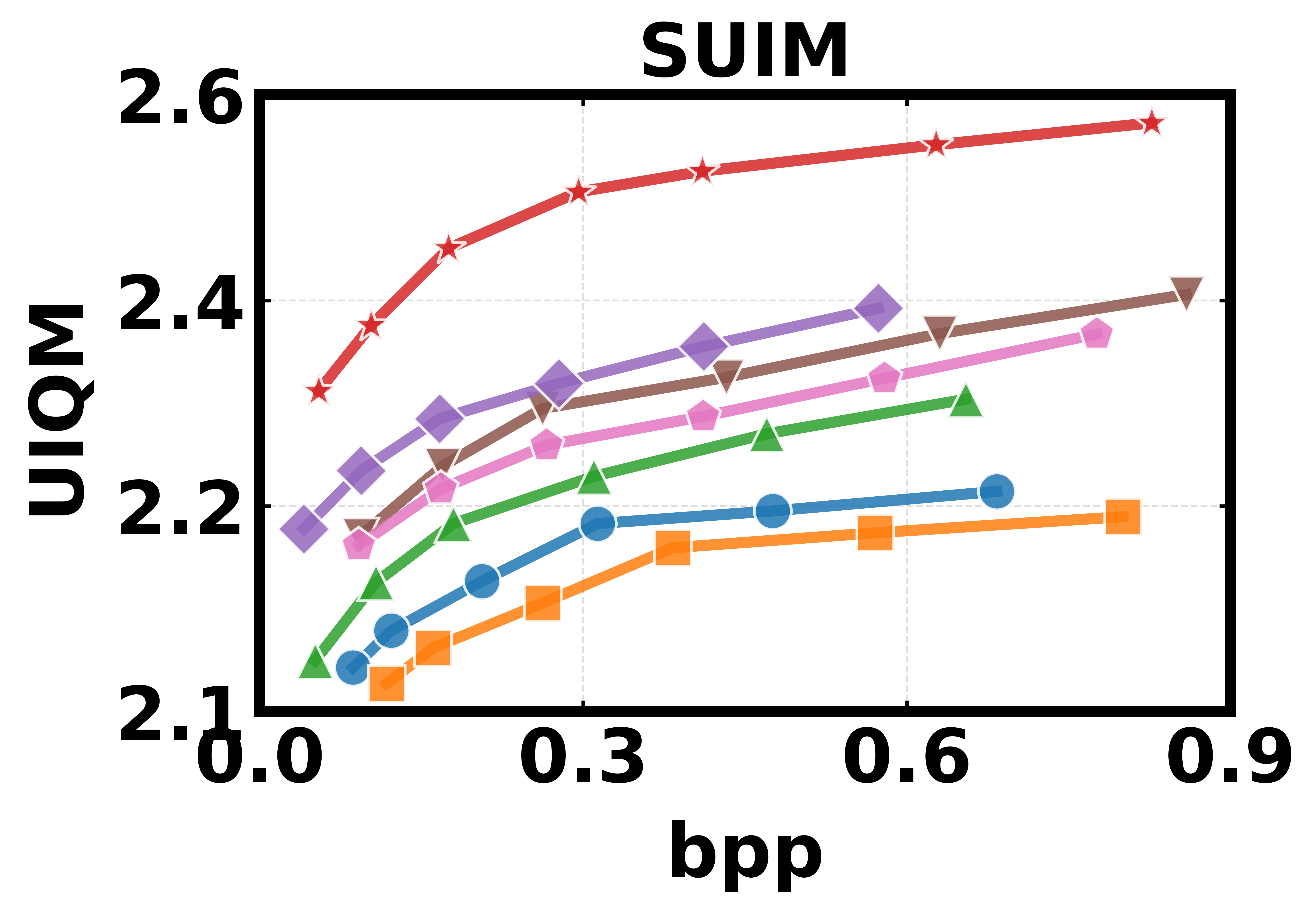}} &
      \multicolumn{1}{c}{\includegraphics[height=0.235\textwidth, width=0.165\textwidth]{fig/blank.png}} \\
      
      \multicolumn{6}{c}{\includegraphics[width=0.99\textwidth]{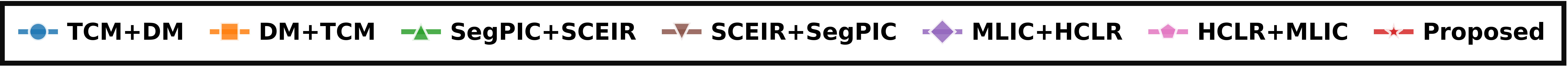}} 
 
    \end{tabular}
  \caption{Coding performance comparison with the state-of-the-art works in terms of UIQM under the datasets of UW2023 and SUIM.}
  \label{curve2}
  \end{figure*}
  
  \begin{figure*}
    \centering
    \small
    \setlength{\tabcolsep}{1pt} 
    \begin{tabular}{c c c c c c}
      
      \multicolumn{2}{c}{\includegraphics[height=0.235\textwidth, width=0.33\textwidth]{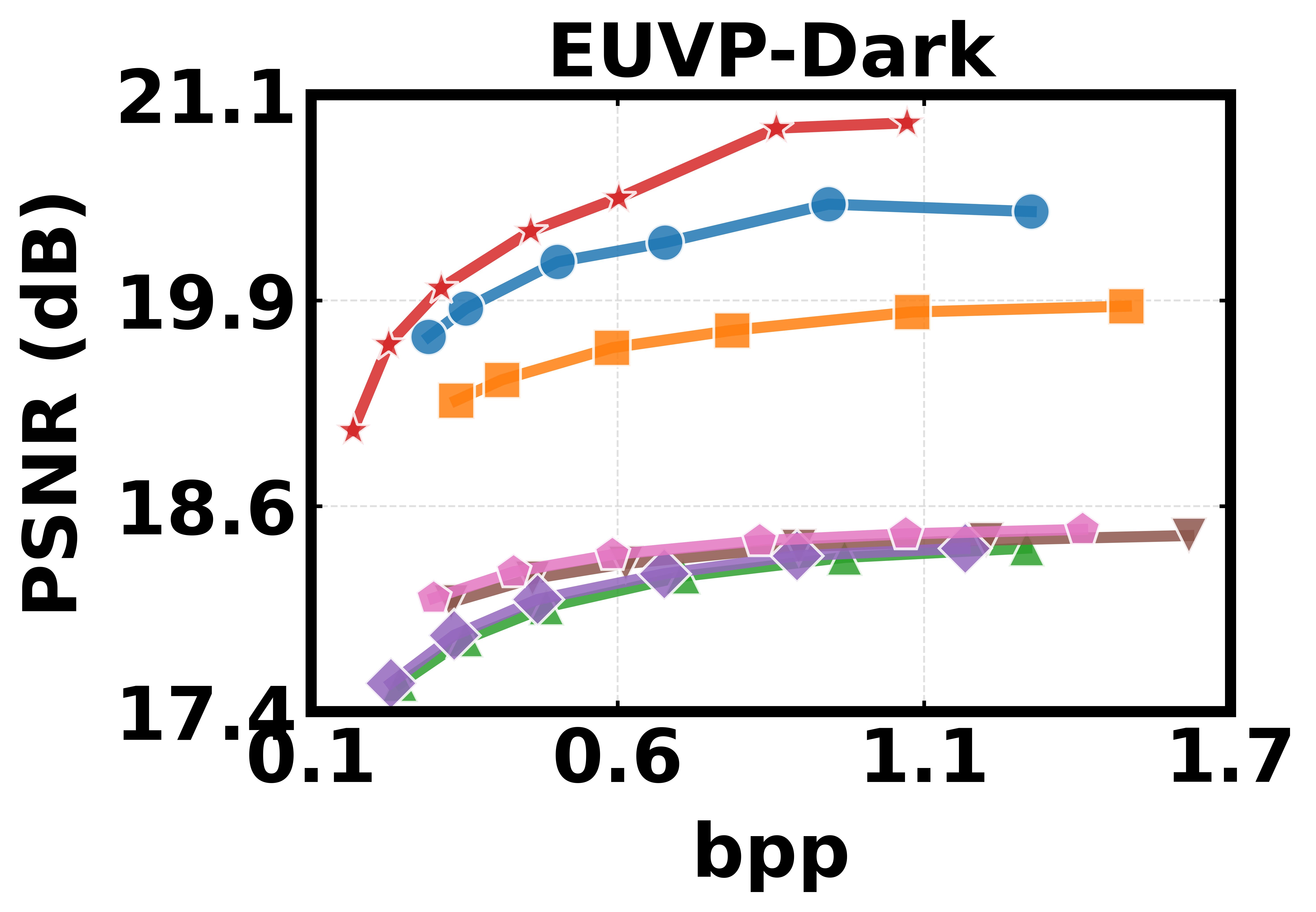}} &
      \multicolumn{2}{c}{\includegraphics[height=0.235\textwidth, width=0.33\textwidth]{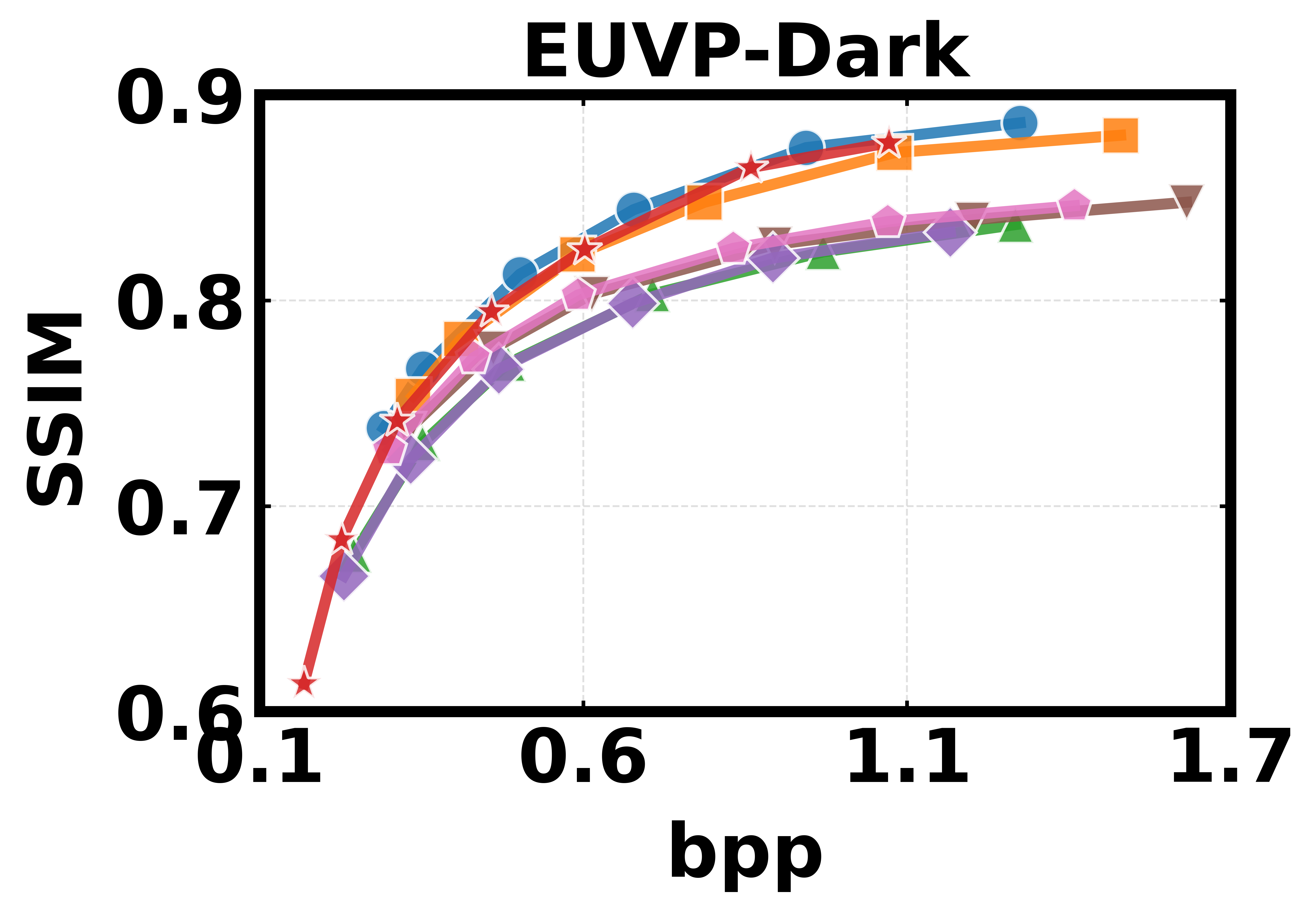}} &
      \multicolumn{2}{c}{\includegraphics[height=0.235\textwidth, width=0.33\textwidth]{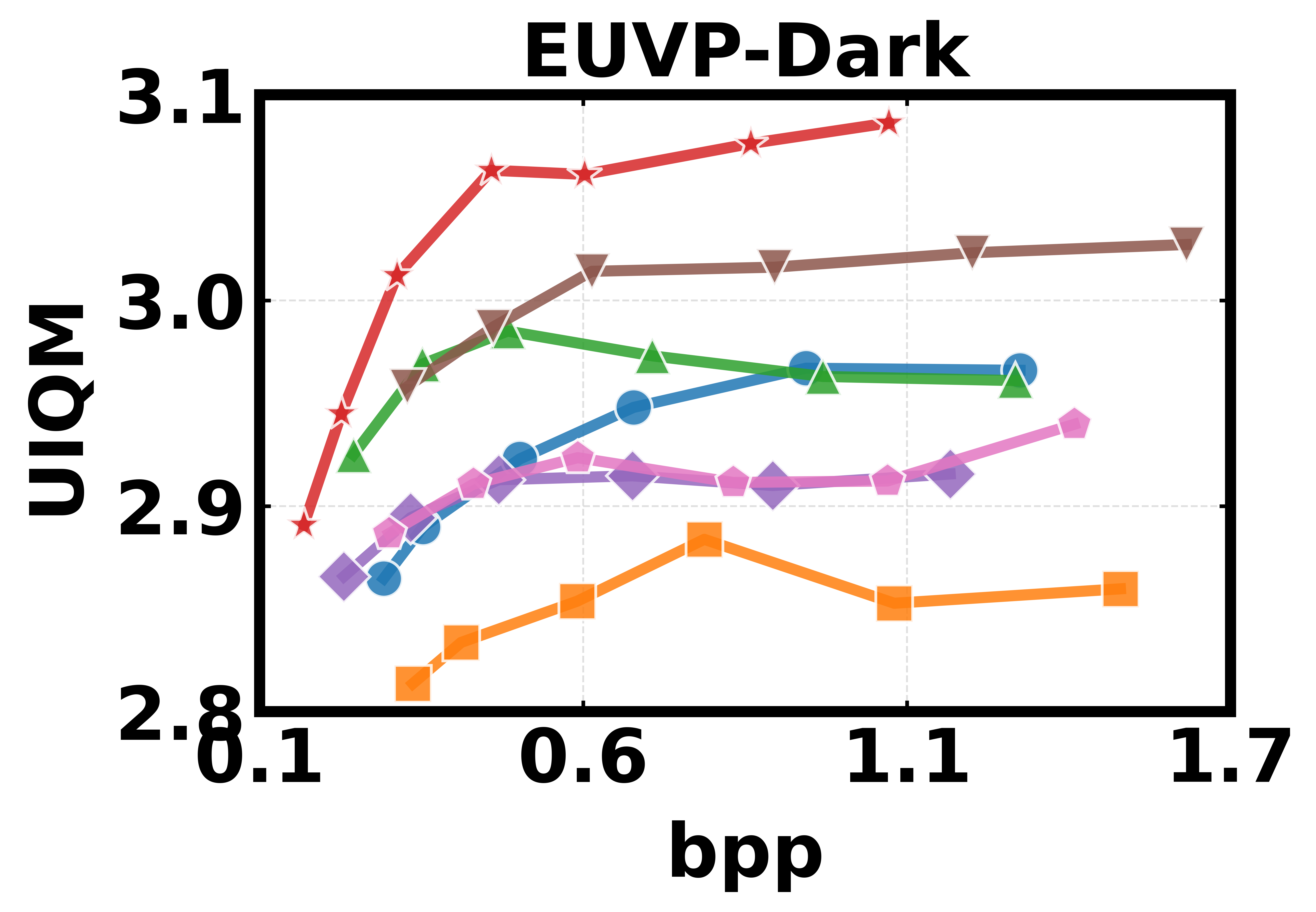}} \\
      
      \multicolumn{2}{c}{\includegraphics[height=0.235\textwidth, width=0.33\textwidth]{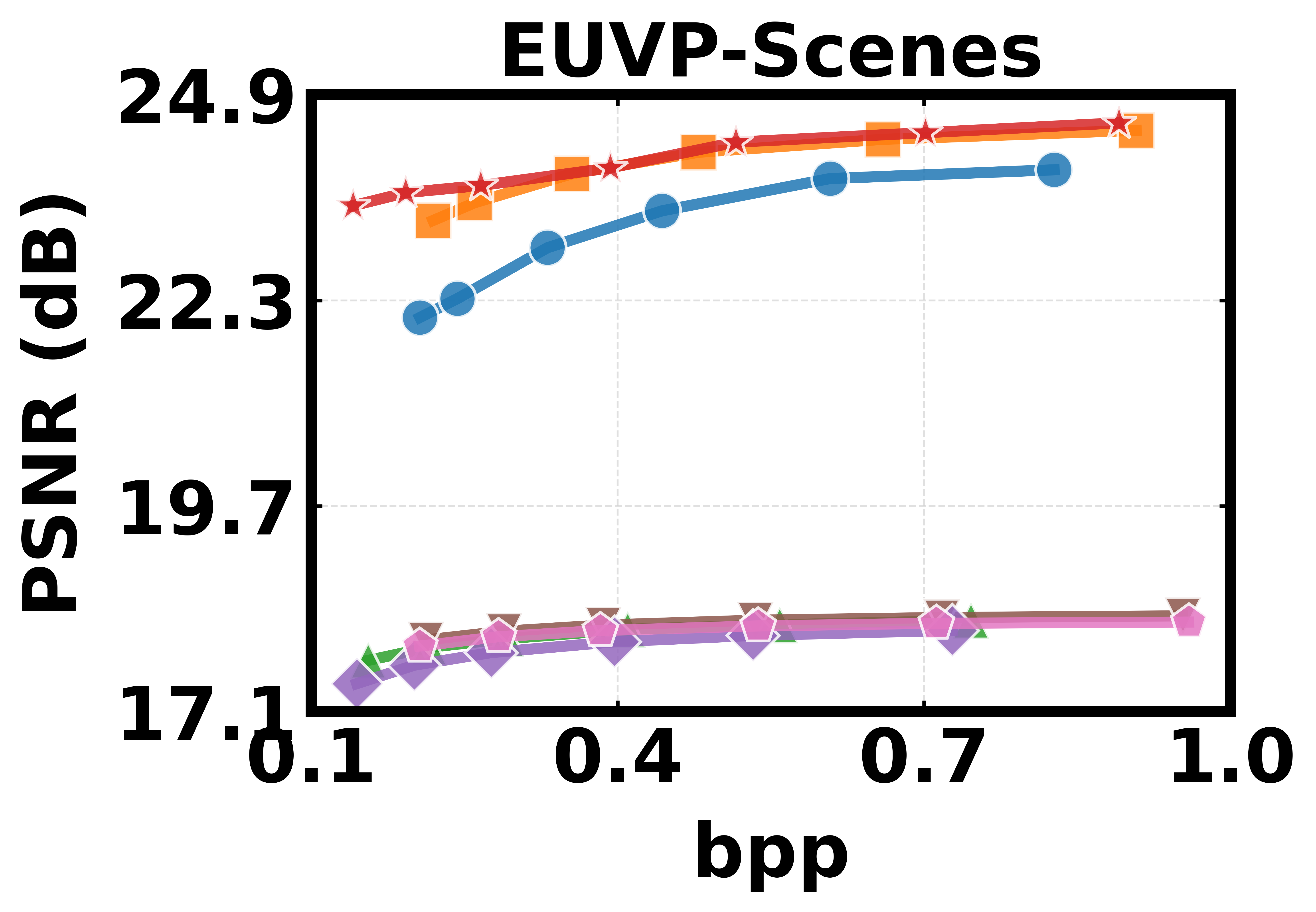}} &
      \multicolumn{2}{c}{\includegraphics[height=0.235\textwidth, width=0.33\textwidth]{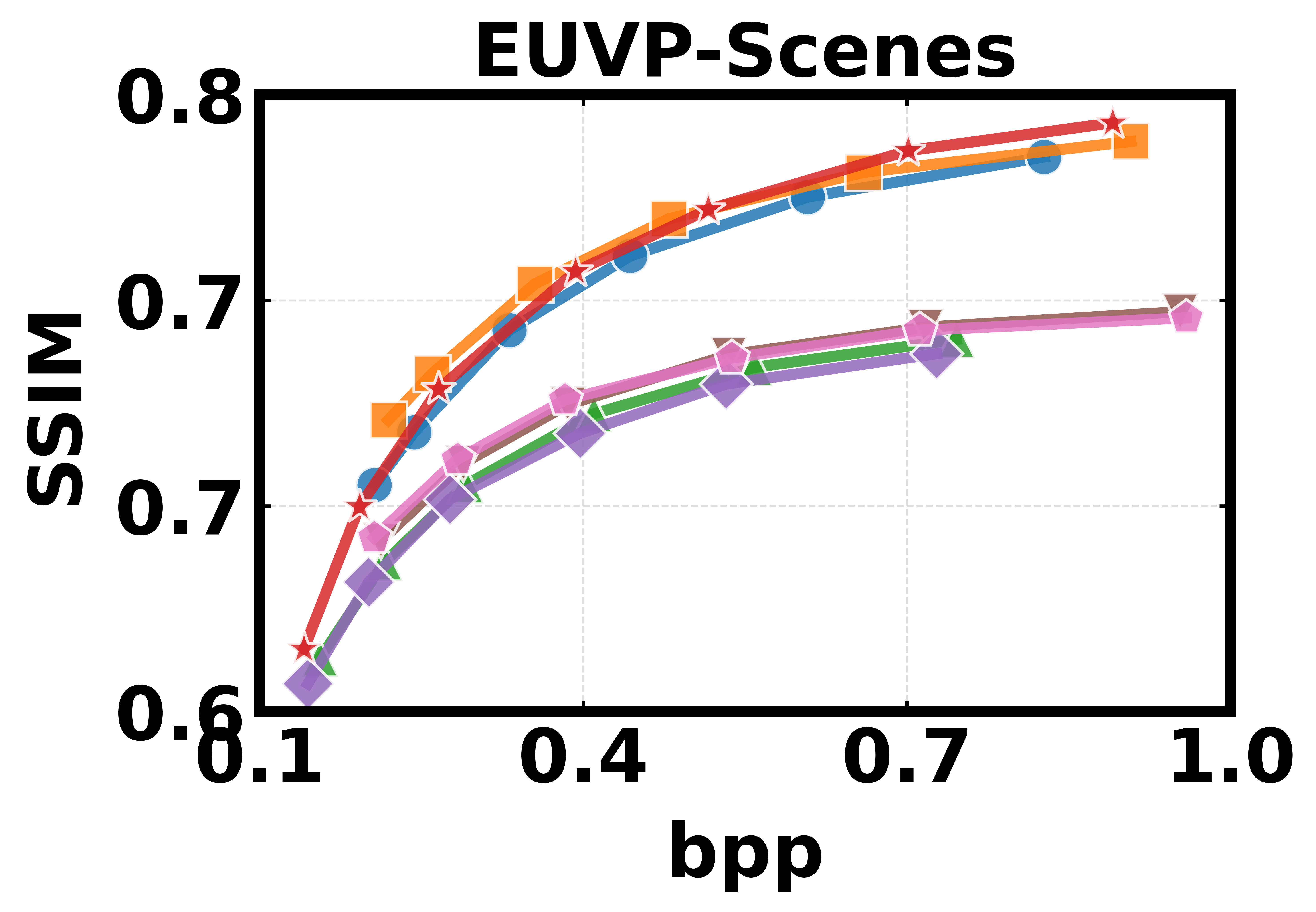}} &
      \multicolumn{2}{c}{\includegraphics[height=0.235\textwidth, width=0.33\textwidth]{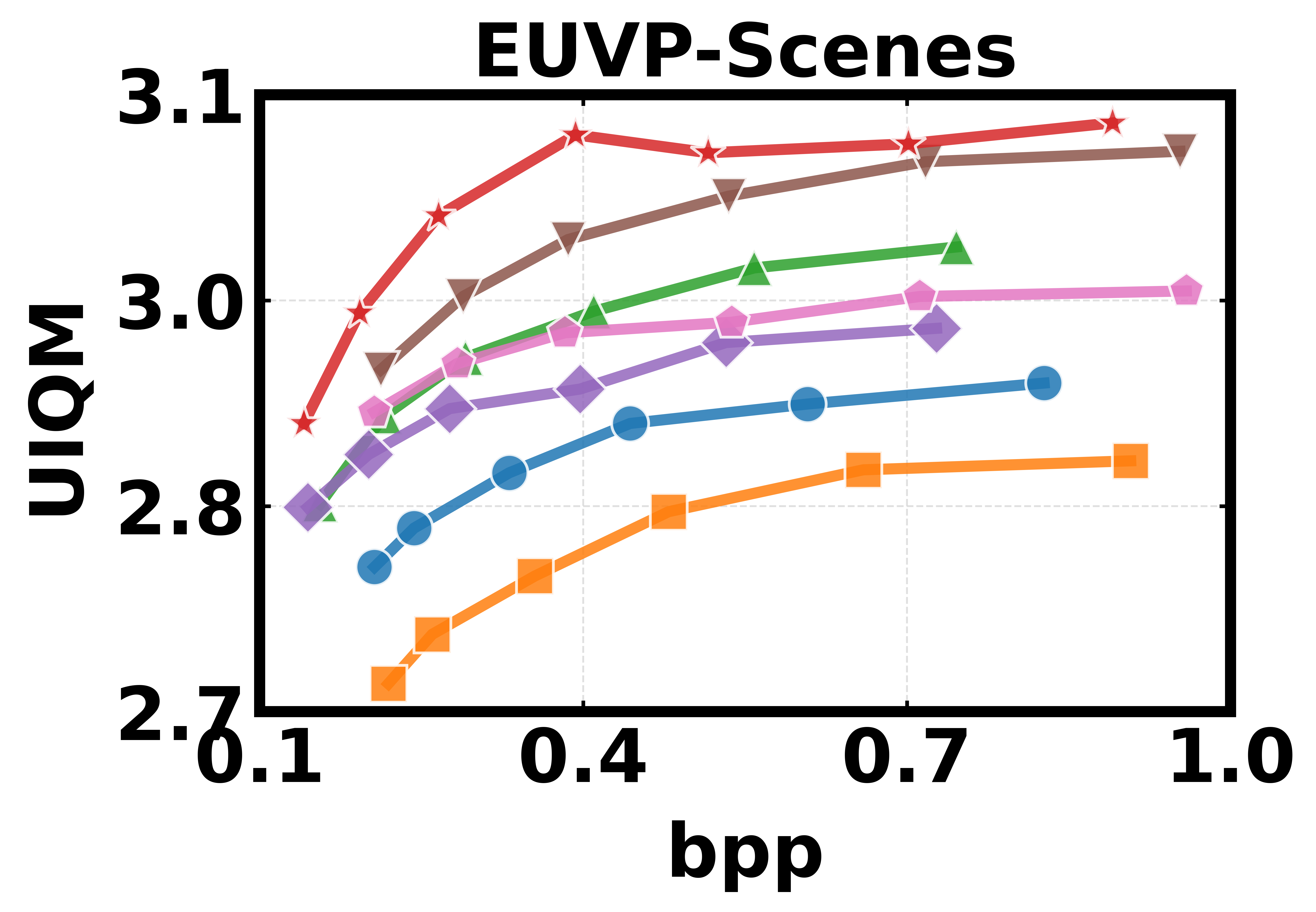}} \\
      
      \multicolumn{2}{c}{\includegraphics[height=0.235\textwidth, width=0.33\textwidth]{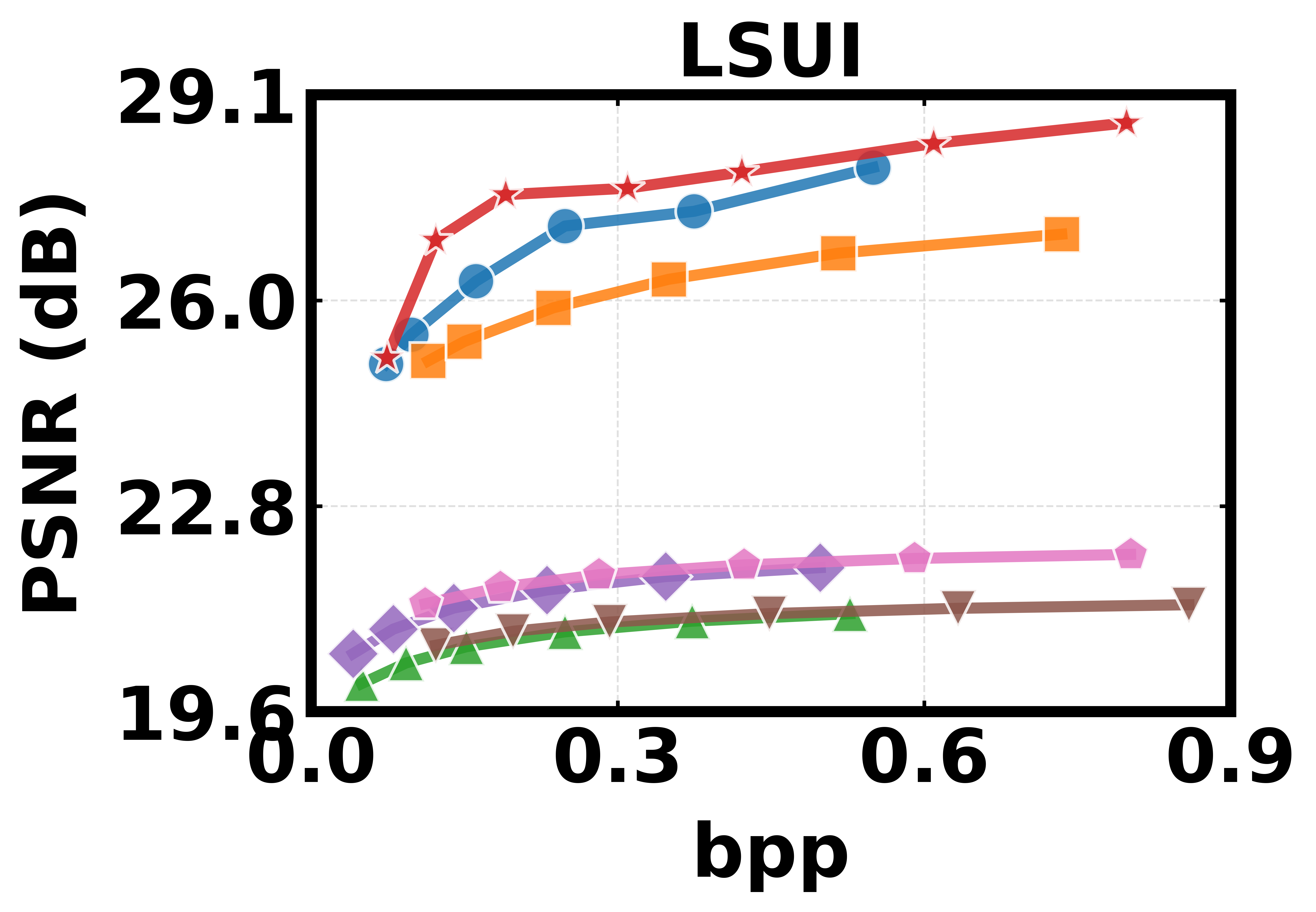}} &
      \multicolumn{2}{c}{\includegraphics[height=0.235\textwidth, width=0.33\textwidth]{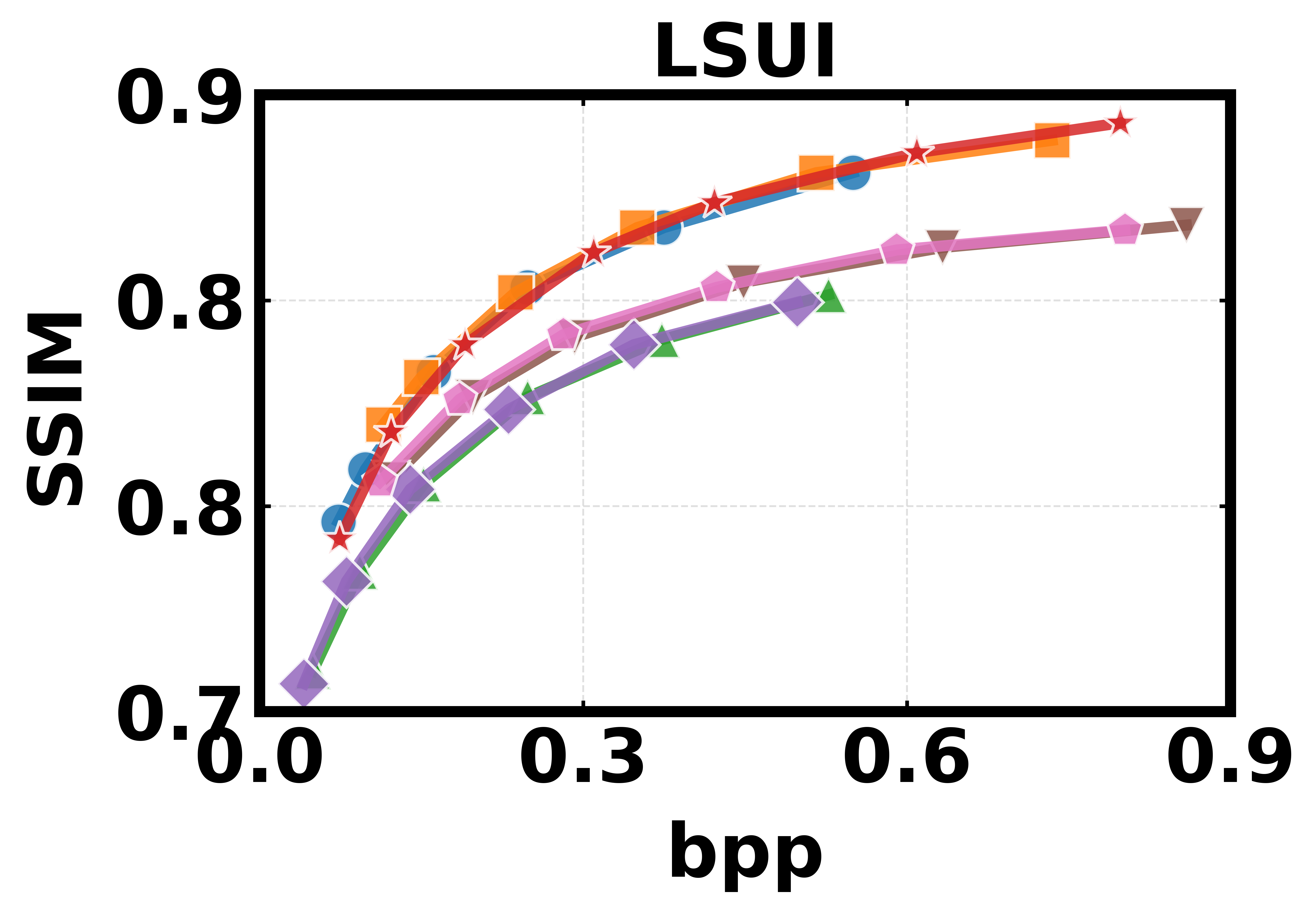}} &
      \multicolumn{2}{c}{\includegraphics[height=0.235\textwidth, width=0.33\textwidth]{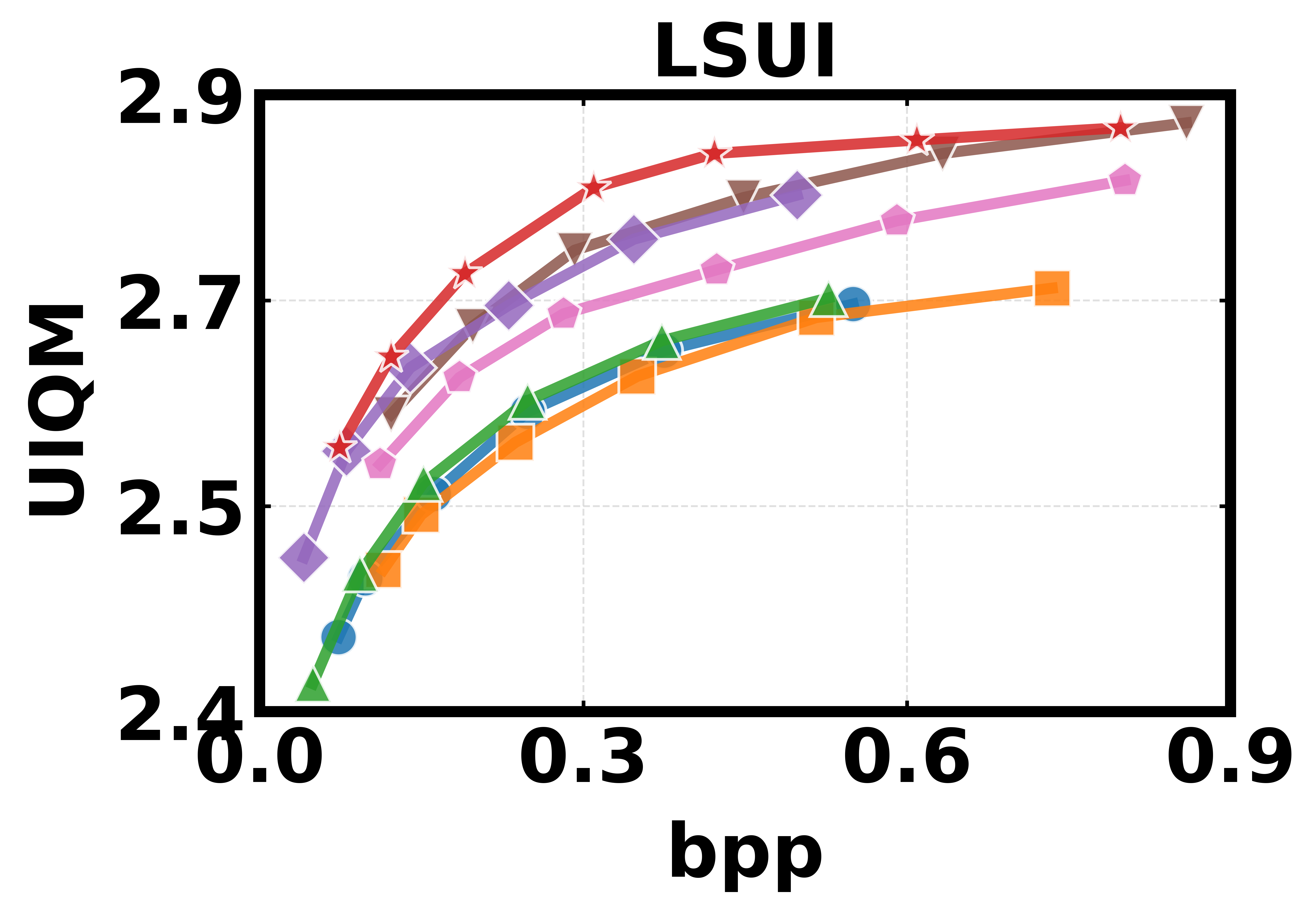}} \\
      
      \multicolumn{6}{c}{\includegraphics[width=0.99\textwidth]{fig/legend.png}} 
 
    \end{tabular}
  \caption{Coding performance comparison with the state-of-the-art works in terms of PSNR, SSIM, and UIQM under the datasets of EUVP-Dark, EUVP-Scenes, and LSUI.}
  \label{curve}
  \end{figure*}
 
\section{Experimental Results}
\subsection{Comparison with the State-of-the-art Works}
To evaluate the performance of the proposed method for both compression and enhancement, three end to end image compression methods (i.e., MLIC++ \cite{MLIC}, TCM \cite{TCM}, SegPIC \cite{SegPIC}) and three underwater image enhancement methods (i.e., HCLR \cite{HCLR}, DM \cite{DM}, SCEIR \cite{SCEIR}) are adopted. Since there are no direct comparative schemes, these image compression methods and underwater image enhancement methods are combined in the cases of Fig. \ref{fig:framework_comparison}(a) and Fig. \ref{fig:framework_comparison}(b). For the framework of compression first and enhancement second, the comparative schemes are MLIC+HCLR, TCM+DM and SegPIC+SCEIR, while for the framework of enhancement first and compression second, the comparative schemes are HCLR+MLIC, DM+TCM and SCEIR+SegPIC. Five large scale underwater image datasets are employed for coding experiment, including EUVP-Dark \cite{EUVP}, UW2023 \cite{10945897}, SUIM \cite{SUIM}, EUVP-Scenes \cite{EUVP} and LSUI \cite{LSUI}. There are 5500, 1000, 1525, 2185 and 4279 underwater images in these datasets, respectively. Here, the reconstructed underwater images after decoding are evaluated by PSNR, SSIM \cite{SSIM} and UIQM \cite{UIQM}. It should be noted that there are no enhanced versions of ground truth for UW2023 and SUIM datasets. Fig. \ref{curve} comprehensively illustrates the rate-distortion performance of the proposed method comparison with the state-of-the-art approaches. The horizontal axis indicates the value of bpp, and the vertical axis indicates the value of PSNR, SSIM or UIQM. Superior coding performance is unambiguously indicated by achieving lower bitrates simultaneously with higher quality scores. Consequently, the optimal method is characterized by its rate-distortion curve occupying the topmost across the evaluated bitrate range. As shown in Fig. \ref{curve}, the proposed method consistently occupies this topmost position in terms of UIQM, demonstrating its superiority over other state-of-the-art methods across diverse underwater image datasets. In terms of SSIM, the proposed method achieves comparative performance when compared with other state-of-the-art works. Moreover, the values of \textbf{Bj{\o}ntegaard Delta Bit Rate} (\textbf{BD-BR}) compared with the state-of-the-art works are presented in Table \ref{table4} and similar results can be observed. 

\begin{figure*}
    \centering
    \small
    \setlength{\tabcolsep}{1pt} 
    \begin{tabular}{c c c c c}
      \includegraphics[height=0.145\textwidth]{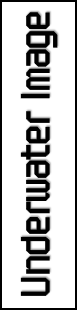} &
      \includegraphics[height=0.145\textwidth, width=0.23\textwidth]{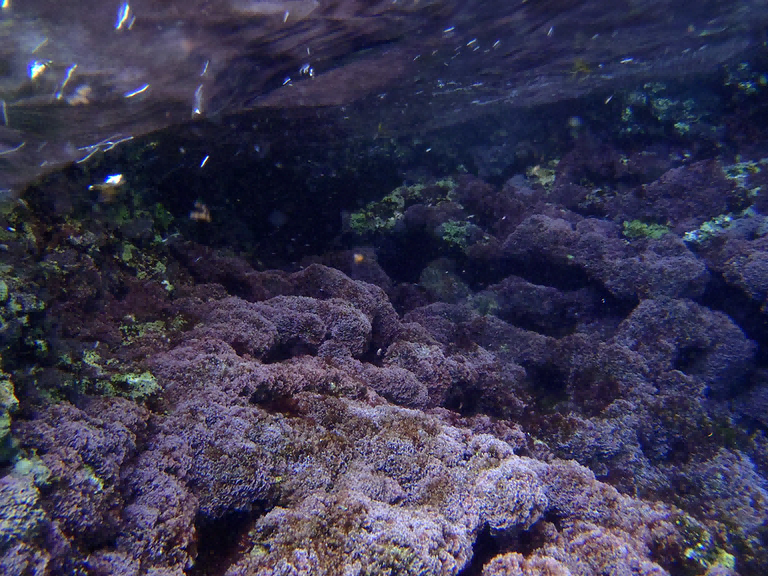} &
      \includegraphics[height=0.145\textwidth, width=0.23\textwidth]{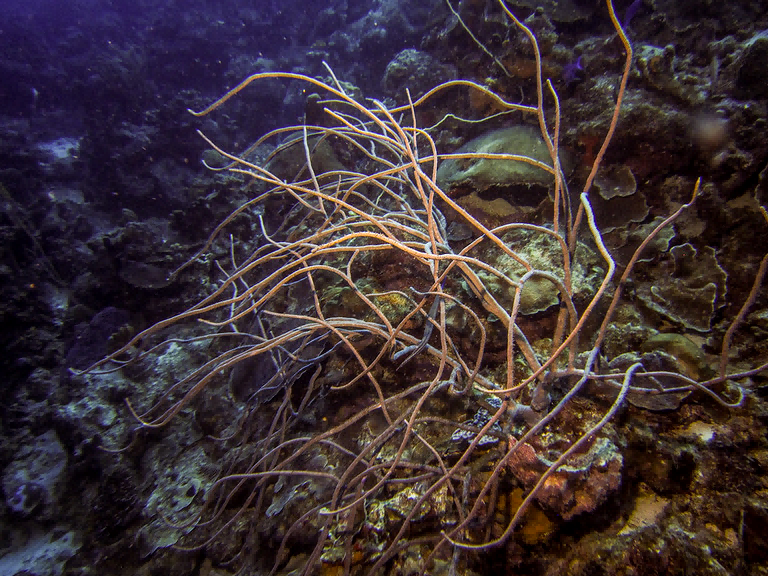} &
      \includegraphics[height=0.145\textwidth, width=0.23\textwidth]{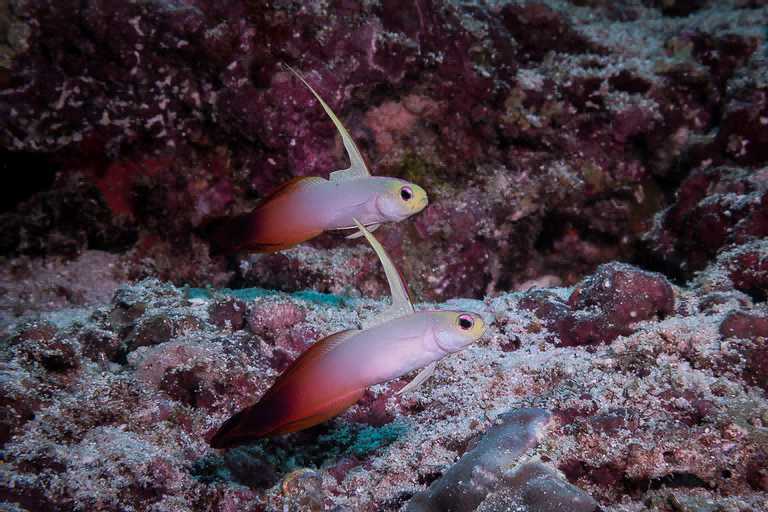} &
      \includegraphics[height=0.145\textwidth, width=0.23\textwidth]{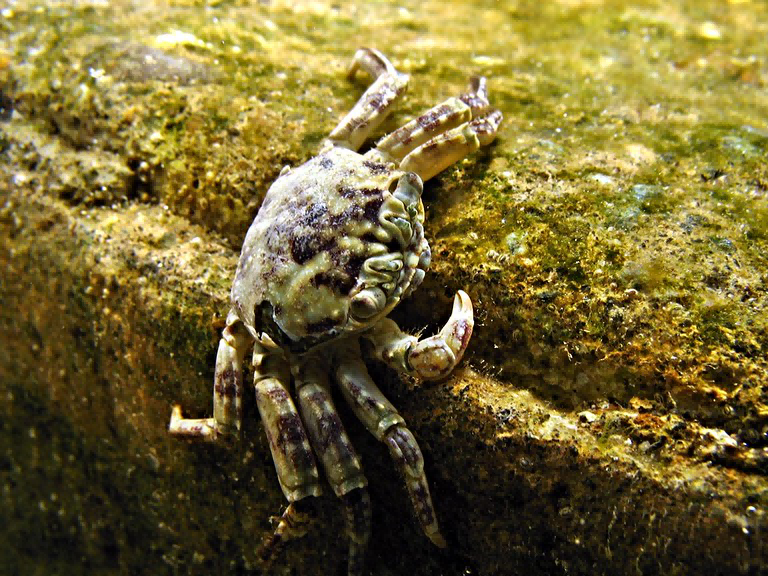} \\
    
      \includegraphics[height=0.145\textwidth]{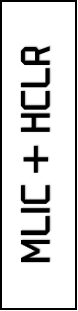} &
      \includegraphics[height=0.145\textwidth, width=0.23\textwidth]{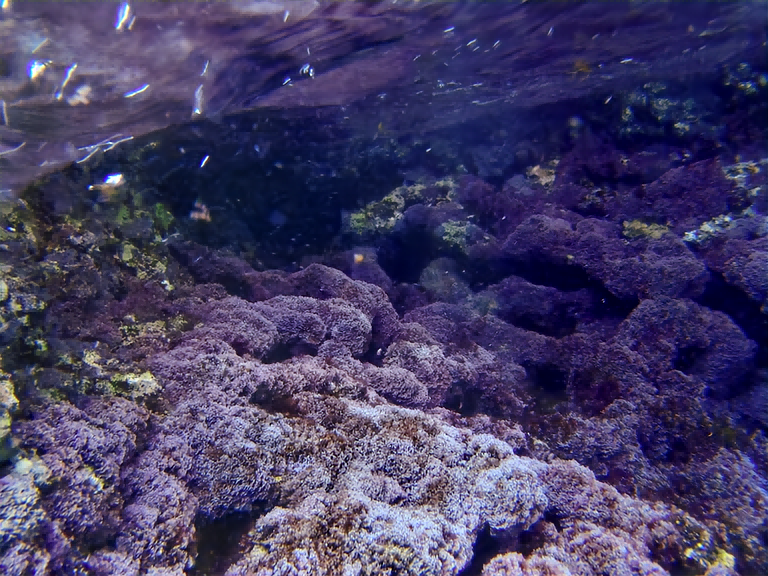} &
      \includegraphics[height=0.145\textwidth, width=0.23\textwidth]{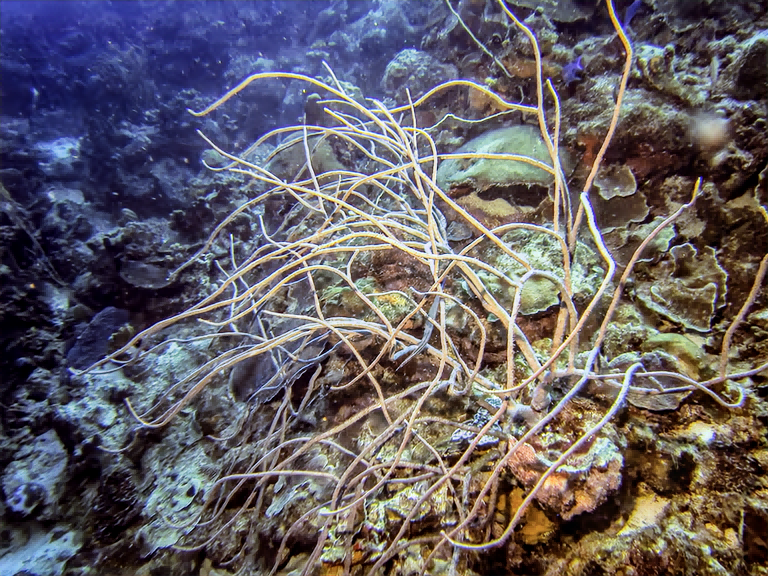} &
      \includegraphics[height=0.145\textwidth, width=0.23\textwidth]{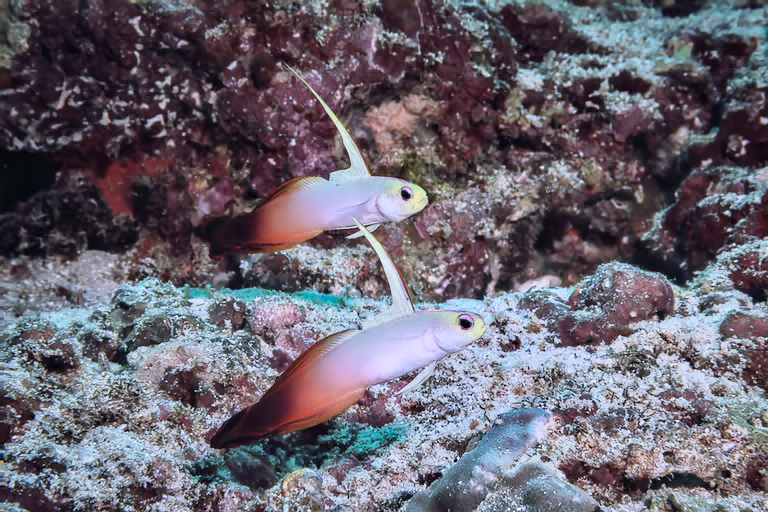} &
      \includegraphics[height=0.145\textwidth, width=0.23\textwidth]{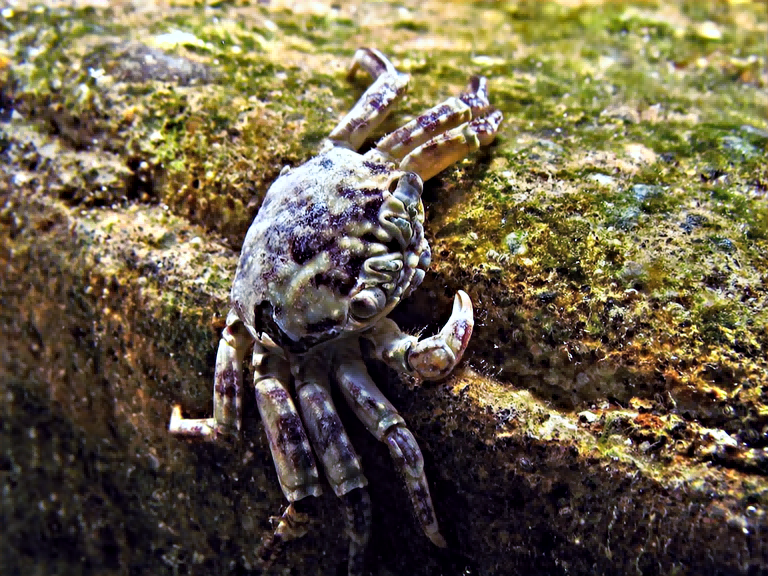} \\

      &bpp: 1.24, UIQM: 2.845 &bpp: 1.28, UIQM: 3.055  & bpp: 1.30, UIQM: 3.057 &  bpp: 1.28, UIQM: 2.420 \\

      \includegraphics[height=0.145\textwidth]{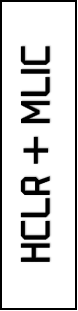} &
      \includegraphics[height=0.145\textwidth, width=0.23\textwidth]{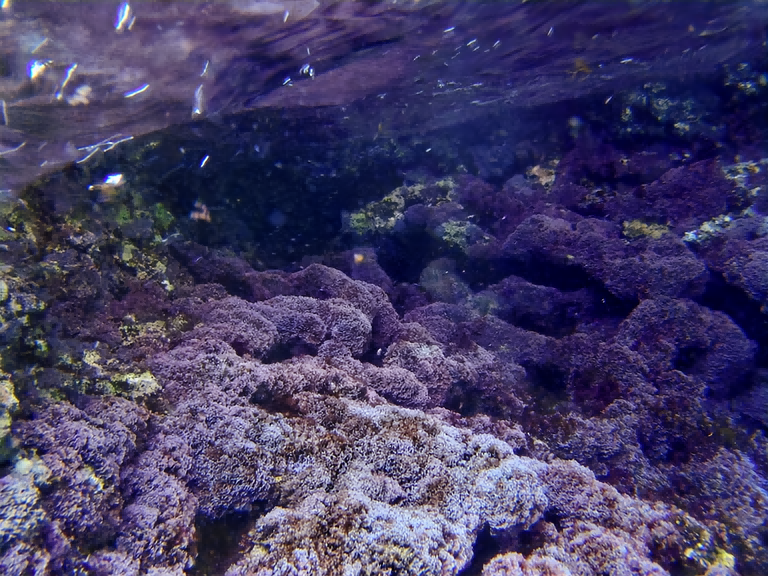} &
      \includegraphics[height=0.145\textwidth, width=0.23\textwidth]{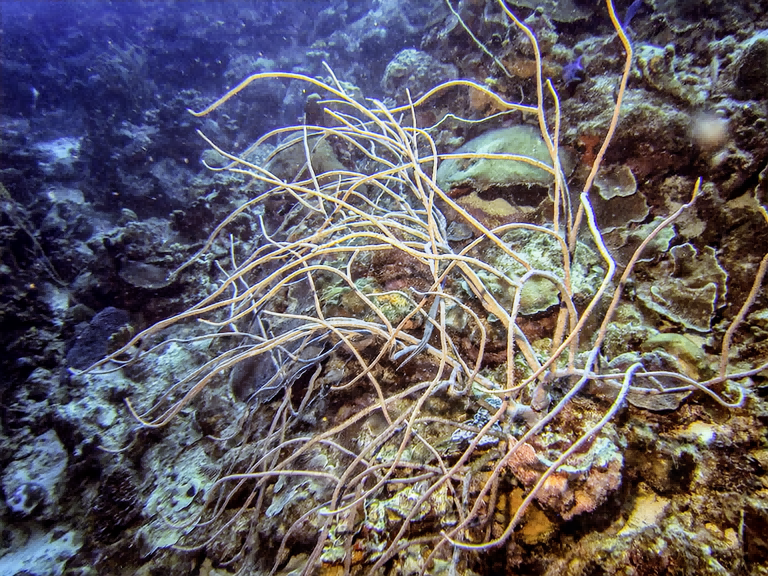} &
      \includegraphics[height=0.145\textwidth, width=0.23\textwidth]{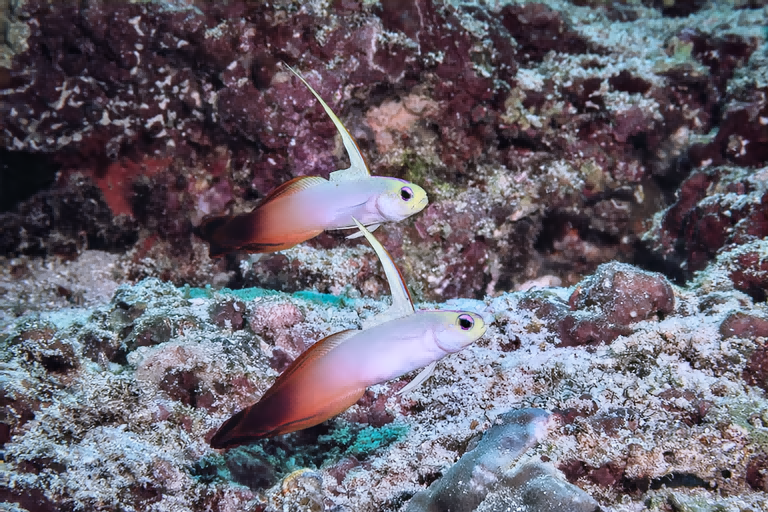} &
      \includegraphics[height=0.145\textwidth, width=0.23\textwidth]{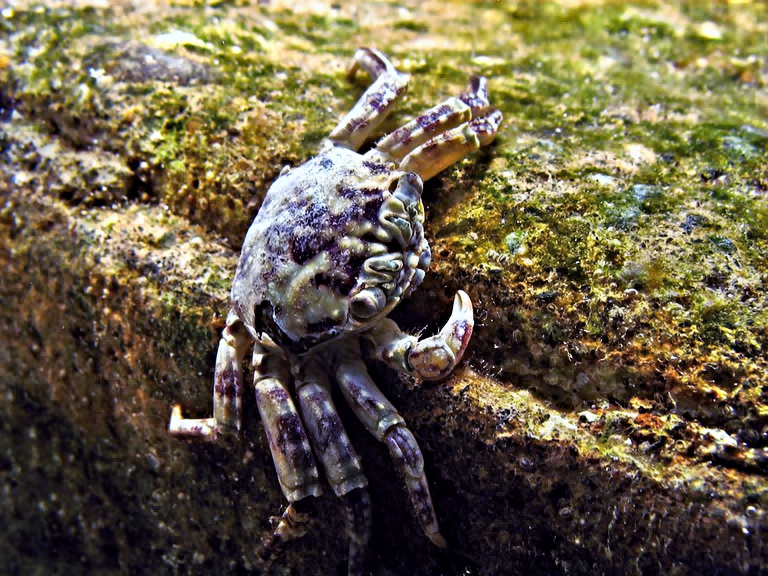} \\

      &bpp: 1.46, UIQM: 2.976 &bpp: 1.90, UIQM: 3.179 &bpp: 1.65, UIQM: {3.497} &bpp: 1.51, UIQM: 2.686 \\

      \includegraphics[height=0.145\textwidth]{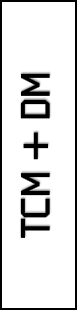} &
      \includegraphics[height=0.145\textwidth, width=0.23\textwidth]{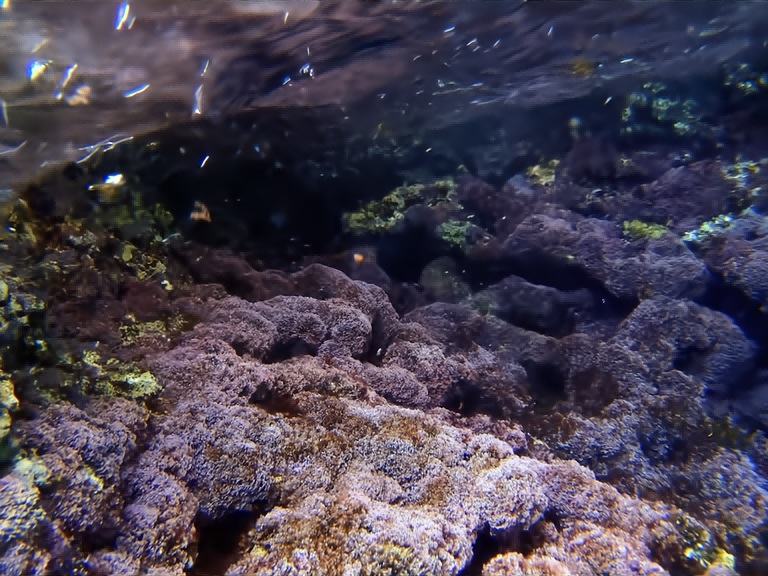} &
      \includegraphics[height=0.145\textwidth, width=0.23\textwidth]{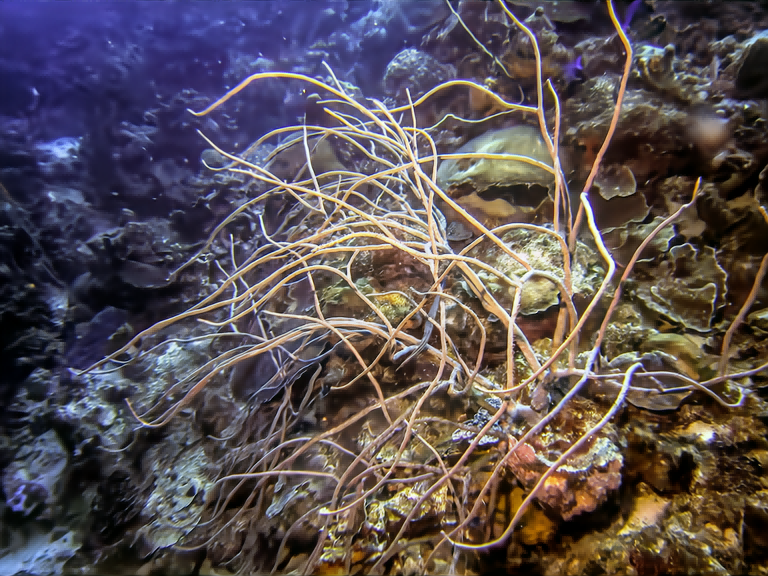} &
      \includegraphics[height=0.145\textwidth, width=0.23\textwidth]{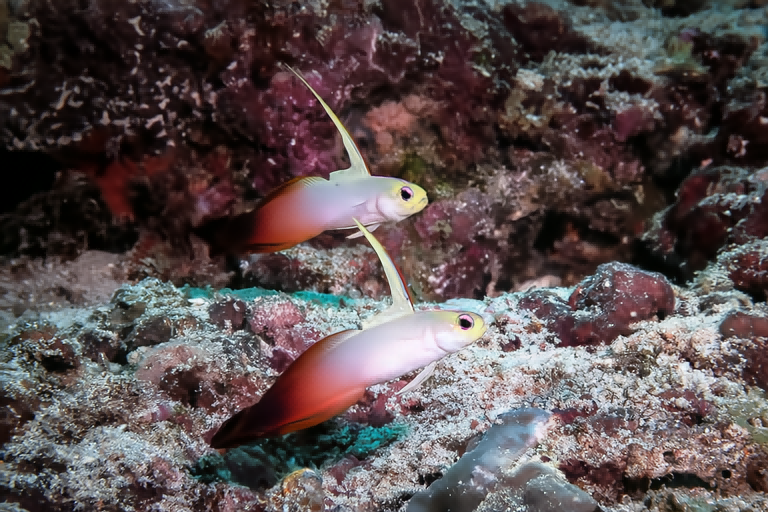} &
      \includegraphics[height=0.145\textwidth, width=0.23\textwidth]{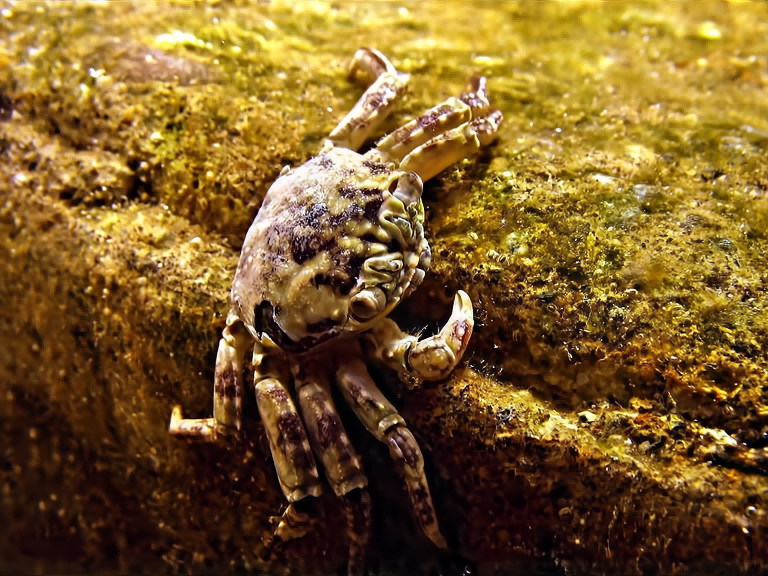} \\

       &bpp: 1.47, UIQM: 3.227 &bpp: 1.50, UIQM: 3.408 & bpp: 1.50, UIQM: 3.273 &  bpp: 1.47, UIQM: 2.493 \\

      \includegraphics[height=0.145\textwidth]{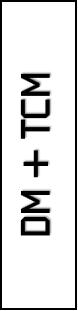} &
      \includegraphics[height=0.145\textwidth, width=0.23\textwidth]{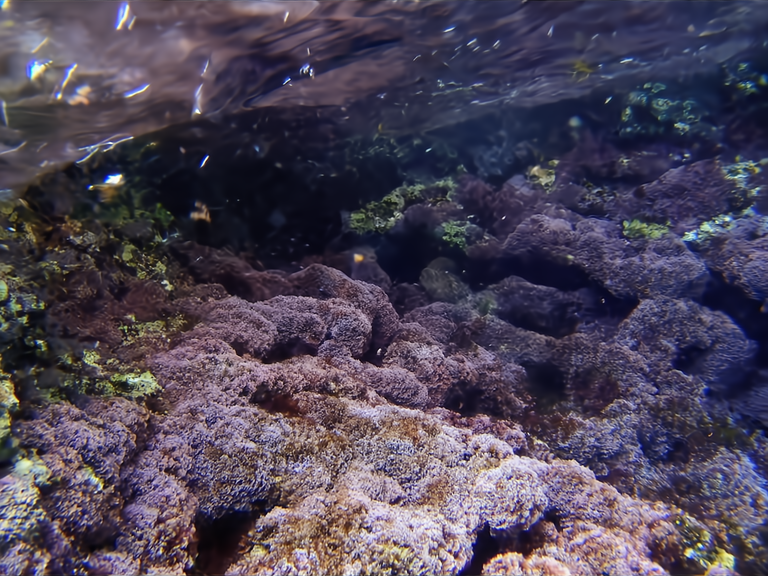} &
      \includegraphics[height=0.145\textwidth, width=0.23\textwidth]{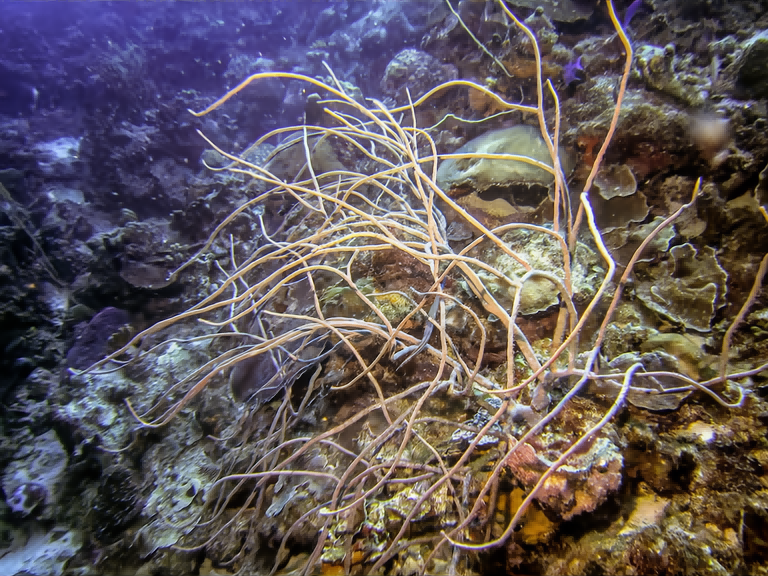} &
      \includegraphics[height=0.145\textwidth, width=0.23\textwidth]{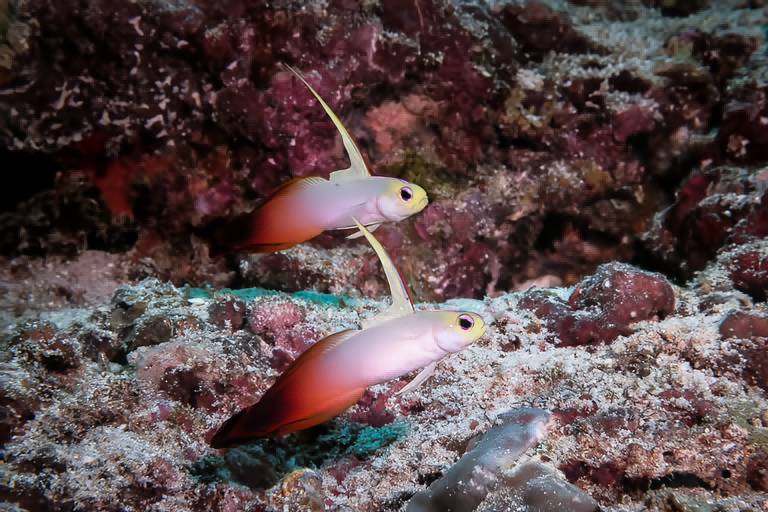} &
      \includegraphics[height=0.145\textwidth, width=0.23\textwidth]{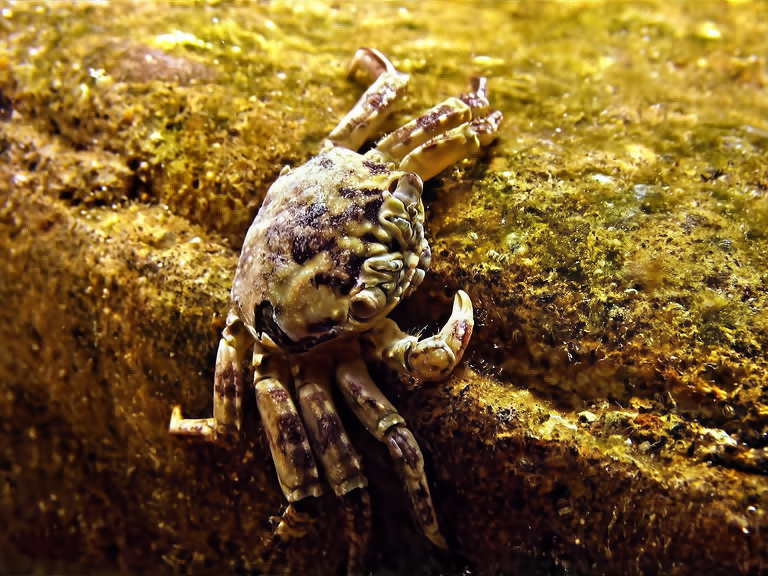} \\

       &bpp: 1.76, UIQM: 3.260&bpp: 2.09, UIQM: 3.391& bpp: 1.68, UIQM: 3.181 &  bpp: 1.59, UIQM: 2.622 \\

      \includegraphics[height=0.145\textwidth]{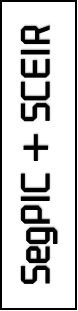} &
      \includegraphics[height=0.145\textwidth, width=0.23\textwidth]{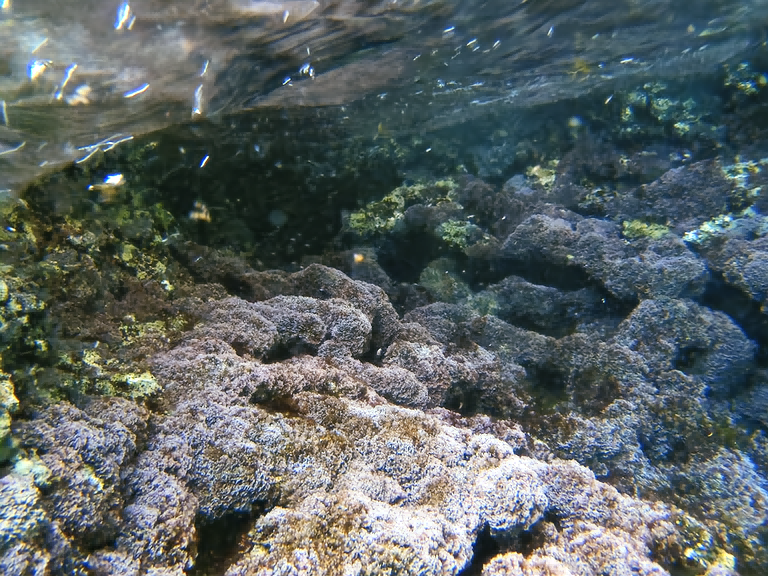} &
      \includegraphics[height=0.145\textwidth, width=0.23\textwidth]{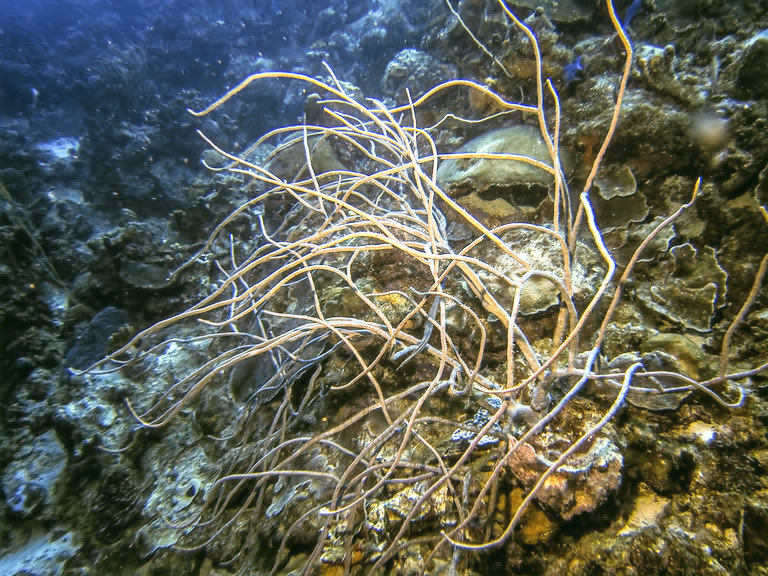} &
      \includegraphics[height=0.145\textwidth, width=0.23\textwidth]{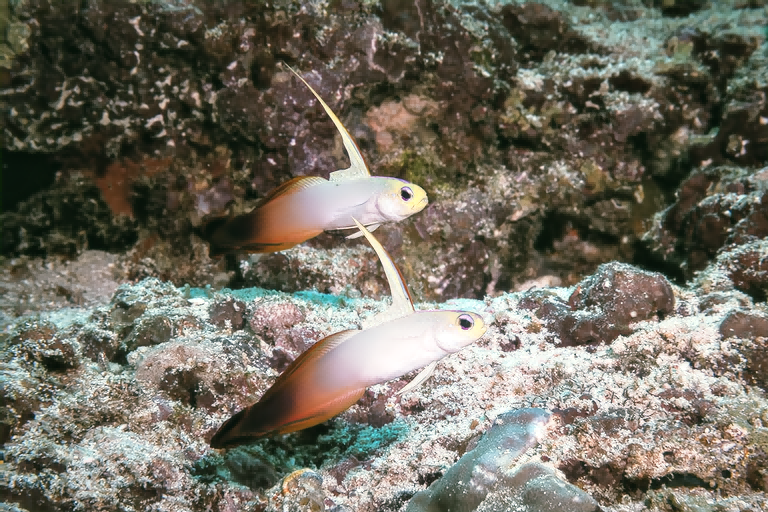} &
      \includegraphics[height=0.145\textwidth, width=0.23\textwidth]{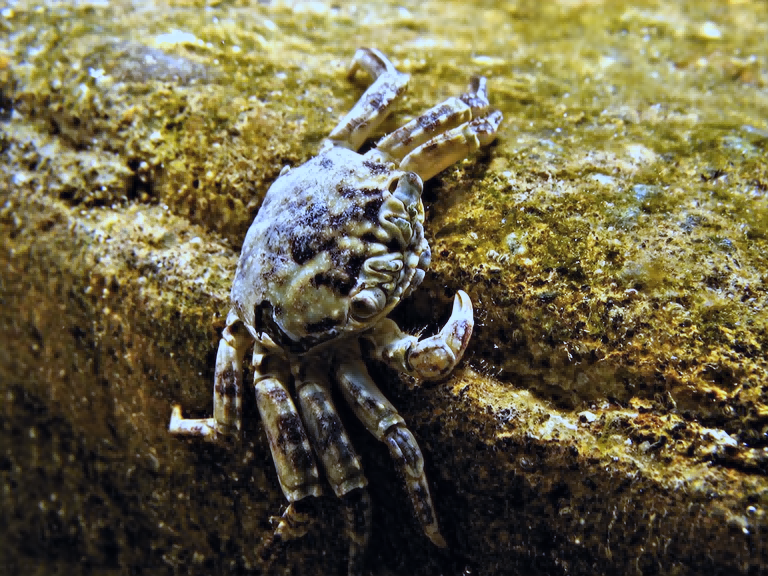} \\

       &bpp: 1.46, UIQM: 3.293&bpp: 1.49, UIQM: 3.483& bpp: 1.51, UIQM: 3.409 &  bpp: 1.49, UIQM: {3.187} \\

      \includegraphics[height=0.145\textwidth]{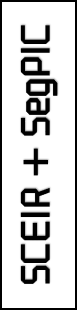} &
      \includegraphics[height=0.145\textwidth, width=0.23\textwidth]{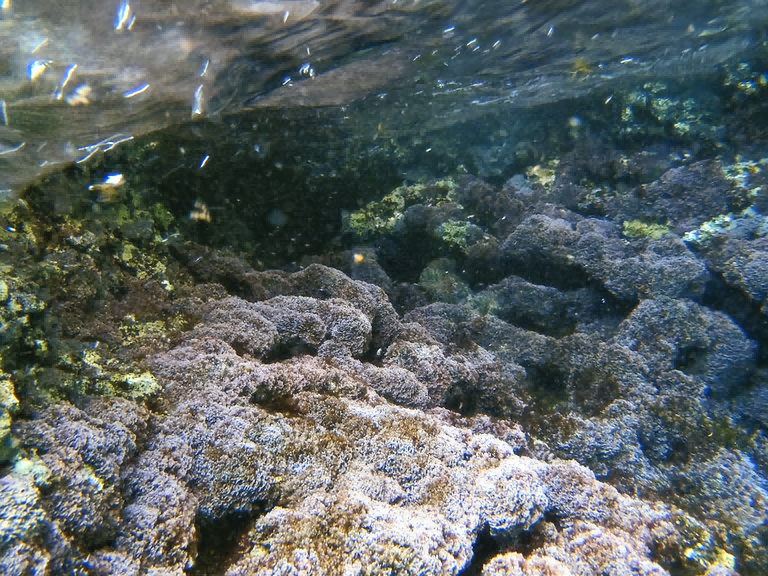} &
      \includegraphics[height=0.145\textwidth, width=0.23\textwidth]{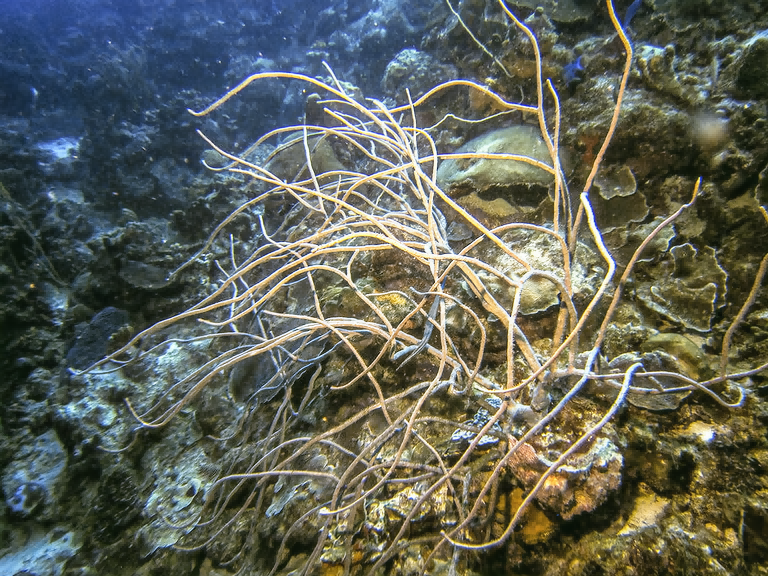} &
      \includegraphics[height=0.145\textwidth, width=0.23\textwidth]{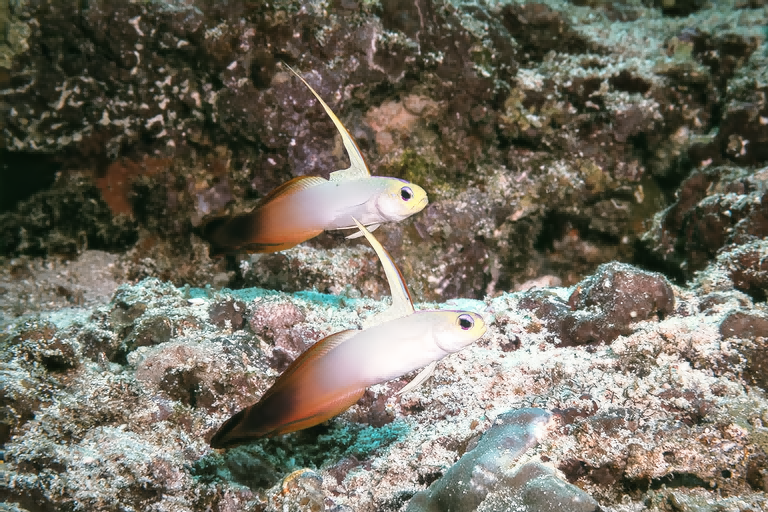} &
      \includegraphics[height=0.145\textwidth, width=0.23\textwidth]{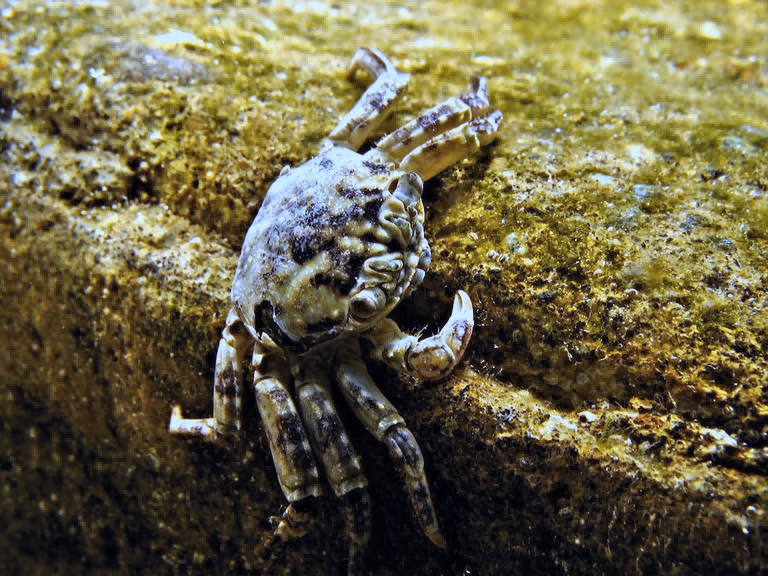} \\

       &bpp: 1.97, UIQM: 3.408&bpp: 2.06, UIQM: 3.517& bpp: 1.82, UIQM: 3.401 &  bpp: 1.50, UIQM: 3.165 \\

      \includegraphics[height=0.145\textwidth]{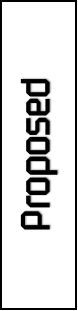} &
      \includegraphics[height=0.145\textwidth, width=0.23\textwidth]{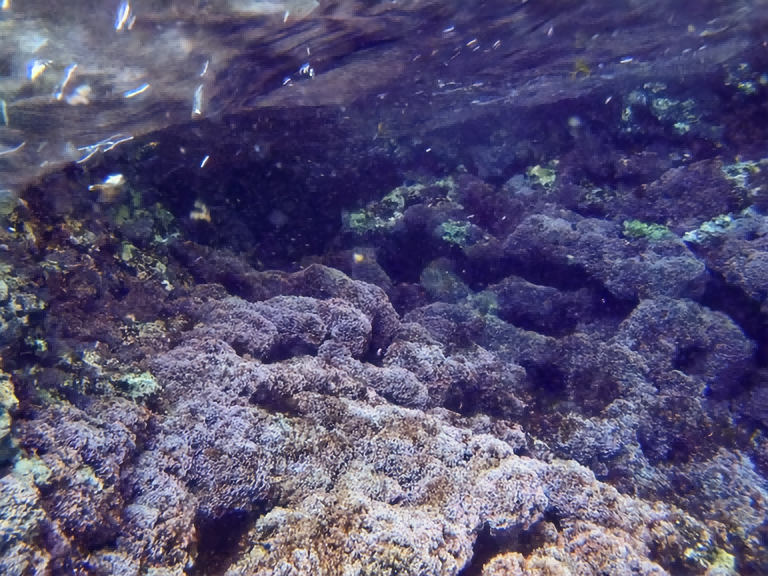} &
      \includegraphics[height=0.145\textwidth, width=0.23\textwidth]{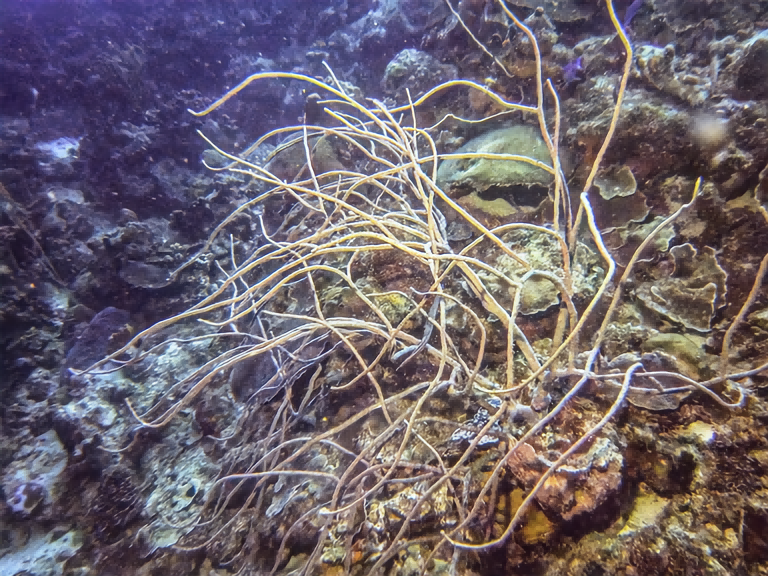} &
      \includegraphics[height=0.145\textwidth, width=0.23\textwidth]{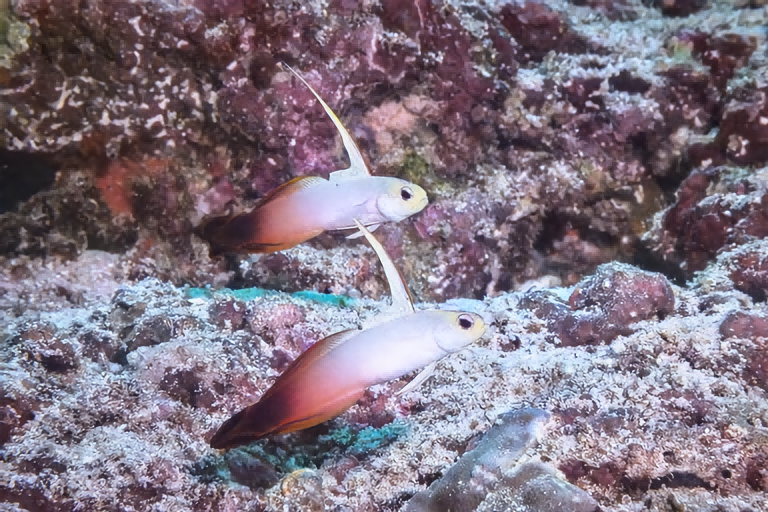} &
      \includegraphics[height=0.145\textwidth, width=0.23\textwidth]{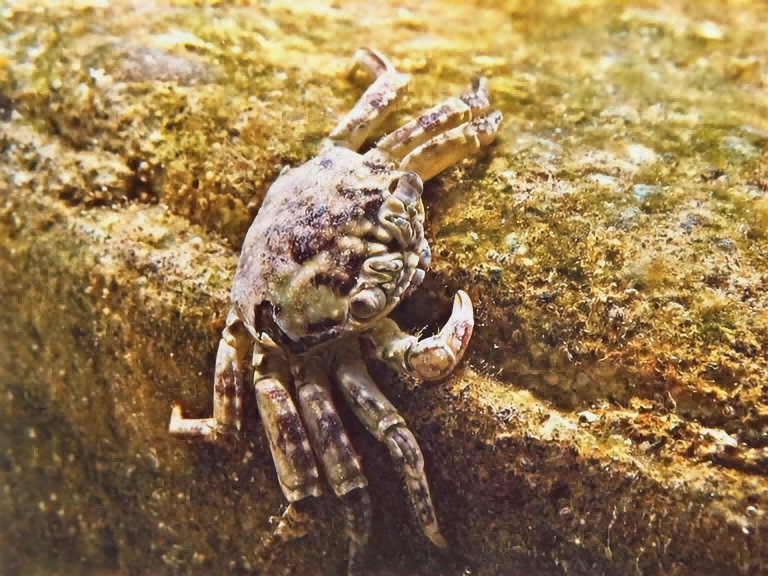} \\

       &bpp: {0.93}, UIQM: {3.548}&bpp: {0.99}, UIQM: {3.609} &bpp: {0.89}, UIQM: 3.429&bpp: {0.79}, UIQM: 3.014 \\

  \end{tabular}
  \caption{Visual comparison with the state-of-the-art works.}
  \label{fig:visual}
  \end{figure*}

\begin{figure*}
    \centering
    \small
    \setlength{\tabcolsep}{1pt} 
    \begin{tabular}{c c c c c c c c c c c c c c c c c c c c c c c c c c c c c}
      \includegraphics[height=0.17\textwidth]{fig/case-org.png} &
      \includegraphics[height=0.17\textwidth]{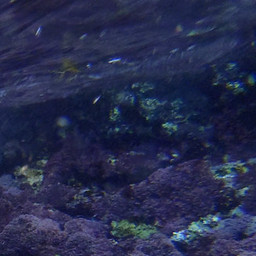} & & & & & & & & &
      \includegraphics[height=0.17\textwidth]{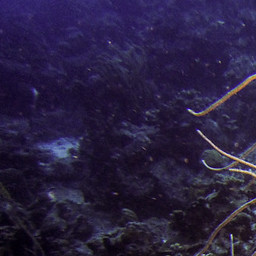} & & & & & & & & &
      \includegraphics[height=0.17\textwidth]{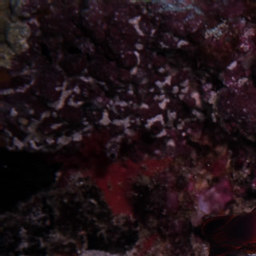} & & & & & & & & &
      \includegraphics[height=0.17\textwidth]{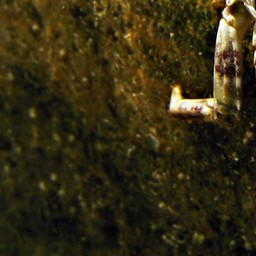} \\
    
      \includegraphics[height=0.17\textwidth]{fig/case1.png} &
      \includegraphics[height=0.17\textwidth]{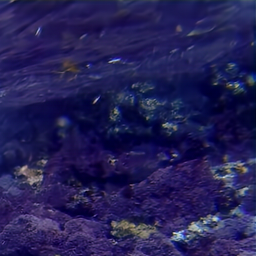} & & & & & & & & &
      \includegraphics[height=0.17\textwidth]{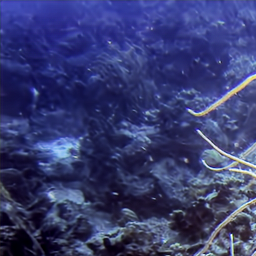} & & & & & & & & &
      \includegraphics[height=0.17\textwidth]{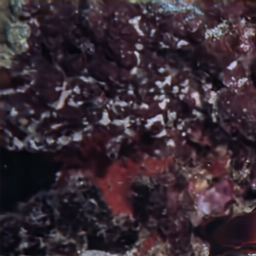} & & & & & & & & &
      \includegraphics[height=0.17\textwidth]{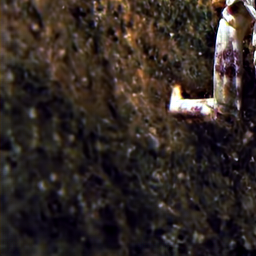} \\

      \includegraphics[height=0.17\textwidth]{fig/case2.png} &
      \includegraphics[height=0.17\textwidth]{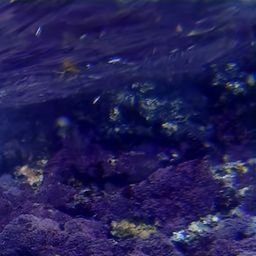} & & & & & & & & &
      \includegraphics[height=0.17\textwidth]{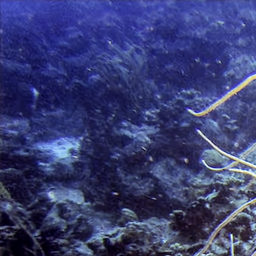} & & & & & & & & &
      \includegraphics[height=0.17\textwidth]{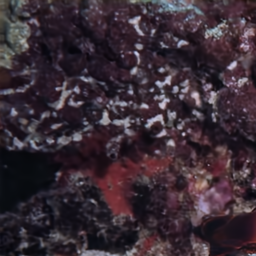} & & & & & & & & &
      \includegraphics[height=0.17\textwidth]{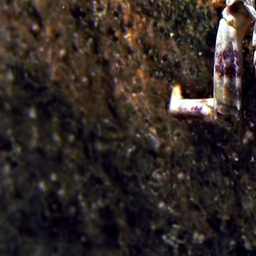} \\

      \includegraphics[height=0.17\textwidth]{fig/case3.png} &
      \includegraphics[height=0.17\textwidth]{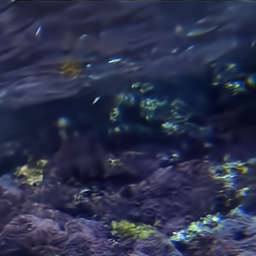} & & & & & & & & &
      \includegraphics[height=0.17\textwidth]{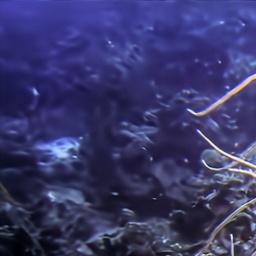} & & & & & & & & &
      \includegraphics[height=0.17\textwidth]{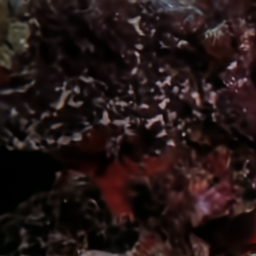} & & & & & & & & &
      \includegraphics[height=0.17\textwidth]{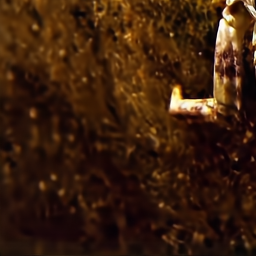} \\

      \includegraphics[height=0.17\textwidth]{fig/case4.png} &
      \includegraphics[height=0.17\textwidth]{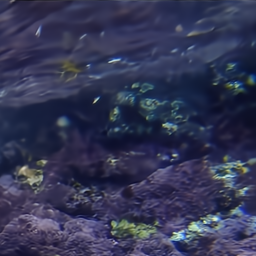} & & & & & & & & &
      \includegraphics[height=0.17\textwidth]{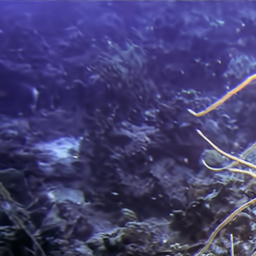} & & & & & & & & &
      \includegraphics[height=0.17\textwidth]{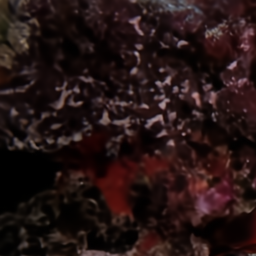} & & & & & & & & &
      \includegraphics[height=0.17\textwidth]{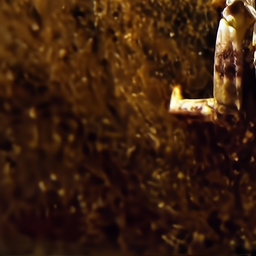} \\

      \includegraphics[height=0.17\textwidth]{fig/case5.png} &
      \includegraphics[height=0.17\textwidth]{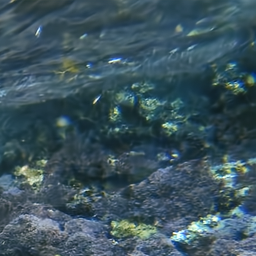} & & & & & & & & &
      \includegraphics[height=0.17\textwidth]{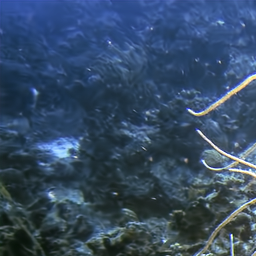} & & & & & & & & &
      \includegraphics[height=0.17\textwidth]{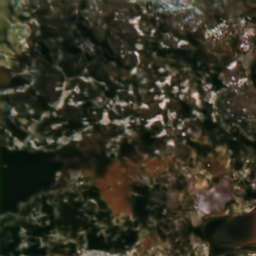} & & & & & & & & &
      \includegraphics[height=0.17\textwidth]{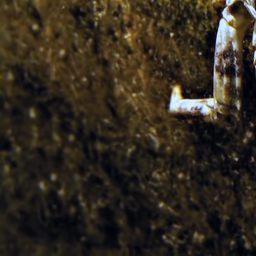} \\

      \includegraphics[height=0.17\textwidth]{fig/case6.png} &
      \includegraphics[height=0.17\textwidth]{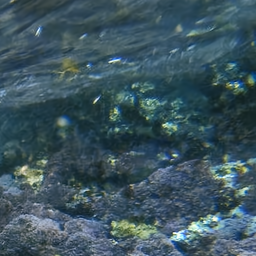} & & & & & & & & &
      \includegraphics[height=0.17\textwidth]{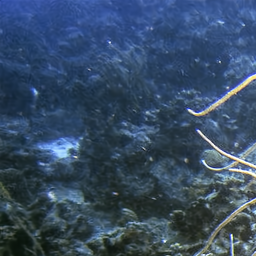} & & & & & & & & &
      \includegraphics[height=0.17\textwidth]{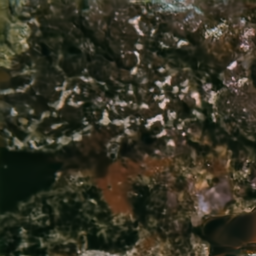} & & & & & & & & &
      \includegraphics[height=0.17\textwidth]{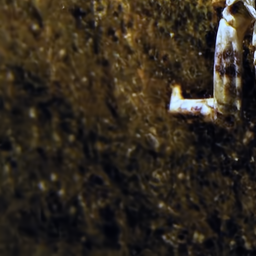} \\

      \includegraphics[height=0.17\textwidth]{fig/case7.png} &
      \includegraphics[height=0.17\textwidth]{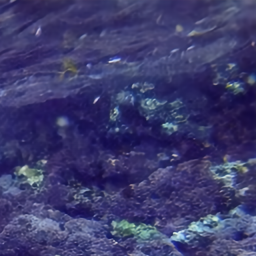} & & & & & & & & &
      \includegraphics[height=0.17\textwidth]{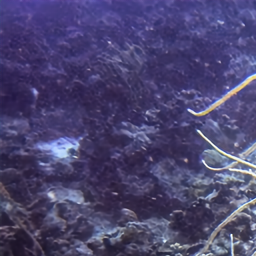} & & & & & & & & &
      \includegraphics[height=0.17\textwidth]{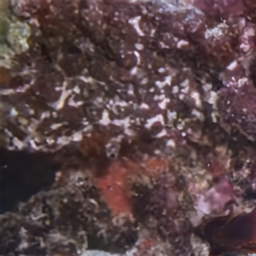} & & & & & & & & &
      \includegraphics[height=0.17\textwidth]{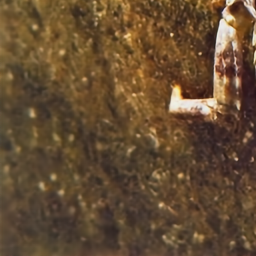} \\

  \end{tabular}
  \caption{Visual comparison (cropped versions of Fig. \ref{fig:visual}).}
  \label{fig:visual2}
  \end{figure*}

Additionally, Figs. \ref{fig:visual} and \ref{fig:visual2} provide visual comparison with the state-of-the-art methods. Four underwater images from the UW2023 dataset are displayed. These images all exhibit very rich textural details, and corresponding bpp and UIQM values are reported as well. The first row shows the original underwater images, while subsequent rows present reconstructed results after decoding. Subjectively, the proposed method achieves superior reconstruction quality compared to existing approaches. For instance, enhanced details are notably visible in the underwater image featuring a crab when compared with the other reconstructed underwater images.

\begin{table*}[t]\caption{Performance comparison in terms of BD-BR (\%). ``NA" indicates that the value cannot be calculated.} \label{table4}
\begin{center}
\footnotesize
    \begin{tabular}{|c|c|c|c|c|c|c|c|c|c|c|c|}
    \cline {1-12} 
    \multirow{2}{*}{Pro. vs. Scheme}&\multicolumn{3}{c|}{EUVP-Dark}&\multicolumn{3}{c|}{EUVP-Scenes}&\multicolumn{3}{c|}{LSUI}&{SUIM}&{UW2023}\\
    \cline {2-12}
    &{PSNR}&{SSIM}&{UIQM}&{PSNR}&{SSIM}&{UIQM}&{PSNR}&{SSIM}&{UIQM}&{UIQM}&{UIQM}  \\
    \hline
    \hline
    {MLIC+HCLR}&NA&-24.43&-68.03&NA&-34.57&-65.81&NA&-34.89&-23.16&-69.99&-97.40  \\
    {TCM+DM}      &-33.11&+6.28&-67.15&-61.65&-5.94&-78.54&-38.52&+1.17&-58.75&-98.78&NA  \\
    {SegPIC+SCEIR}      &NA&-25.70&-81.67&NA&-33.88&-52.95&NA&-36.97&-56.02&-86.72&-96.32  \\
    \hline
    {HCLR+MLIC}&NA&-18.93&-76.53&NA&-34.73&-56.36&NA&-27.82&-44.09&-82.56&-88.85  \\
    {DM+TCM}      &-68.17&-1.48&-78.04&-15.54&+4.79&-88.94&-66.95&+3.67&-63.29&NA&NA  \\
    {SCEIR+SegPIC}     &NA&-23.13&-53.42&NA&-36.41&-35.70&NA&-31.17&-25.54&-77.96&-88.11  \\
    \hline
   \end{tabular}
\end{center}
\end{table*}

\subsection{Computational Complexity Comparison}
As presented in Table \ref{table3}, the computational complexity is analyzed in terms of the number of network parameters and running time. It should be noted that the running time is the average time of processing an underwater image. The results show that the numbers of network parameter for MLIC+HCLR and HCLR+MLIC, TCM+DM and DM+TCM, SegPIC+SCEIR and SCEIR+SegPIC are 70.63M, 55.67M, and 83.64M, respectively. In contrast, the number of network parameters in the proposed method is the smallest, i.e., 32.86M. In terms of running time, the computational complexity is greater than other methods due to the module of sparse representation. However, the proposed method can significantly outperform other methods in terms of coding performance.

\subsection{BL and EL Comparison}
In addition, the performances of BL and EL are compared in detail and the results are illustrated in Table \ref{table2}. Here, the EUVP-Dark dataset is employed and the quality metric is UIQM. The coding results under different compression levels are reported. It should be noted that the value of bpp in BL is $\mathcal{R}_{s}(y) + \mathcal{R}_{s}(z)$ and the value of UIQM in BL is calculated from the output of BL, while the value of bpp in EL is $\mathcal{R}_{s}(y) + \mathcal{R}_{s}(z) + \mathcal{R}_{r}(y) + \mathcal{R}_{r}(z)$ and the value of UIQM in EL is calculated from the output of EL.

For BL, the values of bpp and UIQM change from 0.031 to 0.099 and from 2.753 to 3.035. For EL, the values of bpp and UIQM change from 0.131 to 1.098 and from 2.893 to 3.089. It can be observed that the value of bpp in BL is much smaller than that in EL for each compression level, which benefits from the limited non-zero sparse coefficients compression, while the values of UIQM in BL are close to those in EL.

  \begin{table*}[t]\caption{Computational complexity comparison.} \label{table3}
\footnotesize
\begin{center}
    \begin{tabular}{|c|c|c|c|c|c|c|}
    \cline {1-7} 
    \multirow{2}{*}{Scheme}&\multicolumn{2}{c|}{Parameters (M)}&\multicolumn{4}{c|}{Running Time (s)}\\
    \cline {2-7}
    &{Compression}&{Enhancement}&{Encoding}&{Decoding}&{Enhancement}&{Total}  \\
    \hline
    \hline
    {MLIC+HCLR}&{65.76}&{4.87}&{0.11}&{0.15} &{0.39}&{0.65}  \\
    {TCM+DM}      &{44.96}&{10.71}&{0.07}&{0.08}&{0.14}&{0.29}  \\
    {SegPIC+SCEIR}      &{83.50}&{0.14}&{0.04}&{0.06} &{0.37}&{0.47}  \\
    \hline
    {HCLR+MLIC}&{65.76}&{4.87}&{0.11}&{0.15} &{0.39}&{0.65}  \\
    {DM+TCM}      &{44.96}&{10.71}&{0.07}&{0.07} &{0.14}&{0.28}  \\
    {SCEIR+SegPIC}      &{83.50}&{0.14}&{0.05}&{0.06} &{0.35}&{0.46} \\
    \cline{1-7}
   {Proposed}      &\multicolumn{2}{c|}{32.86}&{2.73}&{1.30} &{---}&{4.03} \\
    \hline
   \end{tabular}
\end{center}
\end{table*}

\begin{table}[t]\caption{BL and EL Comparison} \label{table2}
\footnotesize
\begin{center}
    \begin{tabular}{|c|c|c|c|c|}
    \cline {1-5} 
    \multirow{2}{*}{$\lambda$}&\multicolumn{2}{c|}{BL}&\multicolumn{2}{c|}{EL}\\
    \cline {2-5}
    &{bpp $\downarrow$}&{UIQM $\uparrow$}&{bpp $\downarrow$}&{UIQM $\uparrow$}\\
    \hline
    \hline
    {4}      &0.031&2.753&0.131&2.893\\
    {8}      &0.035&2.799&0.193&2.948\\
    {16}     &0.041&2.872&0.285&3.016\\
    {32}     &0.060&2.922&0.441&3.067\\
    {64}     &0.071&2.967&0.595&3.066\\
    {128}    &0.089&3.030&0.870&3.079\\
    {256}    &0.099&3.035&1.098&3.089\\
    \hline
   \end{tabular}
\end{center}
\end{table}

\subsection{Enhanced Dictionary Evaluation}
Moreover, the performance of enhanced dictionary is evaluated. 500 blocks with size of $16\times16$ are randomly collected from EUVP-Dark dataset. Suppose the current block is $\textbf{I}_c$, the enhanced version of ground truth is $\textbf{G}_c$, the sparse coefficient $\alpha_c^*$ can be written as follows,
\begin{equation}
\alpha_c^* \leftarrow \mathop{\arg \min}\limits_{\alpha}\Biggl\{ ||\textbf{I}_c - \textbf{D}_1\alpha||^2 + \eta ||\alpha||_1 \Biggr\},
\end{equation}
where $\eta$ is employed to control the number of non-zero coefficients and randomly selected from the range from 16 to 128 in this experiment. The performance gain can be calculated as follows,
\begin{equation}
\Delta \mathrm{PSNR} = 10 \log_{10} \left( \frac{\mathbb{MSE}(\mathbf{G}_c, \mathbf{D}_1 \boldsymbol{\alpha}_c^*)}{\mathbb{MSE}(\mathbf{G}_c, \mathbf{D}_2 \boldsymbol{\alpha}_c^*)} \right),
\end{equation}
where $\mathbb{MSE}$() indicates the Mean Squared Error (MSE) value calculation. 

\begin{figure}[!t]
\centering
\includegraphics[width=0.45\textwidth]{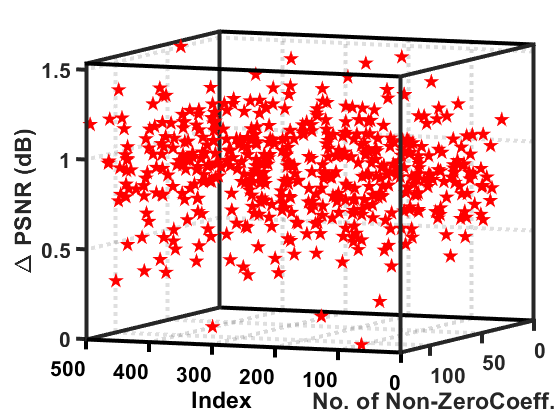}
\caption{Enhanced dictionary evaluation. }
\label{enhanced_dict}
\end{figure}

Fig. \ref{enhanced_dict} presents the results of enhanced dictionary evaluation. The x axis is the index of these 500 blocks, the y axis is the number of non-zero coefficients, while the z axis is the value of $\Delta \mathrm{PSNR}$. It can be observed that most of the values of $\Delta \mathrm{PSNR}$ are greater than 0.5 dB, which means that the sparse reconstruction is more close to the enhanced version of ground truth with the shared sparse coefficients.

\section{Conclusions}

This work aims to address the challenges of transmission bandwidth limitation and severe distortion in underwater imaging, enabling simultaneous compression and enhancement. In the BL, underwater images are efficiently represented by controllable number of non-zero sparse coefficients, achieving significant coding bit savings. Crucially, an enhancement dictionary derived from these shared sparse coefficients ensures that the sparse reconstruction inherently approximates an enhanced version of underwater image. The EL further improves compression efficiency by effective removal of residual redundancies and enhancement of the final reconstruction with dual-branch filter.

Comprehensive evaluations across five large-scale underwater image datasets demonstrate the superiority of the proposed framework. This approach provides a practical and efficient solution for applications demanding high-fidelity underwater visual data transmission and analysis, such as marine exploration and ecological monitoring.

\bibliographystyle{ACM-Reference-Format}
\bibliography{sample-base}

\end{document}